\documentclass [12 pt, one column]{IEEEtran}
\usepackage{mathtools}
\usepackage{amsmath}
\usepackage{amssymb}
\usepackage{amsfonts} 
\usepackage{dsfont}
\usepackage{enumitem} 
\usepackage{amsthm}
\usepackage{marvosym}
\usepackage{setspace}
\usepackage[normalem]{ulem}
\usepackage[T1]{fontenc}
\usepackage{filecontents}
\usepackage{bm}

\def\proveBCRB{}

\usepackage{array}
\newcolumntype{P}[1]{>{\centering\arraybackslash}p{#1}}
\newcolumntype{M}[1]{>{\centering\arraybackslash}m{#1}}

\doublespace
\allowdisplaybreaks

\def\IncludeExplanation{}

\newcommand{\R}{\mathbb{R}}

\newcommand{\C}{\mathbb{C}}
\newcommand{\E}{\mathbb{E}}
\newcommand{\N}{\mathbb{N}}
\newcommand{\Pb}{\mathbb{P}}
\newcommand{\G}{\mathbb{G}}

\newcommand{\Mellin}{\mathcal{M}}
\newcommand{\Laplace}{\mathcal{L}}
\newcommand{\LaplaceL}{\mathcal{L}^{\text{LB}}}
\newcommand{\LaplaceU}{\mathcal{L}^{\text{UB}}}

\newcommand{\indicator}{\mathds{1}}
\newcommand{\indep}{\perp \!\!\! \perp }

\newcommand{\norm}[1]{\left\lVert#1\right\rVert}
\newcommand{\qedwhite}{\hfill \ensuremath{\Box}}
\newcommand{\eqdist}{\stackrel{\mathclap{\normalfont\mbox{d}}}{=}}
\newcommand{\eqlabel}[1]{\stackrel{\mathclap{\normalfont\mbox{(#1)}}}{=}}
\newcommand{\lelabel}[1]{\stackrel{\mathclap{\normalfont\mbox{(#1)}}}{\le}}
\newcommand{\gelabel}[1]{\stackrel{\mathclap{\normalfont\mbox{(#1)}}}{\ge}}

\newcommand{\PhiB}{\Phi_\text{B}}
\newcommand{\PiB}{\Pi_\text{B}}
\newcommand{\PiBa}{\Pi_\text{B,1}}
\newcommand{\PiBb}{\Pi_\text{B,2}}
\newcommand{\PsiB}{\Psi_\text{B}}
\newcommand{\PiBt}{\tilde{\Pi}_\text{B}}
\newcommand{\PhitB}{\tilde{\Phi}_\text{B}}
\newcommand{\PhiL}{\Phi_\text{L}}
\newcommand{\PhiN}{\Phi_\text{N}}
\newcommand{\PiLa}{\Pi_\text{L,1}}
\newcommand{\PiNa}{\Pi_\text{N,1}}
\newcommand{\PiLb}{\Pi_\text{L,2}}
\newcommand{\PiNb}{\Pi_\text{N,2}}

\newcommand{\PhiU}{\Phi_\text{U}}
\newcommand{\PhiS}{\Phi_\text{S}}
\newcommand{\XiB}{\Psi_\text{block}}

\newcommand{\lambdaB}{\lambda_\text{B}}
\newcommand{\lambdaBa}{\lambda_\text{B,1}}
\newcommand{\lambdaBb}{\lambda_\text{B,2}}

\newcommand{\lambdaU}{\lambda_\text{U}}
\newcommand{\lambdaS}{\lambda_\text{S}}
\newcommand{\lambdaL}{\lambda_\text{L}}
\newcommand{\lambdaN}{\lambda_\text{N}}
\newcommand{\lambdaLa}{\lambda_\text{L,1}}
\newcommand{\lambdaNa}{\lambda_\text{N,1}}
\newcommand{\lambdaLb}{\lambda_\text{L,2}}
\newcommand{\lambdaNb}{\lambda_\text{N,2}}

\newcommand{\rhoBa}{\rho^0_\text{B,1}}
\newcommand{\rhoBb}{\rho^0_\text{B,2}}
\newcommand{\rhoLa}{\rho^0_\text{L,1}}
\newcommand{\rhoLb}{\rho^0_\text{L,2}}
\newcommand{\rhoNa}{\rho^0_\text{N,1}}
\newcommand{\rhoNb}{\rho^0_\text{N,2}}
\newcommand{\rhoAk}{\rho^0_\text{a,k}}

\newcommand{\rhoLk}{\rho^0_\text{L,k}}
\newcommand{\rhoNk}{\rho^0_\text{N,k}}

\newcommand{\nuBa}{\nu^0_\text{B,1}}
\newcommand{\nuBb}{\nu^0_\text{B,2}}
\newcommand{\nuLa}{\nu^0_\text{L,1}}
\newcommand{\nuLb}{\nu^0_\text{L,2}}
\newcommand{\nuNa}{\nu^0_\text{N,1}}
\newcommand{\nuNb}{\nu^0_\text{N,2}}
\newcommand{\nuAk}{\nu^0_\text{a,k}}

\newcommand{\nuLk}{\nu^0_\text{L,k}}
\newcommand{\nuNk}{\nu^0_\text{N,k}}

\newcommand{\Ts}{T_\text{s}}
\newcommand{\Tp}{T_\text{CPI}}

\newcommand{\Tofdm}{T_\text{MC}}

\newcommand{\Rc}{C_\text{com}}
\newcommand{\Res}{C_\text{rad}}
\newcommand{\Rest}{\tilde{C}_\text{rad}}
\newcommand{\Resub}{C_\text{rad}^{UB}}

\newcommand{\Pc}{P_{\text{c, JCAS}}}
\newcommand{\Pcr}{P_{\text{c, rad}}}
\newcommand{\Pcru}{P_{\text{c, rad}}^{\text{UB}}}
\newcommand{\Pcrl}{P_{\text{c, rad}}^{\text{LB}}}

\newcommand{\Pcc}{P_{\text{c, com}}}
\newcommand{\Pccu}{P_{\text{c, com}}^{\text{UB}}}
\newcommand{\Pccl}{P_{\text{c, com}}^{\text{LB}}}

\newcommand{\Ec}{E_{\text{c, JCAS}}}
\newcommand{\Ecc}{E_{\text{c, com}}}
\newcommand{\Ecr}{E_{\text{c, rad}}}
\newcommand{\sigmaN}{\sigma_{\text{N}}}

\newcommand{\SNRc}{\normalfont{\text{\fontfamily{cmss}\selectfont  SNR}_\text{com}}}
\newcommand{\SNRr}{\normalfont{\text{\fontfamily{cmss}\selectfont  SNR}_\text{rad}}}

\newcommand{\SINRc}{\normalfont{\text{\fontfamily{cmss}\selectfont  SINR}_\text{com}}}
\newcommand{\SIRc}{\normalfont{\text{\fontfamily{cmss}\selectfont  SIR}_\text{com}}}

\newcommand{\SINRr}{\normalfont{\text{\fontfamily{cmss}\selectfont  SINR}_\text{rad}}}
\newcommand{\SINRt}{\normalfont{\text{\fontfamily{cmss}\selectfont  SINR}_\text{typ}}}

\newcommand{\SINR}{\normalfont{\text{\fontfamily{cmss}\selectfont  SINR}}}

\newcommand{\SINRam}{\normalfont{\text{\fontfamily{cmss}\selectfont  SINR}_\text{AM}}}
\newcommand{\SINRgm}{\normalfont{\text{\fontfamily{cmss}\selectfont  SINR}_\text{GM}}}
\newcommand{\SIRgm}{\normalfont{\text{\fontfamily{cmss}\selectfont  SIR}_\text{GM}}}
\newcommand{\SINRhm}{\normalfont{\text{\fontfamily{cmss}\selectfont  SINR}_\text{HM}}}
\newcommand{\SINRgen}{\normalfont{\text{\fontfamily{cmss}\selectfont  SINR}_\text{gen}}}

\newcommand{\Ila}{I_{\text{L,1}}}
\newcommand{\Ina}{I_{\text{N,1}}}
\newcommand{\Ilb}{I_{\text{L,2}}} 
\newcommand{\Inb}{I_{\text{N,2}}}

\newcommand{\Gbt}{G_\text{B,Tx}}
\newcommand{\Gbr}{G_\text{B,Rx}}

\newcommand{\Gur}{G_\text{U,Rx}}

\newcommand{\thetabt}{\theta_\text{B,Tx}}
\newcommand{\thetabr}{\theta_\text{B,Rx}}

\newcommand{\thetaur}{\theta_\text{U,Rx}}

\newcommand{\xibt}{\xi_\text{B,Tx}}
\newcommand{\xibr}{\xi_\text{B,Rx}}

\newcommand{\xiur}{\xi_\text{U,Rx}}

\newcommand{\pB}{p_\text{B}}
\newcommand{\pBr}{p_\text{B,Rx}}

\newcommand{\Sr}{\bold{S}_\text{rad}}
\newcommand{\Srt}{\bold{\tilde{S}}_\text{rad}}
\newcommand{\Sc}{\bold{S}_\text{com}}

\newcommand{\pl}{p_\text{LoS}}

\newcommand{\gammaL}{\gamma_\text{L}}
\newcommand{\gammaN}{\gamma_\text{N}}
\newcommand{\alphaL}{\alpha_\text{L}}
\newcommand{\alphaN}{\alpha_\text{N}}
\newcommand{\KL}{K_\text{L}}
\newcommand{\KN}{K_\text{N}}
\newcommand{\Llos}[1]{\KL #1^{-\alphaL} e^{-\gammaL #1}}
\newcommand{\Lrad}[1]{\frac{\KL}{4\pi} #1^{-2\alphaL} e^{-2\gammaL #1}}
\newcommand{\Lnlos}[1]{\KN #1^{-\alphaN} e^{-\gammaN #1}}

\newcommand{\gL}{g_\text{L}}
\newcommand{\ga}{g_\text{a}}
\newcommand{\gLr}{g_\text{L,ret}}
\newcommand{\gN}{g_\text{N}}

\newcommand{\fa}{f_\text{a}}

\newcommand{\NL}{N_\text{L}}

\newcommand{\NN}{N_\text{N}}
\newcommand{\Na}{N_\text{a}}

\newcommand{\kc}{\kappa_\text{CS}}

\newcounter{saveenum}

\newcommand{\nuc}{\nu_\text{com}}
\newcommand{\nur}{\nu_\text{rad}}

\newcommand{\sinc}{\text{sinc}}
\newcommand{\lest}{\le_{\text{s.t.}}}

\newcommand{\HL}{H_\text{LB}}
\newcommand{\HU}{H_\text{UB}}

\newcommand{\lambdaEQ}{\lambda_\text{EQ}}

\newcommand{\LambdaEQ}{\Lambda_\text{EQ}}

\newcommand{\AM}{\normalfont{\text{\fontfamily{cmss}\selectfont  AM}}}
\newcommand{\GM}{\normalfont{\text{\fontfamily{cmss}\selectfont  GM}}}
\newcommand{\HM}{\normalfont{\text{\fontfamily{cmss}\selectfont  HM}}}

\begin{document}
\bstctlcite{IEEEexample:BSTcontrol}

\author{Nicholas~R.~Olson,
       Jeffrey~G.~Andrews,
      and~Robert~W.~Heath, Jr.
\thanks{N. R. Olson and J. G. Andrews are with 6G@UT and WNCG at The University of Texas at Austin, Austin, TX, USA (email: nolson@utexas.edu, jandrews@ece.utexas.edu). R. W. Heath Jr. is with North Carolina State University, Raleigh, North Carolina, USA (email: rwheathjr@ncsu.edu). }}

\title{Coverage and Rate of Joint Communication and Parameter Estimation in Wireless Networks}

\maketitle 

\begin{abstract}
From an information theoretic perspective, joint communication and sensing (JCAS) represents a natural generalization of communication network functionality. However, it requires the re-evaluation of network performance from a multi-objective perspective. We develop a novel mathematical framework for characterizing the sensing and communication coverage probability and ergodic rate in JCAS networks. We employ a formulation of sensing parameter estimation based on mutual information to extend the notions of coverage probability and ergodic rate to the radar setting. We define sensing coverage probability as the probability that the rate of information extracted about the parameters of interest associated with a typical radar target exceeds some threshold, and sensing ergodic rate as the spatial average of the aforementioned rate of information. Using this framework, we analyze the downlink sensing and communication coverage and rate of a mmWave JCAS network employing a shared  waveform, directional beamforming, and monostatic sensing. Leveraging tools from stochastic geometry, we derive upper and lower bounds for these quantities. We also develop several general technical results including: i) a generic method for obtaining closed form upper and lower bounds on the Laplace Transform of a shot noise process, ii) a new analog of H{\"o}lder's Inequality to the setting of harmonic means, and iii) a relation between the Laplace and Mellin Transforms of a non-negative random variable. We use the derived bounds to numerically investigate the performance of JCAS networks under varying base station and blockage density. Among several insights, our numerical analysis indicates that network densification improves sensing SINR performance -- in contrast to communications.

\end{abstract}



\section{Introduction}
\label{sec:intro}


There is increasing interest in leveraging communication networks to provide the additional services of user localization and radar sensing -- a concept termed \emph{joint communication and sensing} or {\em JCAS}. The defining feature of JCAS networks is that the network is co-designed to perform the dual functions of communication and sensing: network transceivers, spectrum, and even waveforms are used, potentially simultaneously, to communicate with user equipment (UEs) and to detect, locate, and track objects of interest (which we refer to as {\em sensed objects} or SOs). Armed with these additional services, JCAS networks could enable precision navigation in urban environments, monitor activity in a given coverage area, provide collision avoidance services to autonomous vehicles, enhance remote automation, and facilitate AR/VR applications \cite{DeLima2021}. Moreover, the environmental information obtained through sensing could improve communication performance and reliability by facilitating channel estimation, beam alignment, and user tracking \cite{Liu2021}. One may view these important communication network functions as forms of sensing themselves.

Through the introduction of a parallel sensing objective, JCAS networks require the reconsideration of nearly every network layer from a multi-objective perspective. Waveforms and antenna array codebooks must now be designed to  efficiently convey data for communication and to provide sufficient detection and tracking performance for sensing \cite{Kumari2021}. Scheduling at the MAC layer must now not only consider tradeoffs with respect to traffic flows among UEs, but also tradeoffs with respect to sensing coverage for SOs \cite{Husheng22}. Likewise, network deployments and protocols must be designed with the performance of both functions in mind. One of the key challenges inherent in this design problem is jointly accounting for and mitigating the effects of intercell interference. Developing tractable models which capture the impact of this phenomena on both functions and allow for insight into tradeoffs with respect to each is an important step in addressing this issue. 

\ifdefined\IncludeExplanation
This phenomena has been adeptly addressed in the setting of wireless communication networks by  characterizing notions of network coverage probability and ergodic rate using stochastic geometry. Inspired by this, to address the intercell interference issue in the JCAS setting, we take a macroscopic view and seek to quantify JCAS performance through the lens of coverage and rate. That is, we seek to address the questions {\em ``What fraction of UEs and SOs achieve satisfactory performance?''} (i.e. the coverage probability), and {\em ``What is the average performance of all UEs and SOs?''} (i.e. the ergodic performance).

In particular we focus on downlink communication and parameter estimation via monostatic radar sensing, and employ performance characterizations thereof based on information theoretic objects. Our restriction to parameter estimation, the objective of radar tracking, as opposed to detection is motivated by its lack of study in related prior work, and that, in our view, it is the more challenging task. For communication, we quantify performance via the Shannon rate of the links for each UE. Thus, we consider a UE to be covered if its rate is satisfactorily high, and define the ergodic performance as the spatial average of these rates. For the parameter estimation objective, we exploit a metric based on the mutual information between the measured returns of the SOs and their associated parameters of interest. Dividing this quantity by the time taken to perform the measurement, one obtains an analogous sensing rate: the rate of information gained about the parameters of interest of an SO via a measurement procedure. This sensing rate metric allows for the natural generalization of the concepts of the coverage probability and ergodic rate to the sensing objective. These, in tandem with the communication metrics, lead to a precise notion of JCAS coverage probability and ergodic rate.

\else
Stochastic geometry provides a powerful mathematical framework with which to address network-wide performance. Typically, analysis within this framework is done from the perspective of a snapshot in time representative of temporally stationary network behavior. Then, certain network-wide performance metrics may be obtained using spatial averages. Such an approach allows one to address questions such as {\em "What fraction of users achieve performance above some threshold at a representative instant in time?"} and {\em "What is the average performance among all users at a representative instant in time?"}. Answering similar questions in the setting of JCAS networks would allow one to ascertain the impact of key network parameters and intercell interference on the joint performance of communication and sensing and thereby facilitate network co-design \cite{Zhang2021}. Given the utility of stochastic geometry in providing tractable characterizations of communication network performance -- particularly with respect to issues regarding intercell interference -- we endeavor to extend these methods to characterize the performance of communication and parameter estimation in JCAS networks from a similar perspective. In light of the preceding questions this first requires a precise notion of sensing (and consequently JCAS) performance. We argue for a metric based on the mutual information between the radar return and parameters of interest associated with an SO. This allows for generalizations of the notions of coverage probability and ergodic rate to be extended to sensing.

We note that we are not seeking to establish fundamental performance limits of JCAS networks. Such results would require solving and extending longstanding open problems in network information theory. Rather, we aim to provide a framework for addressing certain notions of JCAS network performance under a set of modeling criteria, which may or may not be optimal. Moreover, we specifically focus on the performance of parameter estimation, or {\em radar tracking}\footnote{In the parlance of radar terminology, this refers to the process of parameter estimation that typically follows some detection procedure.}, as opposed to detection in JCAS networks. Arguably, this process is of increased significance in a network setting. 
\fi

\subsection{Prior Work}

Joint communication and sensing, sometimes referred to as joint radar and communication (JRC), integrated sensing and communication (ISAC), or dual function radar and communication (DFRC), has emerged as a promising potential function for future cellular networks \cite{DeLima2021}, \cite{Wild2021}. In depth surveys of prior work, implementation approaches, and network integration issues may be found in \cite{Liu2021}, \cite{Nguyen2021}, \cite{Zhang2021}. While much of the prior work in this area has focused on signal processing and waveform design issues, for instance in \cite{Kumari2018}, our focus is on network wide performance analysis. Stochastic geometry has been widely used for the analysis of wireless networks in a variety of settings. Notably in \cite{Andrews2011} for the analysis of coverage and rate in cellular networks. This was extended in \cite{Bai2015}, for the analysis of mmWave cellular networks upon which we base some of our system model.

With respect to the analysis of JCAS in wireless networks within a stochastic geometry framework, there have been relatively few works. In \cite{Ping2018}, the authors characterize the performance of radar range detection and communication coverage probability using a time multiplexed system in an ad hoc network. Certain aspects of their analysis of this setting are extended in \cite{Ren2020}. In the setting of an indoor network, \cite{Ram2022} characterizes the detection performance of a radar system amidst clutter, but without considering interference, that is time multiplexed with a communication system. In \cite{Ghozlani2021}, the radar detection performance in a vehicular network employing a shared waveform for communication and sensing is characterized. In \cite{Fang2019}, a radar network is considered which leverages sensing waveforms for communication. The detection probability is characterized when multiple terminals share information in addition to the communication coverage probability. Focusing on exclusively on radar performance, in \cite{Munari2018} the authors characterize the impact of network interference on radar detection and false alarm rate in an ad hoc network. In a similar vein, the performance of a sensor network to detect blockages is characterized in \cite{Park2018}. Finally, in \cite{Skouroumounis2021} the detection performance of a heterogenous cellular network is considered in which the network access is split between radar and communications functions. Performance analysis is conducted for a variety of cooperative methods in which the individual detection hypotheses of multiple receivers are fused according to some hard decision rules. 

While these works offer some insight into JCAS performance, they employ somewhat simplified and/or limited models of radar detection and do not address the performance of the parameter estimation problem inherent in radar tracking. Indeed, many of these models fail to account directly for the impact of the radar waveform on performance and instead simply study a narrowband SINR model at a specific time slot as a proxy for sensing performance.


\subsection{Contributions and Summary}

To the best of our knowledge, our work is the first to present a rigorous analytical framework with which to characterize the joint performance of communication and parameter estimation in JCAS networks. In Sec. \ref{sec:concept_model}, we develop a notion of sensing performance based on the mutual information between the radar return, $Y$, and the SO's parameters of interest, $\Theta$. As summarized earlier, this allows for the natural generalization of the concepts of coverage probability and ergodic rate to be applied to sensing, while still maintaining close correspondence with the more traditional estimation theoretic metric of error covariance. We note further, that the application of mutual information to radar has been widely employed in prior work. Such an approach was first proposed by \cite{Woddward1951} and later extended by \cite{Bell1993}. Recently, it has been used to study waveform design and rate bounds for joint radar and communication in \cite{Bliss2014, Paul2015, Chryath2017, Ouyang23}.

Unlike communications, the sensing mutual information is typically intractable. To address this issue, we establish that the sensing mutual information is approximately lower bounded in terms of the Fisher Information Matrix (FIM), $\boldsymbol{\mathcal{J}}(\Theta)$,  as
\begin{align}
I(Y; \Theta) \gtrsim \frac{1}{2}\log\left(\left \lvert \mathbf{I} + \mathbf{Q}^{\frac{1}{2}} \E_{\Theta}\left[\boldsymbol{\mathcal{J}}(\Theta) \right]\mathbf{Q}^{\frac{1}{2}} \right \rvert\right) - c.
\end{align}
Where $\mathbf{Q}$ is the covariance of $\Theta$ under the reference prior, and $c$ is a non-negative constant depending only on the differential entropy of the prior for $\Theta$ scaled to have identity covariance. Focusing on the setting of a shared multi-carrier waveform, we further establish that the log-determinant term admits upper and lower bounds of the form
\begin{align}
	\frac{1}{2}\log\left(1 + G ~ \SINRr \right) \le \frac{1}{2}\log\left(\left \lvert \mathbf{I} + \mathbf{Q}^{\frac{1}{2}} \E_{\Theta}\left[\boldsymbol{\mathcal{J}}(\Theta) \right]\mathbf{Q}^{\frac{1}{2}} \right \rvert\right) \le \log\left(1 + \frac{G}{2} ~ \SINRr \right).
\end{align}
Where $G$ is a constant derived from the FIM, and $\SINRr$ is an average of the SINRs over the resource elements employed for sensing. We argue that these bounds imply that $\SINRr$ may be used as one would the communication SINR, $\SINRc$, to characterize the coverage and rate performance of parameter estimation. Therefore, we equivalently characterize the JCAS coverage probability as the {\em joint} fraction of UEs and SOs whose corresponding SINR is above some corresponding threshold and the JCAS ergodic rate as the {\em joint} spatial average of the corresponding rate functions.


Leveraging the communication and sensing rate functions, we finally establish that, when the UEs follow a stationary, ergodic point process, $\PhiU$, with intensity $\lambdaU$, and the SOs follow an independent stationary, ergodic point process, $\PhiS$, with intensity, $\lambdaS$, the JCAS coverage probability may be expressed as
\begin{align}
	&\Pc(\tau_{\rm com}, \tau_{\rm rad}) = \frac{\lambdaU}{\lambdaU + \lambdaS}\Pb_{\PhiU}^0(\SINRc \ge \tau_{\rm com}) + \frac{\lambdaS}{\lambdaU + \lambdaS} \Pb_{\PhiS}^0(\SINRr \ge \tau_{\rm rad}),
	\label{eq:cov_prob_overview}
\end{align}
where $\Pb_{\PhiU}^0$ and $\Pb_{\PhiS}^0$ denote the Palm measures associated with $\PhiU$ and $\PhiS$. Similarly the JCAS ergodic rate may be expressed as
\begin{align}
	&\Ec = \frac{\lambdaU}{\lambdaU + \lambdaS}\E^0_{\PhiU}\left[\log(1 + \SINRc)\right] +\frac{\lambdaS}{\lambdaU + \lambdaS}\E^0_{\PhiS}\left[\frac{k}{2}\log\left(1 + \frac{G}{k} ~ \SINRr \right)\right] &k \in \{1, 2\}.
	\label{eq:erg_cap_overview}
\end{align}
Hence, even though the SINR models arise from a network in which communication and sensing are performed simultaneously (and thereby strongly coupled), it suffices to analyze the performance of each function separately in characterizing network-wide JCAS coverage and rate performance.

Having formally developed our notion of JCAS coverage probability and ergodic rate, in Sec. \ref{sec:system_mod} we detail a system model for a mmWave JCAS network performing downlink communication and monostatic sensing of doppler and delay  using a shared multi-carrier waveform. From this model, we induce stochastic expressions for $\SINRc$ and $\SINRr$ with respect to the typical UE and SO. Using these, in Sec. \ref{sec:prelim_res} through Sec. \ref{sec:pcc_analysis} we establish a series of novel results that outline an approach to obtain integral closed form upper and lower bounds and approximations for the JCAS coverage and rate of the network. We detail preliminary results pertaining the development of these bounds in Sec. \ref{sec:prelim_res}. We then detail their derivations in Sec. \ref{sec:pcr_analysis} and Sec. \ref{sec:pcc_analysis}, respectively. Through the course of our analysis, we develop some intermediate results that are of independent interest. These include a generic method for obtaining closed form upper and lower bounds on the Laplace Transform of a shot noise process, a new analog of H{\"o}lder's Inequality to the setting of harmonic means, and a relation between the Laplace and Mellin Transforms of a non-negative random variable. 

Finally, we present a numerical case study of JCAS networks in Sec. \ref{sec:num_res}. Notably, our analysis indicates that the sensing SINR strictly improves with base station density -- in contrast to the communication SINR. We additionally demonstrate that interference has a more pronounced impact on the performance of sensing than communication. We summarize the main conclusions of the paper in Sec. \ref{sec:conclusion}, and end the main body with a discussion of future work that builds off of our framework in Sec. \ref{sec:future-work}. Our analysis requires several intermediate results, the proofs and discussions of which we relegate to Appendices \ref{app:sens-sig-mod} through \ref{app:tech_res} at the end of the paper.


\subsection{Notation}

We employ the following notation. The sets, $\N$, $\R$, $\R_+$, and $\C$ denote the natural numbers, real numbers, non-negative real numbers, and complex numbers, respectively. Deterministic scalars or vectors are denoted by lower case letters and random variables or matrices by upper case letters. 
For a random variable $X$, we denote its Laplace Transform as $\Laplace_{X}(s) = \E[e^{-sX}]$ and its Mellin Transform as $\Mellin_X(p) = \E[X^{p-1}]$. For random variables $X$ and $Y$, the relation $X \indep Y$ denotes their independence. 
The symbol $\lest$ denotes stochastic dominance: $X \lest Y$ if $\Pb(X \ge \tau) \le \Pb(Y \ge \tau)$ for all $\tau \in \R$. For a measure $\nu$ over $\G$, its Laplace Transform is denoted as $\Laplace_{\nu}(s) = \int_{\G}e^{-sx}\nu(dx)$ and its Mellin Transform $\Mellin_{\nu}(p) = \int_{\G}x^{p-1}\nu(dx)$. The inverse Mellin Transform is denoted as $\Mellin^{-1}\{ \cdot \}$. For some function $f: \G \rightarrow \G'$, $\nu \circ f$ denotes image of $\nu$ by $f$. For $x \in \G$, $S_x$ denotes the shift operator on $\nu$. That is, for some measurable set, $A$, $S_x \nu(A) = \nu(A + x)$. 
For $k \in \N$, $[k]$ denotes the interval $\{1, \dots, k\}$. $\indicator\{ \cdot \}$ denotes the indicator function. It returns one if the specified condition is true, and zero otherwise. $\Gamma(s)$ denotes the Gamma function. Finally, the speed of light is denoted as $c_0$.

\section{Characterization of JCAS Performance Metrics}
\label{sec:concept_model}

Before enumerating the complete attributes of our system model, we first develop the communication and sensing rate functions in more detail and rigorously characterize our notion of JCAS coverage probability and ergodic rate. 
We restrict our focus to the setting where multi-carrier waveform is simultaneously used communication and sensing, and hence develop the aforementioned rate functions with this in mind. Complete details of the multi-carrier signal model, and how it is used for communication and parameter estimation are discussed in Appendix \ref{app:sens-sig-mod}. 

Generally, as a consequence of scheduling considerations and related tradeoffs in the temporal domain, over a period of $N_{\rm s}$ symbols with $N_{\rm c}$ available sub-carriers a particular UE is allocated some subset of the multi-carrier resource elements -- say $\Sc \in \{0, 1\}^{N_{\rm s} \times N_{\rm c}}$. Similarly, a particular SO receives measurement effort on some potentially non-disjoint subset of these resource elements -- say $\Sr$. As scheduling considerations are beyond the scope of our work, we treat these as fixed quantities. The communication and sensing rate functions are characterized in terms of these matrices, and hence the coverage and rate results should be interpreted as conditional upon certain prior allocation decisions. As will be discussed, the resource-element allocation matrices play an important role in the formulation of sensing rate in particular.

\subsection{Communication Performance Metrics}
	
	In the downlink setting, base stations (BSs) in the network transmit data to UEs using the multi-carrier waveform. Consider a particular UE that is scheduled to receive data on some subset of resource elements, $\Sc$. Let $\SINR_{\rm m, n}$ denote the SINR on resource element $(m,n) \in \Sc$ and $T_{\rm MC}$ denote the multi-carrier symbol period. Then, treating the interference as noise, the Shannon capacity of the link in bits per second for a fixed power allocation is (following \cite{Bolcskei2002})
\begin{align}
&\Rc = \frac{1}{T_{\rm MC} \lvert \{m : (m,\cdot) \in \Sc \} \rvert} \sum_{(m,n) \in \Sc}\log_2(1 + \SINR_{\rm m,n}).
\end{align}
Note that $T_{\rm MC} \Rc$ is the spectral efficiency of the link, which we refer to as {\em communication efficiency} hereafter.

$\Sc$ has the principal effect of modulating the set of available sub-channels for communication. Following \cite{Ahsan2019} and \cite{Lin2015}, we employ a notion of coverage probability defined with respect an {\em arbitrary} resource element, which abstracts the impact of $\Sc$. Such an approach holds under the reasonable condition that $\{\SINR_{\rm m,n}\}_{(m,n) \in \Sc}$ are probabilistically stationary over resource elements -- which does not precluded some underlying adaptivity between communication and sensing. Therefore, we take $\SINRc = \SINR_{\rm m,n}$ (for some arbitrary $(m,n) \in \Sc$) the communication coverage probability reduces to
\begin{align*}
	\Pcc(\tau) = \Pb^0_{\PhiU}\left(\SINRc \ge \tau \right).
\end{align*}
Similarly, the communication ergodic rate is simply the spatial average of the channel capacity normalized by the number of resource elements employed
\begin{align}
	\Ecc = \frac{1}{T_{\rm MC} \lvert \Sc \rvert} \sum_{(m,n) \in \Sc} \E^0_{\PhiU} \left[\log_2\left(1 + \SINR_{\rm m,n} \right) \right] = \frac{1}{T_{\rm MC}} \E^0_{\PhiU} \left[\log_2\left(1 + \SINRc \right) \right].
\end{align}
Where, equality follows from the stationarity of $\{\SINR_{\rm m,n}\}_{(m,n) \in \Sc}$.

\subsection{Development of Approximate Lower Bound on the Sensing Rate}

In a similar vein, one may take an analogous, information theoretic view of radar tracking. At a high level, mutual information in this context provides a metric with which to characterize the information gain provided by a measurement for the purposes of parameter estimation. Simultaneous with communication, over the coherent processing interval $T_{\rm CPI}$, BSs in the network monitor the return of the transmitted waveform, $Y$, to track the parameters of interest, $\Theta$, associated with SOs in their vicinity. For an arbitrary SO, about which the network's prior belief regarding its parameters of interest is the distribution $\Pb_{\Theta}$, the informativeness of this sensing procedure -- irrespective of the tracking filter or estimator employed -- may be expressed in terms of the mutual information $I(Y; \Theta)$. Dividing this by $T_{\rm CPI}$ we obtain the rate of information gain or {\em sensing rate} as defined in \cite{Paul2015} and \cite{Chryath2017}
\begin{align}
 \Res = \frac{I(Y; \Theta)}{T_{\rm CPI}}.
 \label{eq:r_est_def}
\end{align}


Ideally, we would use the mutual information directly to define notions of coverage probability and ergodic rate for sensing. However, in general, an explicit characterization (\ref{eq:r_est_def}) is intractable. Thus we resort to bounds and approximations. To that end, we exploit the following lower bound.

{\bf Proposition 1: Lower Bound for Mutual Information using Minimum Mean Square Error (MMSE) Covariance.} {\em Let $Y$ and $\Theta$ be random elements defined on a common probability space, and let $\Theta$ take values $\R^{n}$. Let the distribution of $Y$ and $\Theta$, $\Pb_{Y, \Theta}$, be such that the conditional distribution of $\Theta$ given $Y$, $\Pb_{\Theta \vert Y}$, and the marginal distribution of $\Theta$, $\Pb_{\Theta}$, admit a density with respect to the Lebesgue measure over $\R^{n}$ $\Pb_{Y, \Theta}$-almost surely. 
Additionally let the MMSE covariance be denoted as
	{\em \begin{align}
	\mathbf{R}_{\rm MMSE} = \E\left[ \left(\Theta - \E[\Theta \vert Y] \right)\left(\Theta - \E[\Theta \vert Y] \right)^{\rm T}\right],
	\end{align}}
	and the marginal covariance of $\Theta$ be denoted as
	{\em \begin{align}
	\mathbf{Q} = \E\left[ \left(\Theta - \E[\Theta] \right)\left(\Theta - \E[\Theta] \right)^{\rm T}\right],
	\end{align}}
Then the mutual information $I(Y; \Theta)$ is lower bounded as
\begin{align}
 I(Y; \Theta) \ge \frac{1}{2} \log\left(\left \lvert \mathbf{Q}^{\frac{1}{2}} \mathbf{R}^{-1}_{\rm MMSE}\mathbf{Q}^{\frac{1}{2}} \right \rvert \right) - \left(\frac{n}{2} \log(2 \pi e) - h\left(\mathbf{Q}^{\frac{-1}{2}} \Theta \right) \right)
\end{align}
Where $h\left(\mathbf{Q}^{\frac{-1}{2}} \Theta \right)$ denotes the differential entropy of $\Pb_{\Theta} \circ (\mathbf{Q}^{\frac{1}{2}} u)$.}

{\em Proof:} See Appendix \ref{app:ofdm_rad}.

To further simplify this expression, we resort to approximating the MMSE covariance in terms of the Bayesian FIM. Under some mild regularity and support conditions on $\Pb_{Y, \Theta}$, $\mathbf{R}^{-1}_{\rm MMSE}$ is upper bounded by the Bayesian FIM in the positive definite sense. That is, the Bayesian Cram{\'e}r Rao Lower Bound (BCRB) holds:
\begin{align}
 \mathbf{R}^{-1}_{\rm MMSE} \le \E[\boldsymbol{\mathcal{J}}(\Theta)]	 + \E\left[\nabla_{\theta}\log(p(\Theta)) \nabla_{\theta}\log(p(\Theta))^{\rm T} \right],
\end{align}
where $\boldsymbol{\mathcal{J}}(\theta)$ is the FIM and $p(\theta)$ is the PDF of $\Theta$. For further details, see Appendix \ref{app:bcrb}.

Putting these together, we arrive at the following approximate lower bound on the sensing mutual information.
\begin{align}
I(Y; \Theta) + \frac{n}{2} &\log(2 \pi e) - h\left(\mathbf{Q}^{\frac{-1}{2}} \Theta \right) \ge \frac{1}{2} \log\left(\left \lvert \mathbf{Q}^{\frac{1}{2}} \mathbf{R}^{-1}_{\rm MMSE}\mathbf{Q}^{\frac{1}{2}} \right \rvert \right) \nonumber
\\
& \stackrel{(a)}{\approx} \frac{1}{2} \log \left( \left \lvert \mathbf{Q}^{\frac{1}{2}}\left( \E\left[\nabla_{\theta}\log(p(\Theta)) \nabla_{\theta}\log(p(\Theta))^{\rm T} \right] +  \E\left[\boldsymbol{\mathcal{J}}(\Theta) \right] \right) \mathbf{Q}^{\frac{1}{2}}\right \rvert \right)
\\
&\stackrel{(b)}{\ge} \frac{1}{2} \log \left( \left \lvert \mathbf{I} + \mathbf{Q}^{\frac{1}{2}} \E\left[\boldsymbol{\mathcal{J}}(\Theta) \right]\mathbf{Q}^{\frac{1}{2}} \right \rvert \right),
\end{align}
where (a) follows from the BCRB, and (b) follows from the fact that $\mathbf{Q}^{-1} \le \E\left[\nabla_{\theta}\log(p(\Theta)) \nabla_{\theta}\log(p(\Theta))^{\rm T} \right]$ (which is a corollary of the BCRB as stated in Proposition 4 in Appendix \ref{app:bcrb}).

In light of these considerations, we arrive at the our approximate lower bound of the sensing rate by discarding the $\left(\frac{n}{2} \log(2 \pi e) - h\left(\mathbf{Q}^{\frac{-1}{2}} \Theta \right) \right)$ term. This  term is necessarily non-positive due to the maximum entropy principle, but is invariant to the sensing channel conditions and the choice of waveform. Thus, we  may disregard it for the purposes of characterizing sensing coverage probability and ergodic rate: the addition of a constant to the sensing rate would have the effect of shifting the coverage threshold and adding a constant to the ergodic rate. Our approximate lower bound on the sensing rate is thus
\begin{align}
	\Res \gtrsim \Rest = \frac{1}{2 T_{\rm CPI}} \log_2 \left( \left \lvert \mathbf{I} + \mathbf{Q}^{\frac{1}{2}} \E\left[\boldsymbol{\mathcal{J}}(\Theta) \right]\mathbf{Q}^{\frac{1}{2}} \right \rvert \right),
	\label{eq:R_est_formula1}
\end{align}
where we have changed the logarithm to be base $2$. 

In keeping with the communication case, we refer to the quantity $\Tp \Rest$ as the {\em sensing efficiency}. We note that, although we have obtained this expression through the exploitation of information theoretic objects, it is nonetheless closely related to more conventional sensing performance metrics by inclusion of the FIM. Indeed, one may equivalently interpret this expression as a measure of the reduction of the covariance of the parameters of interest following a sensing measurement. The FIM is the key feature of interest in this expression which depends on waveform parameters (including the pulse repetition interval), other exogenous features (for instance interference and clutter), and the specific parameters of interest in question.

\subsection{Characterization of Sensing Performance Metrics}

In the particular case of tracking doppler and delay via monostatic sensing using a multi-carrier waveform, the formulation of the sensing rate in (\ref{eq:R_est_formula1}) admits upper and lower bounds in terms of an SINR-type function. We relegate a more in depth discussion of the sensing signal model to Appendix \ref{app:sens-sig-mod}, but note that the signal model for the radar return from a single target may be expressed as
\begin{align}
&(\bold{F})_{m,n} = e^{-j 2 \pi \Delta f \tau n} e^{j 2 \pi \Tofdm f_D m} + Z_{m,n}. &\bold{Z} \sim \mathcal{CN}(0, \text{diag}\{\lvert X_{m,n} \rvert^{-2} \SINR_{m,n}^{-1}\}_{(m,n) \in \Sr}),
\label{eq:ofdm_radar_est_prob}
\end{align}
where $\Theta = (\tau, f_D)$ is the doppler and delay of an arbitrary SO, $\Delta f$ is the sub-carrier spacing, $X_{m,n}$ is the data symbol corresponding to the $(m,n)^{th}$ resource element, and $\mathcal{CN}(\boldsymbol{\mu}, \mathbf{R})$ denotes the complex normal distribution with mean $\boldsymbol{\mu}$ and covariance $\mathbf{R}$. Leveraging the FIM obtained from this signal model, we bound the sensing rate by way of the following theorem.

{\bf Theorem 1: Sensing Rate Bounds in the Multi-carrier Setting} {\em Let $k_1 = 2\Delta f/ c_0$ and $k_2 = 2 \Tofdm	 f_\text{c}/ c_0$, and define $G$ and $\eta_{m,n}$ as
	\begin{align}
		G &= \sum_{(m,n) \in \Sr} 8 \pi^2  \left( \left( k_1 (\bold{Q}^{\frac{1}{2}})_{1,1} n - k_2 (\bold{Q}^{\frac{1}{2}})_{1,2} m\right)^2 + \left( k_1 (\bold{Q}^{\frac{1}{2}})_{1,2} n - k_2 (\bold{Q}^{\frac{1}{2}})_{2,2} m\right)^2\right),
		\label{eq:G-def}
	\end{align}
	\begin{align}
		\eta_{m,n} &= G^{-1}  \left( \left( k_1 (\bold{Q}^{\frac{1}{2}})_{1,1} n - k_2 (\bold{Q}^{\frac{1}{2}})_{1,2} m\right)^2 + \left( k_1 (\bold{Q}^{\frac{1}{2}})_{1,2} n - k_2 (\bold{Q}^{\frac{1}{2}})_{2,2} m\right)^2\right).
	\end{align}
	Further, define $\SINRr$ as
	\begin{align}
		\SINRr = \sum_{(m,n) \in \Sr} \eta_{m,n} \lvert X_{m,n} \rvert^2 \SINR_{m,n}.
		\label{eq:sinr-rad-def}
	\end{align}
	Then the sensing rate expression in (\ref{eq:R_est_formula1}) may be bounded as
	\begin{align}
	\frac{1}{2} \log_2\left(1 + G ~ \SINRr\right) \le \Tp \Rest \le  \log_2\left(1 +\frac{1}{2}G ~\SINRr\right).
	\label{eq:est_rate_bd1}
	\end{align}}

{\em Proof:} See Appendix \ref{app:ofdm_rad}.

Theorem 1 forms the basis for our characterization of sensing coverage probability and ergodic rate. It implies that we may obtain necessary and sufficient conditions for sensing coverage in terms of different thresholds on $\SINRr$, and that we may express bounds on the ergodic sensing rate in terms of said logarithmic functions of $\SINRr$.

In contrast to the communication case, the impact of the $\Sr$ may not be abstracted -- indeed it is a key component of the FIM. Thus, we define the sensing coverage probability as
\begin{align}
\Pcr(\tau) = \Pb^0_{\PhiS}(\SINRr \ge \tau),
\end{align}
and the ergodic sensing rate as 
\begin{align}
&\Ecr = \frac{k}{2 T_{\rm CPI}} \E^0_{\PhiS}\left [\log_2\left(1 + \frac{G}{k} ~ \SINRr \right) \right] &k \in \{1, 2\}.
\end{align}

\subsection{JCAS Coverage and Rate and the Ergodic Theorem}
Having characterized the communication and sensing coverage conditions and rate functions, we combine them to obtain a unified definition of JCAS coverage probability and ergodic rate. 

Consider first the coverage case. Recall that we qualitatively defined the JCAS coverage probability as the joint fraction of UEs and SOs whose coverage conditions are satisfied. This may be expressed formally as follows. As stated in the introduction, let $\PhiU$ and $\PhiS$ denote independent, ergodic point processes representing the UEs and SOs served by the network with respective intensities $\lambdaU$ and $\lambdaS$. For $Y_k \in \PhiU + \PhiS$ let $\SINR(Y_k)$ denote its corresponding communication or radar SINR and let $\tau(Y_k) \in \{\tau_{\rm com}, \tau_{\rm rad}\}$ denote the target threshold for the coverage condition to be satisfied. Then, letting $\{A_n\}_{n \in \N}$ denote a convex averaging sequence, the JCAS coverage probability may be expressed in terms of a spatial average as
\begin{align}
	\Pc(\tau_{com}, \tau_{rad}) &= \E\left[ \lim_{n \rightarrow \infty} \frac{1}{\PhiU (A_n) + \PhiS (A_n)} \sum_{Y_k \in (\PhiU + \PhiS) \cap A_n} \indicator\{\SINR(Y_k) \ge \tau(Y_k)\} \right] \nonumber
	\\
	&\eqlabel{a} \Pb_{\PhiU + \PhiS}^0 \left(\SINR(\mathbf{0}) \ge \tau(\mathbf{0})\right) \nonumber
	\\
	&\eqlabel{b} \frac{\lambdaU}{\lambdaU + \lambdaS}\Pb_{\PhiU}^0\left(\SINRc \ge \tau_\text{com}\right) + \frac{\lambdaS}{\lambdaU + \lambdaS}\Pb_{\PhiS}^0\left(\SINRr \ge \tau_\text{rad}\right) \nonumber
	\\
	&= \frac{\lambdaU}{\lambdaU + \lambdaS}\Pcc(\tau_{\rm com}) + \frac{\lambdaS}{\lambdaU + \lambdaS}\Pcr(\tau_{\rm rad}),
	\label{eq:JCAS_cov}
\end{align}
where (a) follows from the Ergodic Theorem of Random Measures \cite[Theorem 8.3.4]{baccelliSG} and (b) follows from the independence of $\PhiU$ and $\PhiS$ and the Superposition Theorem for stationary processes \cite[Proposition 6.3.5]{baccelliSG}. By a similar argument, we may obtained the JCAS ergodic rate as
\begin{align}
	\Ec = \frac{\lambdaU}{\lambdaU + \lambdaS}\Ecc+ \frac{\lambdaS}{\lambdaU + \lambdaS}\Ecr.
	\label{eq:JCAS_cap}
\end{align}

Note that both expressions decompose into individual communication and sensing coverage probabilities and ergodic rates. Therefore, we may characterize the JCAS coverage probability and ergodic rate by characterizing the communication and sensing coverage probabilities and ergodic capacities separately. We emphasize, however, that in the above derivations, we have made no assumptions regarding the generating model for $\SINR(Y_k)$, and thus (\ref{eq:JCAS_cov}) and (\ref{eq:JCAS_cap}) hold even for the case where a common waveform is used for communication and sensing -- which would induce strong spatial correlations between both functions. Hence, even though our analysis is conducted separately, the results apply to a network in which communication and sensing are performed jointly. It is therefore worthwhile to analyze the performance of both functions in a single framework: our results allow one to ascertain the impact of {\em common} network parameters on {\em joint} performance. Indeed, in Sec. \ref{sec:num_res} we combine our analytical results for communication and sensing to investigate such trends.

\section{System Model}
\label{sec:system_mod}

Having characterize precisely our notion of JCAS coverage and rate, we now present a detailed system model upon which $\SINRr$ and $\SINRc$ are constructed.  The key assumptions of the model are enumerated as {\bf An)}, with {\bf n} being the assumption number. We begin with a development of the spatial attributes of the model; discuss the path-loss and fading models for communication and sensing; detail the beamforming models; define the BS association policy for the typical UE and the BS selection policy for the typical SO; and then state the stochastic model for the communication SINR, $\SINRc$. Following this, we make  a few additional assumptions regarding the sensing model and then conclude the section with the statement of the sensing SINR model, $\SINRr$.

\subsection{Spatial Attributes}

In general, the spatial components of the JCAS network in question are represented by the tuple,
\begin{align}
	\{\PhiB, \PhiU, \PhiS, \XiB\}
\end{align}
Where, $\PhiB$ is a point process on $\R^2$ modeling the locations of BSs in the JCAS network, $\PhiU$ is a point process on $\R^2$ modeling the locations of UEs in the JCAS network, and $\PhiS$ is a point process on $\R^2$ modeling the locations of SOs in the JCAS network.
$\XiB$ is a set process on $\R^2$ representing the locations and shapes of blockages in the network. The explicit consideration of blockages is necessary as we are focused on a mmWave JCAS network. 

We impose the following constraints.

\begin{enumerate}[label={\bf A\arabic*)}]
	\item We take $\PhiB$, $\PhiU$, and $\PhiS$ are ergodic, mutually independent point processes with intensities $\lambdaB$, $\lambdaU$, and $\lambdaS$, respectively. We assume that $\lambdaU$ and $\lambdaS$ are much greater than $\lambdaB$. Moreover, we take $\PhiB$ to be a Poisson Point Process.
\setcounter{saveenum}{\value{enumi}}
\end{enumerate}
Note that the independence of $\PhiU$ and $\PhiS$ implies that the SOs and UEs are mutually exclusive. Hence, without loss of generality, we assume the typical UE and SO to be located at the origin. As the SINR models in these expressions depend only on the typical SO and UE, the Palm measures in (\ref{eq:JCAS_cov}) and (\ref{eq:JCAS_cap}) may both be taken to be the nominal measure.

\begin{enumerate}[label={\bf A\arabic*)}]
\setcounter{enumi}{\value{saveenum}}
	\item With respect to the blockage process, we take $\XiB$ to be a Boolean line process, and simplify its impact by approximating the induced LoS regions using the independent, exponential blockage model \cite{Bai2014}. That is, a given link of length $r$ is LoS or NLoS (independent of the others) with probability 
	\begin{align}
	\pl(r) = \exp(-\beta r).
\end{align} 
	Where $\beta$ is a parameter related to the density of blockages in $\XiB$ and their corresponding geometry. 

	\item For the scenario in question, we consider communication to occur over either LoS or NLoS links, but that sensing is LoS only. This is motivated primarily by the fact that NLoS delay and doppler estimation is both more challenging and typically less informative than the LoS case, and requires some underlying knowledge of the geometry of the reflections of the NLoS paths.
\setcounter{saveenum}{\value{enumi}}
\end{enumerate}

We note further that, while multistatic sensing exploiting NLoS paths is certainly viable, and of interest for JCAS networks, we consider LoS only monostatic sensing as an initial investigation. In addition to being unexplored using this approach, the problem of characterizing the coverage and rate of JCAS networks employing monostatic setting is an interesting problem in its own right. For instance, it is one the application scenarios considered in \cite{DeLima2021} and \cite{Wild2021}. Hence, we leave extensions to multi-static sensing in JCAS networks to future work.

In light of these assumptions regarding the spatial attributes of the network, we employ the following notation. Let $X_k$ denote the canonical enumeration of the atoms of $\PhiB$, and let $M_k$ be a mark associated with $X_k$ denoting whether or not the link from $X_k$ to the origin is LoS. Then $\{M_k\}_{k \in \N}$ are conditionally independent and $M_k \sim \text{Bern}(\pl\left(\norm{X_k}_2\right))$. We take $\PhitB$ to denote the marked PPP $\{(X_k, M_k)\}_{k \in \N}$, accounting for both the locations and LoS/NLoS statuses of BSs. In addition, where useful we take $\PhiL$ to denote the set of points in $\PhitB$ that are LoS and $\PhiN$ denote the set of points in $\PhiB$ that are NLoS.

\subsection{Channel and Radar Cross Section Models}
As noted in the introduction, we consider a mmWave network. This is primarily motivated by the fact that mmWave band posses greater bandwidths, and thus facilitate great delay resolution than conventional frequencies. Consequently, we employ the following path loss models.
\begin{enumerate}[label={\bf A\arabic*)}]
\setcounter{enumi}{\value{saveenum}}
	\item The LoS/NLoS one-way path loss functions are modeled as
	\begin{align}
		&\gL(r) = \Llos{r} \text{\quad when LoS} &\gN(r) = \Lnlos{r} \text{\quad when NLoS.}
	\end{align}
	Which are applicable to mmWave and sub-THz channels, \cite{Journet2011}. Thus the path loss to the origin of $X_k \in \PhitB$ is 
	\begin{align}
		L(\norm{X_k}_2) = \gL(\norm{X_k}_2) M_k + \gN(\norm{X_k}_2)(1 - M_k).
	\end{align}
	\item  In a similar manner, the path loss of the LoS radar return is given by the two-way monostatic path loss model \cite{Canan2020}
	\if 0
	\begin{align}
	\gLr(r) =
	\begin{cases}
		\Lrad{r} &r \ge 1
		\\
		\frac{\KL}{4 \pi} r^{-\alphaL}e^{-\gammaL r} &\text{otherwise}.
	\end{cases}
	\end{align}
	\else
	\begin{align}
	\gLr(r) = \Lrad{r}.
	\end{align}
	\fi
\setcounter{saveenum}{\value{enumi}} 
\end{enumerate}

\begin{enumerate}[label={\bf A\arabic*)}]
\setcounter{enumi}{\value{saveenum}}
	\item To avoid aberrant behavior of the path loss function for small $r$, we further assume that the path loss functions are zero for distances such that they would exceed 1. Thus the actual LoS path loss is $\gL(r) \indicator\{\gL(r) \le 1\}$, and similar forms hold for the other pathloss models. 
\setcounter{saveenum}{\value{enumi}} 
\end{enumerate}
Such a transformation follows those detailed in \cite{Haenggi2012}, and effectively removes interferers that are unrealistically close to the typical UE or serving BS for the typical SO. 


As $\SINRr$ is a weighted average over a set of allocated resource elements, $\Sr$, we require the distribution of the fading terms over each of the $(m,n) \in \Sr$ resource elements for each BS $X_k \in \PhiB$, $\{|H^k_{m,n}|^2\}$. To that end, we stipulate the following.

	
\begin{enumerate}[label={\bf A\arabic*)}]
\setcounter{enumi}{\value{saveenum}}
	\item We assume that the sensing procedure occurs within a coherence interval, and thus that $|H^k_{m,n}|^2$ do not vary in time. Hence we drop the $m$ subscript.  
	\item Additionally, for interfering links, the fading terms are distributed as follows
	\begin{align}
	&|H_n^k|^2 \sim \text{Gamma}\left(\NL, \NL\right) \text{\quad when $X_k$ is LoS to the receiver}
	\\
	&|H_n^k|^2 \sim \text{Gamma}\left(\NN, \NN\right) \text{\quad when $X_k$ is NLoS to the receiver},
	\end{align}
	where, $N_L, N_N \ge 1$ denotes the order of the fading model. Note that $N_L, N_N = 1$ reduces to Rayleigh fading. We further assume that $\{|H^k_n|^2\}$ are IID. These general Nakagami fading models for LoS and NLoS links are applicable to mmWave channels \cite{Bai2015}.
\setcounter{saveenum}{\value{enumi}} 
\end{enumerate}
The independence assumption is valid when the subcarriers used for sensing are separated by intervals greater than a coherence bandwidth. Practically, non-contiguous sub-carrier allocation is useful for ensuring a greater unambiguous range resolution \cite{Braun2014ofdm}. We note further that relaxing this assumption would only affect our results for the Laplace Transforms of the aggregate fading terms detailed in Sec. \ref{subsec:int-fad-LT}.
\begin{enumerate}[label={\bf A\arabic*)}]
\setcounter{enumi}{\value{saveenum}}
	\item Finally, with respect to fading on the desired signal, we assume that the typical UE experiences Rayleigh fading. That is $\lvert H^0 \rvert^2 \sim \text{Exp}(1)$. This assumption is less realistic for mmWave channel models, but facilitates analytical tractability. We note, however, that this assumption has a minor impact on the overall coverage probability in the event of no fading, or other fading or shadowing distributions \cite{Hmamouche2021}.
	\item In the sensing case, we assume no fading on desired signal in light of the LoS assumption, but that the radar cross section of the typical SO, $\kc$, is exponentially distributed (denoted as the Swerling III model in radar settings). That is, $\kc \sim \text{Exp}(1)$. The cross section is further assumed to be constant over the excitation duration.  
\setcounter{saveenum}{\value{enumi}} 
\end{enumerate}

\subsection{Beamforming Models}
We consider a JCAS network in which BSs and UEs perform directional beamforming. Such a model representative of near-term network capability and is commonly employed in prior work, for instance \cite{DeLima2021} and \cite{Wild2021}. 
\begin{enumerate}[label={\bf A\arabic*)}]
\setcounter{enumi}{\value{saveenum}}
	\item Thus, for a communication link, the BS and UE select the beam directions which maximize the received power. Since sensing is LoS only, the BS points its transmit and receive beams in the direction of the desired SO(s).
\setcounter{saveenum}{\value{enumi}} 
\end{enumerate}
We note that, as we consider a monostatic scenario for sensing, the serving BS is both the transmitter and receiver. This requires a full duplex transceiver at the base station in addition to antenna separation \cite{Wild2021}. Prior work in the area of full duplex has demonstrated its feasibility, \cite{Roberts22}, especially true at mmWave.
	
\begin{enumerate}[label={\bf A\arabic*)}]
\setcounter{enumi}{\value{saveenum}}
	\item For all links, we model the antenna patterns of BSs/UEs using the sectored model. The antenna patterns for BSs and UEs are parameterized by the terms in the following table.\\

	\begin{center}
	\begin{tabular}{ |p{2cm}||p{3cm}|p{3cm}|p{3.5cm}| }
	\hline
	 Type &Main lobe Gain &3 dB Beam width &Front to Back Ratio \\
	 \hline \hline
	 BS, Tx &$\Gbt$ &$\thetabt$ &$\xibt$ \\
	 \hline
	 BS, Rx &$\Gbr$ &$\thetabt$ &$\xibr$ \\
	 \hline
	 UE, Rx &$\Gur$ &$\thetaur$ &$\xiur$ \\
	 \hline
	\end{tabular}
	\end{center}
	
	{\quad }\\[-6mm]
	
	Thus, the total antenna gain for the desired signal of the typical UE is $\Gbt \Gur$, and the total antenna gain for the radar return is $\Gbt \Gbr$.
	\item With respect to interfering links, the antenna gains are taken to be independent random variables. For the $k^{th}$ interferer, the normalized antenna gain over the $m^{th}$  symbol is modeled as
	\begin{align}
		B_{m}^k \sim 
		\begin{cases}
			1 &\text{ with prob. }  \pB = \frac{\thetabt}{2 \pi}
			\\
			\xibt &\text{ with prob. } 1 - \frac{\thetabt}{2 \pi}.
		\end{cases}
	\end{align}
	
	\item In a similar manner, since the BS process is isotropic, the receive gain between the typical UE and $k^{th}$ interferer is modeled as
	\begin{align}
		Z_{U}^k \sim 
		\begin{cases}
			1 &\text{ with prob. } \frac{\thetaur}{2 \pi}
			\\
			\xiur &\text{ with prob. } 1 - \frac{\thetaur}{2 \pi}.
		\end{cases}
	\end{align}
	
\setcounter{saveenum}{\value{enumi}} 
\end{enumerate}

	These models are justified under the assumption that the UE and SO densities are much greater than the BS density. Additionally, note that one may define analogous random variables for the overall antenna gain between the serving BS for the typical SO and the $k^{th}$ interfering BS, $Z_B^k$. However, as a consequence of the SO selection policy (which will be discussed in the following subsection), the process of interfering BSs in non-isotropic. Hence, it does not admit an expression of the same form as $Z_{U}^k$. We derive the distribution of $Z_B^k$ in the Sec. \ref{sec:palm_int_derivation}.
	
\begin{enumerate}[label={\bf A\arabic*)}]
\setcounter{enumi}{\value{saveenum}}
\item Completing the beamforing models, transmit beam directions are assumed to be fixed within a time slot, which is composed of multiple symbols, so as to facilitate communication. As a simplifying assumption, $B^k_m \indep B^k_{s}$ if $m$ and $s$ fall in different time slots; $B^k_m = B^k_{s}$ otherwise.
\setcounter{saveenum}{\value{enumi}} 
\end{enumerate}

Using the notation developed thus far, the total antenna gain over the $k^{th}$ interfering link for the typical UE is $\Gbt \Gur B^k Z_U^k$, and total antenna gain over the $k^{th}$ interfering link on the $m^{th}$  symbol for the typical SO is $\Gbt \Gbr B_{m}^k Z_B^k$.



\subsection{UE Association and SO Selection Policies}
As we are considering a cellular network, the association of UEs to BSs and a the selection of BSs for SO measurement plays an important role in coverage and rate.
\begin{enumerate}[label={\bf A\arabic*)}]
\setcounter{enumi}{\value{saveenum}}
	\item The typical UE is associated with BS that has the lowest path loss. Letting $X_0$ denote the location of the serving BS associated with the typical UE, we have
	\begin{align}
		X_0 = \arg \sup\{L(\norm{X_k}_2) : X_k \in \PhitB\}.
	\end{align}
	
	\item The typical SO is measured by the nearest LoS BS. Letting $X_0$ denote the location of the serving BS associated with the typical SO. Given $\PhiL \ne \emptyset$, we have
	\begin{align}
		X_0 = \arg\inf\{\norm{X_k}_2 : X_k \in \PhiL\}
	\end{align}
	\setcounter{saveenum}{\value{enumi}} 
\end{enumerate}

	We note that, in contrast to the UE association policy, the BS selection policy for the SO is not necessarily explicitly conducted by the network. Rather, such a process represents a simple form of sensor fusion. In theory, due to the cooperation facilitated by the cellular network, the measurements from all BSs could be used to revise the beliefs about the typical SO's parameters of interest. By restricting our focus to the nearest BS, we are simply quantifying a lower bound on this theoretical information gain by considering only the measurements obtained by the BS at $X_0$. In making this assumption, however, we note that in practice measurements from this BS may not always be available -- for instance detection uncertainty may result in the BS in question failing to make informative measurements of the typical SO. We leave relaxations of this assumption that account for these non-idealities to future work, but note that similar assumptions were made in prior work, such as \cite{Skouroumounis2021}.

\subsection{Communication SINR Model}
Putting these assumptions together, we obtain the stochastic model for $\SINRc$. Let the normalized noise power at the typical UE be defined as
\begin{align}
	\nuc = \frac{\sigmaN^2}{\Gbt \Gur}.
\end{align}
Then, we define $\SINRc$ as
\begin{align}
	\SINRc = \frac{\lvert H^0 \rvert^2 L(\norm{X_0}_2)}{\sum_{X_k \in \PhitB \setminus \{X_0\}} \lvert H^k \rvert^2 B^k Z_U^k L(\norm{X_k}) + \nuc}.
\end{align}
 In the subsequent section, we shall take $F_L = \lvert H^k \rvert^2 B^k Z_U^k$ for interfering LoS BSs and $F_N = \lvert H^k \rvert^2 B^k Z_U^k$ for interfering NLoS BS, which succinctly capture the aggregate fading and beam alignment terms.
 \subsection{Sensing SINR Model}
Before stating $\SINRr$, we make the following further assumptions.
\begin{enumerate}[label={\bf A\arabic*)}]
\setcounter{enumi}{\value{saveenum}}
\item We assume that the transmitted symbols over the radar excitation signal are constant modulus. That is $\lvert X_{m,n} \rvert = 1$ for all $(m,n) \in \Sr.$
\setcounter{saveenum}{\value{enumi}} 
\end{enumerate}
This assumption is made to facilitate tractability. Although the variability of transmit symbols is an important factor in practical JCAS waveform design, their impact on the network-wide sensing coverage probability defined with respect to $\SINRr$ is limited, as verified by simulation. We note further that the variability of transmit symbols may be approximated by scaling the noise variance in our measurement model in (\ref{eq:ofdm_radar_est_prob}), \cite{Braun2014ofdm}. This would result in attenuating the $G$ factor in (\ref{eq:G-def}), but change nothing else.

In light of A15 and A18, the terms involved in the expression for $\SINRr$ vary only over time slots rather than  symbols. Hence, we may simplify the expression somewhat by reducing the dimensionality of $\Sr$ to the time slot level. Specifically, let the radar excitation take place over $K$ time slots, and define
\begin{align}
	&\Sr^t = \{(m,n) \in \Sr: m \in \text{$t^{th}$ time slot}\} &\forall t\in [K].
\end{align}
Then, we define the reduced radar resource allocation matrix over time slots and sub-carriers as
\begin{align}
	&\Srt = \{(t,n) \in \{0,1\}^{T \times N_c}: (\cdot, n) \in \Sr^t \}.
\end{align}
The corresponding weights, $\{\eta_{m,n}\}_{(m,n) \in \Sr}$, which define the average over resource elements in the definition of $\SINRr$ in (\ref{eq:sinr-rad-def}) may be combined as 
\begin{align}
	\theta_{t,n} = \sum_{(m,n) \in \Sr^t} \eta_{m,n} &&(t,n) \in \Srt.
\end{align}

Though not necessary for the statement of $\SINRr$, in the following analysis, we shall further employ make use of the following derived quantities
\begin{align}
	&w_t = \sum_{(\cdot, n) \in \Srt} \theta_{t,n},
	&q_n = \sum_{(t, \cdot) \in \Srt}  \theta_{t,n}.
\end{align}
We define the supports of these measures as $\text{support}(\bold{w}) = T \le K$ and $\text{support}(\bold{q}) = N \le N_c$.

\begin{enumerate}[label={\bf A\arabic*)}]
\setcounter{enumi}{\value{saveenum}}
\item Finally, we assume that the LoS/NLoS status of interfering BSs with respect to the serving BS is independent of their LoS/NLoS status with respect to the typical SO at the origin. Such a stipulation is necessary as the serving BS in the sensing scenario is both the transmitter and receiver.
\setcounter{saveenum}{\value{enumi}} 
\end{enumerate}

Putting these together, we obtain the stochastic SINR model for sensing, $\SINRr$. Let the normalized noise variance at the serving BS for the typical SO be defined as
\begin{align}
	\nur = \frac{\sigmaN^2}{\Gbt \Gbr}.
\end{align}
Additionally, let the product of the fading term on the $n^{th}$ subcarrier and normalized transmit antenna gain at the $t^{th}$ time slot for the $k^{th}$ interfering BS be denoted as $F_{t,n}^k = \lvert H^k_{n} \rvert^2 B^k_t$. Then $\SINRr$ is defined as
\begin{align}
		&\SINRr =\indicator\{\PhiL(\R^2) >0\} \sum_{(t,n) \in \Srt} \theta_{t,n} \frac{ \kc  \gLr(\norm{X_0})  }{ \sum_{X_k \in \PhitB \setminus \{X_0\}} F_{t,n}^k Z_B^k L(\norm{X_k - X_0}) + \nur}.
		\label{eq:sinr-rad-mod}
\end{align}

Note that the LoS marks implicit in $L(\norm{X_k - X_0})$ are the LoS marks of $X_k$ with respect to the serving BS, rather than with respect to the origin. These follow the same distribution as detailed in A2 in light of A19. In addition, the indicator term, $\indicator\{\PhiL(\R^2) >0\}$, denotes that fact that sensing is LoS only. Thus, if there are no LoS BSs with respect to the typical SO, its SINR (and consequently its sensing rate) is zero.

\section{Preliminary Results}
\label{sec:prelim_res}
Armed with the SINR models for communication and sensing, we now proceed with the analysis of JCAS coverage probability and ergodic rate. Unfortunately, however, we found characterizing the distribution of $\SINRr$, as stated in (\ref{eq:sinr-rad-mod}), to be analytically prohibitive. Consequently, we establish upper and lower bounding sensing SINR models as well as an approximate sensing SINR model that are analytically tractable. These models may be expressed as instantiations of a generic radar SINR model, $\SINRgen$, which is defined as
\begin{align}
	\SINRgen = \indicator\{\PhiL(\R^2) > 0\} \frac{ \kc  \gLr(\norm{X_0})  }{ \sum_{X_k \in \PhitB \setminus \{X_0\}} F_k Z_B^k L(\norm{X_k - X_0}) + \nur},
	\label{eq:sinr-gen-mod}
\end{align}
where $\{F_k\}$ are arbitrary non-negative random variables corresponding to different aggregate interference fading terms. 

Given this common form, we present characterizations of the CCDFs of the alternate radar SINR models in terms of functionals which correspond to upper and lower bounds on the CCDF of $\SINRgen$. These are summarized in Theorem 3 to follow.

\subsection{Alternate Radar SINR Models and Their Relation to $\SINRr$}

The principal challenge in characterizing the distribution of $\SINRr$ lies in the fact that it is expressed as a mean of the inverse interference experienced over the allocated resource elements in $\Sr$. One may recognize the denominator as a harmonic mean. Our alternate SINR models make use of the relationship between arithmetic, geometric, and harmonic means, which are defined as follows.

\textbf{Definition: (Arithmetic, Geometric, and Harmonic Means)} {\em Let $X$ be a non-negative random variable over some countable state space $\mathcal{X}$ with PMF $p$. The arithmetic mean of $X$ with respect to $p$ is
\begin{align}
	\AM(X, p)  = \sum_{x \in \mathcal{X}}p_X(x) ~ x = \E[X].
\end{align}
The geometric mean of $X$ with respect to $p$ is
\begin{align}
	\GM(X, p) = \prod_{x \in \mathcal{X}}x^{p_X(x)} = \exp(\E[\log(X)]).
\end{align}
The harmonic mean of $X$ with respect to $p$ is
\begin{align}
	\HM(X, p) = \left(\sum_{x \in \mathcal{X}}p_X(x) ~ x^{-1}\right)^{-1} = \E[X^{-1}]^{-1}.
\end{align}}
Moreover, Jensen's Inequality yields the (the AM-GM-HM Inequality):
\begin{align}
	\HM(X, p) \le \GM(X,p) \le \AM(X,p).
\end{align}

Using this notation and letting $\boldsymbol{\theta}$ denote the ensemble of weights over $\Srt$, we may express $\SINRr$ as
\begin{align}
	\SINRr = \indicator\{\PhiL(\R^2) > 0\} \frac{ \kc  \gLr(\norm{X_0})  }{ \HM\left( \left\{\sum_{X_k \in \PhitB \setminus \{X_0\}} F_{t,n}^k Z_B^k L(\norm{X_k - X_0}) + \nur\right\}_{(t,n) \in \Srt}, \boldsymbol{\theta}\right)}.
\end{align}
Rather than dealing with this expression directly, we consider the following three SINR models: $\SINRam$, $\SINRgm$, and $\SINRhm$. Letting $\mathbf{F_k}$ denote the ensemble of fading terms $\{F_{t,n}^k\}_{(t,n) \in \Srt}$, these are defined as
\begin{align}
	\SINRam = \indicator\{\PhiL(\R^2) > 0\} \frac{ \kc  \gLr(\norm{X_0})  }{ \sum_{X_k \in \PhitB \setminus \{X_0\}} \AM\left(\mathbf{F_k}, \boldsymbol{\theta}\right)Z_B^k L(\norm{X_k - X_0}) + \nur},
	\\
	\SINRgm = \indicator\{\PhiL(\R^2) > 0\} \frac{ \kc  \gLr(\norm{X_0})  }{ \sum_{X_k \in \PhitB \setminus \{X_0\}} \GM\left(\mathbf{F_k}, \boldsymbol{\theta}\right)Z_B^k L(\norm{X_k - X_0}) + \nur},
	\\
	\SINRhm = \indicator\{\PhiL(\R^2) > 0\} \frac{ \kc  \gLr(\norm{X_0})  }{ \sum_{X_k \in \PhitB \setminus \{X_0\}} \HM\left(\mathbf{F_k}, \boldsymbol{\theta}\right)Z_B^k L(\norm{X_k - X_0}) + \nur}.
\end{align}
We relate these to the exact model, $\SINRr$, by the following proposition.

\textbf{Proposition 2: (Relation Between Proposed SINR Models)} {The following stochastic orderings hold:
\begin{align}
	\SINRam \lest \SINRr \lest \SINRhm,
\end{align}
and
\begin{align}
	\SINRam \lest \SINRgm \lest \SINRhm,
\end{align}}
{\em Proof:} See Appendix \ref{app:sinr_aprx}.

These orderings arise from the AM-GM-HM Inequality, H{\"o}lder's Inequality, and a generalization of H{\"o}lder's Inequality to harmonic means (detailed in Appendix \ref{app:tech_res}). Overall, these orderings indicate that both $\SINRr$ and $\SINRgm$ are lower bounded by $\SINRam$ and upper bounded by $\SINRhm$. This motivates $\SINRgm$ for use as an approximation of $\SINRr$, and establishes $\SINRam$ and $\SINRhm$ as lower and upper bounding models, respectively. In the sequel we provide further analytical justification for the use of $\SINRgm$ as an approximate model by obtaining an upper bound on the CCDF of $\SINRhm$ and a lower bound on the CCDF of $\SINRam$. These in turn imply bounds on the approximation error: $\left \lvert \Pb(\SINRr \ge \tau) - \Pb(\SINRgm \ge \tau) \right \rvert$.

\subsection{Bounds on the CCDF of the Generic Radar SINR Model}
Each of the proposed alternate models differ only in the aggregate fading terms -- $\AM\left(\mathbf{F_k}, \boldsymbol{\theta}\right)$, $\GM\left(\mathbf{F_k}, \boldsymbol{\theta}\right)$, or $\HM\left(\mathbf{F_k}, \boldsymbol{\theta}\right)$ -- which correspond to the $\{F_k\}$ terms in the expression for $\SINRgen$ in (\ref{eq:sinr-gen-mod}). Hence, they may be considered as instantiations of this general model. 
In light of this, before delving into the full technical details regarding the characterization of these alternate radar SINR models, we first summarize tractable upper and lower bounds on the CCDF of $\SINRgen$.

These bounds are expressed in terms of functionals which depend on the following objects.
\begin{enumerate}
	\item The PDF of the distance distribution of the serving base station, $\lvert \lvert X_0 \rvert \rvert$, denoted as $f_{R_0}$. This is obtained in (\ref{eq:dist-distr}) in Lemma 1.
	\item The Mellin Transforms of the path loss functions with respect to the intensity measure of the Palm Process of the distances of interfering BSs. Letting, $\PhiB^{!X_0}$, denote the reduced Palm Process of interferers, the Palm Process of of the distances of interferers is defined as
	\begin{align}
	\PiB^0 = \left\{ \norm{X_k - X_0}_2 : X_k \in \PhiB^{!X_0} \right\}.
\end{align}
This process may be further thinned into Points that are LoS or NLoS with respect to the serving BS at $X_0$ and those which are inside or outside the main lobe of the receive beam at the serving BS. These processes are denoted as  $\PiLa^0$, $\PiNa^0$, $\PiLb^0$, and $\PiNb^0$.

In Lemma 8 in Appendix \ref{app:palm_int_ub} we establish upper bounding intensity measures, $\rhoAk$, $a \in \{L,N\}, k \in \{1,2\}$, and lower bounding intensity measures, $\nuAk$, $a \in \{L,N\}, k \in \{1,2\}$. We then obtain upper and lower bounds on the Mellin Transforms which we define as
\begin{align}
	&\Mellin_{\nuAk \circ g_{a}^{-1}}(p; A, R_0) = \int_{A} g_a(r)^{_p-1} \nuAk(r; R_0) dr & a \in \{L, N\}, k \in \{1, 2\},
\end{align}
and
\begin{align}
	&\Mellin_{\rhoAk \circ g_{a}^{-1}}(p; A, R_0) = \int_{A} g_a(r)^{_p-1} \rhoAk(r; R_0) dr & a \in \{L, N\}, k \in \{1, 2\}.
\end{align}
Where $A$ is a measurable set of $\R_+$, $p$ is an arbitrary complex number (although we only require the positive real branch), and $R_0$ is the distance of the serving BS (upon which the intensity measures depend).
These are established in Lemma 9 in Appendix \ref{app:int_mt}.

\item The Laplace Transforms of the aggregate fading terms. These are dependent on the LoS/NLoS status of the interfering base stations and denoted as $\Laplace_{F_L}$ and $\Laplace_{F_N}$, respectively.
\item Special functionals $H_{UB}$ and $H_{LB}$ which correspond to finite order approximations of the interference process which may be used to obtain closed form upper and lower bounds on the Laplace Transform of a Shot Noise Process in terms of its Mellin Transform. This is stated formally in Theorem 2 in Appendix \ref{app:LST_bds}. Explicit expressions for these functionals are stated in (\ref{eq:H_UB_def}) and (\ref{eq:H_LB_def}) in Appendix \ref{app:LST_bds}.
\end{enumerate}

The bounds are expressed as follows.

{\bf Theorem 3: (Bounds on the CCDF of $\SINRgen$)} {\em Let $N_w \in \N$ denote an arbitrary number of windows employed in the bounds for interference Laplace Transform in Theorem 2. Additionally, let $\{d^L_i\}_{i = 0}^{N_w+1}$ denote arbitrary, ordered boundaries for the windows used to partition $\R_+$ for the LoS interference terms and $\{d^N_i\}_{i = 0}^{N_w+1}$ denote arbitrary, ordered boundaries for the windows used to partition $\R_+$ for the NLoS interference terms. Without loss of generality, take $d^L_0, d^N_0 = 0$ and $d^L_{N_w}, d^N_{N_w + 1} = \infty$. For conciseness, let $W^L_i = [d^L_{i-1}, d^L_i]$ and $W^N_i = [d^N_{i-1}, d^N_i]$. Finally, define $\xi_1 = 1$ and $\xi_2 = \xibr$. The following bounds hold.

\begin{enumerate}[label=\roman*)]
	\item The CCDF of $\SINRgen$ is upper bounded as $\Pb (\SINRgen \ge \tau) \le \Pcru\left(\tau; \Laplace_{F_L}, \Laplace_{F_N}\right)$, where
	\begin{align}
 \Pcru(\tau; &\Laplace_{F_L}, \Laplace_{F_N}) = \left(1 - \exp\left(\frac{-2 \pi \lambdaB}{\beta^2}\right) \right)  \int_{\R_+} f_{R_0}(u) \exp\left({\frac{-\tau \nur}{\gLr(u)}}\right) \nonumber
 \\
 &\prod_{k = 1}^{2}\exp\left(-\sum_{i = 1}^{N_w +1} \HU\left(\frac{\tau \xi_k}{\gLr(u)}, W^L_i; 1 -\Laplace_{F_L}, \Mellin_{\nuLk \circ \gL^{-1}}(\cdot; \cdot, u)\right) \right) \nonumber
\\
 &\prod_{k = 1}^{2}\exp\left(-\sum_{i = 1}^{N_w} \HU\left(\frac{\tau \xi_k}{\gLr(u)}, W^N_i; 1 -\Laplace_{F_N}, \Mellin_{\nuNk \circ \gN^{-1}}(\cdot; \cdot, u)\right) \right) du.
 \label{eq:pcr-gen-ub}
\end{align}

\item The CCDF of $\SINRgen$ is lower bounded as $\Pb (\SINRgen \ge \tau) \ge \Pcrl (\tau; \Laplace_{F_L}, \Laplace_{F_N})$, where
	\begin{align}
 \Pcrl (\tau; &\Laplace_{F_L}, \Laplace_{F_N}) = \left(1 - \exp\left(\frac{-2 \pi \lambdaB}{\beta^2}\right) \right)  \int_{\R_+} f_{R_0}(u) \exp\left({\frac{-\tau \nur}{\gLr(u)}}\right) \nonumber
 \\
 &\prod_{k = 1}^{2}\exp\left(-\sum_{i = 1}^{N_w +1} \HL\left(\frac{\tau \xi_k}{\gLr(u)}, W^L_i; 1 -\Laplace_{F_L}, \Mellin_{\rhoLk \circ \gL^{-1}}(\cdot; \cdot, u)\right) \right) \nonumber
\\
 &\prod_{k = 1}^{2}\exp\left(-\sum_{i = 1}^{N_w} \HL\left(\frac{\tau \xi_k}{\gLr(u)}, W^N_i; 1 -\Laplace_{F_N}, \Mellin_{\rhoNk \circ \gN^{-1}}(\cdot; \cdot, u)\right) \right. &\nonumber
 \\
 &\hspace{0.25\linewidth} \left.+ \frac{\tau \xi_k}{\gLr(u)} \E[F_N] \Mellin_{\rhoNk \circ \gN^{-1}}(2; W^N_{N_w + 1}, u) \right) du.
 \label{eq:pcr-gen-lb}
\end{align}
\end{enumerate}}

{\em Proof:} See Appendix \ref{app:main_res}.

Both bounds require the evaluation of only a single integral. Moreover, while we have established Theorem 3 in the specific setting of the $\SINRgen$, it may be readily extended to other settings. Generally, when the path loss Mellin Transforms of the interference process may be obtained in closed form, Theorem 3 provides a robust framework for obtaining integral closed form upper and lower bounds for SINR models with Rayleigh fading on the desired signal. Using an approach similar to \cite{Bai2015}, this framework may be exploited for Nakagami fading as well.

In the statement of the theorem, we have left the partitions used in the bound for the interference Laplace Transforms as arbitrary parameters. Choosing these terms correctly is important for obtaining tight bounds. The problem of finding optimal partitions is generally difficult (it is akin to a generalization of weighted k-means clustering to a continuum), hence we propose a heuristic method base on uniformly sampling the reciprocal of the path loss functions.

{\bf Definition 3: (Uniform Path Gain Windowing)} {\em Let $N_w \in \N$ and choose a desired end point, $d_{N_w} \in \R_+$. Further, for $a \in \{L, N\}$ define $\fa(r) = 1/\ga(r)$ and define $d_{min} = \ga^{-1}(1)$. Then the partitions in Theorem 3 may be constructed as follows 
\begin{align}
	&\Delta_a = \frac{\fa^{-1}(d_{N_w})- \fa^{-1}(d_{min})}{N_w}
	\\
	&d^a_i = 
	\begin{cases}
		\fa\left(\fa^{-1}(d_{min}) +\Delta_a i \right) & i \le N_w
		\\
		\infty &\text{otherwise}
	\end{cases}
	\\
	&W^a_i = [d^a_{i-1}, d^a_i].
\end{align}}
While suboptimal, this method attempts to capture the tradeoff between the relative strength of closer interferers with the fact that the density of interferers (on $\R_+$) increases with distance. Indeed, as we demonstrate in Section \ref{sec:num_res}, this windowing method leads to sharp bounds on coverage and rate.

\subsection{Bounds on Ergodic Sensing Efficiency Using the Generic Model}
Finally, we note that the upper and lower bounds from Theorem 3, in tandem with Theorem 1, imply the following bounds on the ergodic sensing efficiency.

{\bf Corollary 3: (Bounds on the Ergodic Sensing Efficiency)} {\em The ergodic sensing efficiency defined with respect to the generic SINR model, $\SINRgen$, is lower bounded as
\begin{align}
	\Tp \E[\Resub] \ge \E \left[ \frac{1}{2} \log_2\left(1 + G~ \SINRr\right) \right] \ge \frac{G}{2 \ln(2)}\int_{\R_+} \Pcrl \left(\tau; \Laplace_{F_L}, \Laplace_{F_N} \right) \left(1 + G\tau\right)^{-1} d\tau.
\end{align}
Similarly, it is upper bounded as
\begin{align}
	\Tp \E[\Resub] \le \E \left[  \log_2\left(1 + \frac{1}{2}G~ \SINRr\right) \right] \le \frac{G}{2 \ln(2)}\int_{\R_+} \Pcru \left(\tau; \Laplace_{F_L}, \Laplace_{F_N} \right) \left(1 + \frac{G}{2}\tau\right)^{-1} d\tau.
\end{align}}
{\em Proof:} The terms inside the expectations follow from Theorem 1. Using that bounds from Theorem 3, one may show that the expectations are lower bounded by the integral terms in a manner similar to \cite[Theorem 2]{Olson2022}. $\qedwhite$

\section{Sensing Coverage Probability and Ergodic Efficiency}
\label{sec:pcr_analysis}
In light of these results, to complete our characterization of the sensing coverage probability and ergodic efficiency we must obtain the distance distribution of the serving BS, the pathloss Mellin Transforms of the Palm Interference Process, and the Laplace Transforms of the aggregate fading terms: $\AM\left(\mathbf{F_k}, \boldsymbol{\theta}\right)$, $\GM\left(\mathbf{F_k}, \boldsymbol{\theta}\right)$, and $\HM\left(\mathbf{F_k}, \boldsymbol{\theta}\right)$.
Using Theorem 3, the above objects immediately imply upper and lower bounds on the CCDFs of each of the alternate models, which in turn lead to bounds and approximations on the sensing coverage probability in light of Proposition 2. These are summarized in Theorem 5, to follow. In tandem with Corollary 3, we further obtain ergodic sensing efficiency bounds and approximations.

\subsection{Characterization of the Distribution of the Desired Signal and Conditional Interference PPP}
\label{sec:palm_int_derivation}
We first derive the distribution of the serving BS location, $X_0 = (R_0, \phi_0)$. Following this we characterize the reduced Palm distribution of the point process of interferers, $\PhiB^{!X_0}$, which allows us to characterize the point process of the distances of interfering BSs with respect to the serving BS conditioned on $R_0$:
\begin{align}
	\PiB^0 = \left\{ \norm{X_k - X_0}_2 : X_k \in \PhiB^{!X_0} \right\}.
\end{align}
In so doing, we further characterize the distribution of the receive antenna gain variables, $\{Z_B^k\}_{k \in \PiB^0}$, as marks with respect to $\PiB^0$. This characterization, in turn leads to bounds on the path loss Mellin Transforms required for the application of Theorem 3.

The following lemma establishes the form of the distribution of $(R_0, \phi_0)$.

\textbf{Lemma 1: (Distribution of the Location of the Serving Base Station)} {\em $R_0$ and $\phi_0$ are independent with $\phi_0 \sim \text{Uniform}([0, 2\pi))$ and $R_0$ distributed as 
\begin{align}
	R_0 \sim 
	\begin{cases}
		\frac{f_{R_0}(r)}{1 - \exp(\frac{-2 \pi \lambdaB}{\beta^2})} &\text{w.p. } 1 - \exp(\frac{-2 \pi \lambdaB}{\beta^2})
		\\
		\infty &\text{w.p. } \exp(\frac{-2 \pi \lambdaB}{\beta^2})
	\end{cases}
\end{align} 
where
\begin{align}
	f_{R_0}(r) = 2 \pi \lambdaB \exp\left(\frac{-2 \pi \lambdaB}{\beta^2}(1 - e^{-\beta r}(\beta  r + 1))\right).
	\label{eq:dist-distr}
\end{align} }
{\em Proof:} This may be proved in a manner similar to that of \cite[Theorem 10]{Bai2014}. $\qedwhite$

Note that $R_0$ takes the value of $\infty$ when $\PhiL(\R^2) = 0$. Since $\SINRgen$ is $0$ when $\PhiL(\R^2) = 0$, we shall restrict our attention to the case where $R_0 < \infty$.

We now characterize $\PiB^0$.

\textbf{Lemma 2: (Characterization of the Palm Interference Process)} {\em Condition on $X_0 = (r_0, \phi_0)$. Define the function $J(r; r_0, z)$ as 
\begin{align}
	J(r; r_0, z) = \int_{z}^1(1 - u^2)^{-1/2} \exp\left(-\beta\sqrt{r^2 - 2r r_0 u + r_0^2} \right)du.
\end{align}

Then, for all $\phi_0$, $\PiB^0$ is a PPP on $\R_+$ with intensity function
\begin{align}
	\lambdaB^0(r; r_0) = 2\lambdaB r\left( \pi -  J\left(r; r_0, \frac{r}{2 r_0} \right) \indicator\{r \le 2r_0\}\right).
	\label{eq:palm-int-meas}
\end{align}
Moreover, the normalized receive antenna gain variables $\{Z_B^k\}_{k \in \PiB^0}$ may be treated as independent marks of $\PiB^0$. $\{Z_B^k\}_{k \in \PiB^0}$ take values in $\{\xibr, 1\}$, and the probability that $Z_B^k = 1$ is given by $\pBr(R_k)$ where
\begin{align}
	\pBr(R_k) = \frac{\frac{\thetabr}{2} - J\left(R_k; r_0, \max\left\{\cos\left(\frac{\thetabr}{2}\right), \frac{R_k}{2 r_0}\right\} \right) \indicator\{R_k \le 2 r_0  )\}} {\pi -  J\left(r; R_0, \frac{R_k}{2 r_0} \right) \indicator\{R_k \le 2r_0\}}
	\label{eq:palm-int-rx-pr}
\end{align}}
{\em Proof:} See Appendix \ref{app:palm_int}.

Note that $\PiB^0$ represents the point process of the distances of interfering base stations with respect to the serving base station located at a distance $R_0$ from the origin. In contrast to typical stochastic geometry based analyses of cellular networks, there is now a non-zero probability of interfering base stations being closer to the radar receiver (which is the serving base station in this case) than the typical SO. This "exclusion region" (that is the circle of radius $R_0$ about the typical SO) is an important aspect of this process which has a significant impact on the sensing coverage probability of the network.

 Note also that we have discarded the marks for the LoS and NLoS base stations with respect to the origin. This is due to the fact that the serving base station is also the receiver, and hence observes different LoS/NLoS base stations. As a simplification, we have assumed that the LoS/NLoS status of interfering base stations with respect to the serving base station are independent of the LoS and NLoS status of the base stations observed by the typical SO following A2 and A19. 
 
 Thus, under this assumption, we represent the interfering BSs as a marked point process
\begin{align}
	\PiBt^0 = \{(Z_B^k, M_k, R_k) : R_k \in \PiB, M_k \sim \text{Bern}(\pl(R_k)), Z_B^k \sim (1 - \xibr)\text{ Bern}(\pBr(R_k)) + \xibr \text{ IID} \}.
\end{align}
Equivalently, by the Independent Thinning Theorem the receive antenna gain and LoS marks partition this process into four independent classes of points: those that inside/outside of the receive beam and further those which are LoS/NLoS. These are denoted as $\PiLa^0$, $\PiNa^0$, $\PiLb^0$, and $\PiNb^0$, respectively. 

In light of (\ref{eq:palm-int-meas}) and (\ref{eq:palm-int-rx-pr}), the intensity measures for these processes are not available in closed form. We consequently derived closed form upper and lower bounds for the intensity measures, which are summarized in Lemma 8 in Appendix \ref{app:palm_int_ub}. Armed with these bounds, we then obtain closed form upper and lower bounds on the pathloss Mellin Transforms. These are stated in Lemma 9 in Appendix  \ref{app:int_mt}.

\subsection{Characterizing the Laplace Transform of the Interference Fading Terms}
\label{subsec:int-fad-LT}
To complete our characterization of sensing coverage and rate, we characterize the Laplace Transforms of the aggregate fading terms: $\AM(\bold{F}, \boldsymbol{\theta})$, $\GM(\bold{F}, \boldsymbol{\theta})$, and $\HM(\bold{F}, \boldsymbol{\theta})$. Due to the structure of the means involved in their definitions, these are non-trivial. We provide bounds and approximations for these quantities in the following series of lemmas.

Since the fading terms are dependent on the LoS/NLoS status of the BS in question, wherever necessary we use $a \in \{L, N\}$ as a proxy for subscripts denoting LoS/NLoS parameters. Specifically, we shall denote the dependence of the LoS/NLoS fading parameters as $\Laplace_{\AM(\bold{F}, \boldsymbol{\theta})}(s; N_a)$, $\Laplace_{\GM(\bold{F}, \boldsymbol{\theta})}(s; N_a)$, and $\Laplace_{\HM(\bold{F}, \boldsymbol{\theta})}(s; N_a)$

We begin with the Laplace Transform of $\GM(\bold{F}, \boldsymbol{\theta})$. Notably, we make use of the follow relation between Laplace and Mellin Transforms.

{\bf Proposition 3: (Relation between Mellin and Laplace Transforms of Nonnegative Random Variables)} {\em Let $X$ be a non-negative random variable. Then
\begin{align}
	\Laplace_X(s) = \Mellin^{-1}\{\Mellin_X(1-p)\Gamma(p)\}(s).
\end{align}}
{\em Proof:} See Appendix \ref{app:tech_res}.

{\bf Lemma 10: (Exact Form and Approximation for the Laplace Transform of $\GM(\bold{F}, \boldsymbol{\theta})$)} {\em The following hold.
\begin{enumerate}[label=\roman*)]
	\item $\Laplace_{\GM(\bold{F}, \boldsymbol{\theta})}(s; N_a)$ may be expressed as
	\begin{align}
	\Laplace_{\GM(\bold{F}, \boldsymbol{\theta})}(s; N_a) = \sum_{r = 0}^{T} \pB^{T - r} (1- \pB)^r \sum_{j = 1}^{T \choose r} \Laplace_{\GM(\bold{H}, \bold{q})}\left( s \prod_{i \in T_j^r} \xibt^{w_{t_i}} ; N_a\right),
	\label{eq:lap-gm}
\end{align}
where $T^r_j$ denotes the $j^{th}$ possible permutation of the indices of $T$ time slots such that $r$ of them have $B_t = \xibt$, and
\begin{align}
	\Laplace_{\GM(\bold{H}, \bold{q})}\left( s  ; N_a \right) = \frac{1}{\Gamma(\Na)^N} H^{1, N_c}_{N_c, 1}\left(\frac{s}{\Na} \Bigg\vert \begin{matrix} &(1 - N_a, q_{1}) \dots (1 - N_a, q_{N_c}) \\&(0,1)\end{matrix} \right)
\end{align}
with $H^{m,n}_{p,q}(\cdot \vert \cdot)$ denoting Fox's H function.

\item $\GM(\bold{H}, \bold{q})$ is closely approximated by a gamma random variable, \\ $H' \sim \text{Gamma}(\alpha_0(\bold{q}, \Na), \beta_0(\bold{q}, \Na))$, where $\alpha_0(\bold{q}, \Na)$ and  $\beta_0(\bold{q}, \Na)$ are such that the moments of $H'$ and matched to $GM(\bold{H}, \bold{q})$:
\begin{align}
	&\frac{\alpha_0(\bold{q}, \Na)}{\beta_0(\bold{q}, \Na)} = \Na^{-1}\prod_{n = 1}^{N} \frac{\Gamma(q_n + \Na)}{\Gamma(\Na)}
	\\
	&\frac{\alpha_0(\bold{q}, \Na) (\alpha_0(\bold{q}, \Na) + 1)}{\beta_0(\bold{q}, \Na)^2} = \Na^{-2}\prod_{n = 1}^{N} \frac{\Gamma(2q_n + \Na)}{\Gamma(\Na)}.
\end{align}
Thus,
\begin{align}
	\Laplace_{\GM(\bold{F}, \boldsymbol{\theta})}(s; N_a) \approx \sum_{r = 0}^{T} \pB^{T - r} (1- \pB)^r \sum_{j = 1}^{T \choose r} \left( 1 + \frac{s  \xibt^{\sum_{i \in T_j^r}w_{t_i}}}{\beta_0(\bold{q}, \Na)} \right)^{-\alpha_0(\bold{q}, \Na)}.
	\label{eq:gm_LT_exact}
\end{align}
\end{enumerate}}

{\em Proof:} See Appendix \ref{app:int_fad}.

Though exact, part i) is difficult to use as Fox's H function is challenging to implement in a numerically efficient manner. Hence we employ the gamma approximation in part ii). The gamma approximation is justified in that, for large $N$, $\GM(\bold{H}, \bold{q})$ is closely approximated by a log-normal random variable (a consequence of the Central Limit Theorem). A Log-normal random variable is, in turn, is closely approximated by a gamma random variable.

While (\ref{eq:gm_LT_exact}) may be evaluated efficiently for small $T$, for large $T$ its evaluation becomes cumbersome. Hence, we derive upper and lower bounds in the following lemma.

{\bf Lemma 11: (Bounds on the Laplace Transform of $\GM(\bold{F}, \boldsymbol{\theta})$)} {\em For $s \in \R_+$, the approximation of $\Laplace_{\GM(\bold{F}, \boldsymbol{\theta})}(s; N_a)$ in (\ref{eq:gm_LT_exact}) is upper bounded by $\LaplaceU_{\GM(\bold{F}, \boldsymbol{\theta})}(s; N_a)$, where
\begin{align}
	\LaplaceU_{\GM(\bold{F}, \boldsymbol{\theta})}(s; N_a) = \prod_{t = 1}^{T} \left(\pB + (1- \pB)\left( 1 + \frac{s   \xibt}{\beta_0(\bold{q}, \Na)} \right)^{-\alpha_0(\bold{q}, \Na){w_{t}}}\right).
	\label{eq:lap-gm-lbd}
\end{align}
It is lower bounded by $\LaplaceU_{\GM(\bold{F}, \boldsymbol{\theta})}(s; N_a)$
\begin{align}
	\LaplaceL_{\GM(\bold{F}, \boldsymbol{\theta})}(s; N_a) = \sum_{r = 0}^{T} {T \choose r} \pB^{T - r} (1- \pB)^r \left( 1 + \frac{s  {T \choose r}^{-1} \sum_{j = 1}^{T \choose r}\xibt^{\sum_{i \in T_j^r}w_{t_i}}}{\beta_0(\bold{q}, \Na)} \right)^{-\alpha_0(\bold{q}, \Na)}.
	\label{eq:lap-gm-ubd}
\end{align}}
{\em Proof:} See Appendix \ref{app:int_fad}.

Finally, we now consider the Laplace Transforms of $\AM(\bold{F}, \boldsymbol{\theta})$ and $\HM(\bold{F}, \boldsymbol{\theta})$. Since these correspond to lower and upper bounding models, we derive lower and upper bounds, respectively. 

The Laplace Transform of $\AM(\bold{F}, \boldsymbol{\theta})$ may be lower bounded as follows.

{\bf Lemma 12: (Lower Bound on the Laplace Transform of $\AM(\bold{F}, \boldsymbol{\theta})$)} {\em For $s \in \R_+$, the Laplace Transform of $\AM(\bold{F}, \boldsymbol{\theta})$ is lower bounded by $\LaplaceL_{\AM(\bold{F}, \boldsymbol{\theta})}(s; N_a)$, where
\begin{align}
	\LaplaceL_{\AM(\bold{F}, \boldsymbol{\theta})}(s; N_a) = \sum_{r = 0}^{T} {T \choose r} \pB^{T - r} (1- \pB)^r \left( 1 + \frac{s  {T \choose r}^{-1}\sum_{j = 1}^{T \choose r}\left((\xibt - 1){\sum_{i \in T_j^r}w_{t_i}} + 1\right)}{N \Na} \right)^{-N \Na}.
	\label{eq:lap-am-bd}
\end{align}}
{\em Proof:} See Appendix \ref{app:int_fad}.

The Laplace Transform of $\HM(\bold{F}, \boldsymbol{\theta})$ may be upper bounded as follows.

{\bf Lemma 13: (Lower Bound on the Laplace Transform of $\HM(\bold{F}, \boldsymbol{\theta})$)} {\em For $s \in \R_+$, the Laplace Transform of $\HM(\bold{F}, \boldsymbol{\theta})$ is upper bounded by $\LaplaceU_{\HM(\bold{F}, \boldsymbol{\theta})}(s; N_a)$. Depending on $N_a$ this may be expressed as follows.
\begin{enumerate}[label=\roman*)]
	\item If $N_a = 1$,
	\begin{align}
	\LaplaceU_{\HM(\bold{F}, \boldsymbol{\theta})}(s; N_a) = \pB^{T} \left( 1 + \frac{s}{\left(\sum_{n = 1}^{N} q_n \right)^{2}}  \right)^{-1} + (1- \pB^T) \left( 1 + \frac{\xibt s}{\left(\sum_{n = 1}^{N} q_n \right)^{2}}  \right)^{-1}.
\end{align}
\item Otherwise, for $N_a > 1$, define
\begin{align}
	&m_1 = \left(\left(\pB + (1 - \pB)\xibt^{-1}\right)\frac{\Na}{\Na - 1}\right)^{-1},
	\\
	&m_2 = \left(\sum_{r = 0}^{T} \pB^{T - r} (1- \pB)^r \sum_{j = 1}^{T \choose r} \left( 1 - \left(\sum_{i \in T_j^r}w_{t_i}\right) +  \left(\sum_{i \in T_j^r}w_{t_i}\right) \xibt^{-1} \right)^{-2} \right) \Na^{-2}\prod_{n = 1}^{N} \frac{\Gamma(2q_n + \Na)}{\Gamma(\Na)}
\end{align}
Then,
\begin{align}
	\LaplaceU_{\HM(\bold{F}, \boldsymbol{\theta})}(s;N_a) = \frac{m_2 - m_1^2}{m_2} + \frac{m_1^2}{m_2} \exp\left(-s\frac{m_2}{m_1} \right).
	\label{eq:lap-hm-bd}
\end{align}
\end{enumerate}}
{\em Proof:} See Appendix \ref{app:int_fad}.

In closing, recall that the lower bound in Theorem 3 requires the mean of the NLoS fading term. Hence, we further require $\E\left[ \GM(\bold{F}, \boldsymbol{\theta}) \right]$ and $\E\left[ \AM(\bold{F}, \boldsymbol{\theta}) \right]$. For $\GM(\bold{F}, \boldsymbol{\theta})$ this follows from part ii) of Lemma 10. The mean of $\AM(\bold{F}, \boldsymbol{\theta})$ may be readily obtained as $\E\left[ \AM(\bold{F}, \boldsymbol{\theta}) \right] = \E\left[\lvert H_{n} \rvert^2 B_t\right]$ noting linearity of the arithmetic mean.

\subsection{Statements for Radar Coverage Probability }

Having characterized the Laplace Transforms of the interference fading terms of the alternate SINR models, we arrive at our main result: bounds and approximations for the sensing coverage probability.

{\bf Theorem 4: (Characterization of Sensing Coverage Probability)} {\em The following bounds and approximations hold on the sensing coverage probability, $\Pcr(\tau)$.
\begin{enumerate}[label=\roman*)]
	\item The sensing coverage probability is lower bounded using (\ref{eq:pcr-gen-lb}) and (\ref{eq:lap-am-bd}) as
	\begin{align}
		\Pcr(\tau) \ge \Pcrl\left(\tau; \LaplaceL_{\AM(\bold{F}, \boldsymbol{\theta})}(s; \NL), \LaplaceL_{\AM(\bold{F}, \boldsymbol{\theta})}(s; \NN) \right).
	\end{align}
	\item The sensing coverage probability is upper bounded using (\ref{eq:pcr-gen-ub}) and (\ref{eq:lap-hm-bd})
	\begin{align}
		\Pcr(\tau) \le \Pcru\left(\tau; \LaplaceU_{\HM(\bold{F}, \boldsymbol{\theta})}(s; \NL), \LaplaceU_{\HM(\bold{F}, \boldsymbol{\theta})}(s; \NN) \right).
	\end{align}
	\item The sensing coverage probability is approximately lower bounded using (\ref{eq:pcr-gen-lb}) and (\ref{eq:lap-gm}) as
	\begin{align}
		\Pcr(\tau) \approx \Pb(\SINRgm \ge \tau) \ge \Pcrl\left(\tau; \Laplace_{\GM(\bold{F}, \boldsymbol{\theta})}(s; \NL), \Laplace_{\GM(\bold{F}, \boldsymbol{\theta})}(s; \NN) \right).
	\end{align}
	\item The sensing coverage probability is approximately upper bounded using (\ref{eq:pcr-gen-ub}) and (\ref{eq:lap-gm}) as
	\begin{align}
		\Pcr(\tau) \approx \Pb(\SINRgm \ge \tau) \le \Pcru\left(\tau; \Laplace_{\GM(\bold{F}, \boldsymbol{\theta})}(s; \NL), \Laplace_{\GM(\bold{F}, \boldsymbol{\theta})}(s; \NN) \right).
	\end{align}
\end{enumerate}}

{\em Proof:}  This follows from the generic coverage probability expressions in Theorem 3 and the stochastic orderings of the SINR models in Proposition 2. The fact that the bounds for the interference Laplace Transform in Theorem 2 are increasing in the Laplace Transform of the fading terms concludes the proof, leveraging Lemmas 10 through 13. $\qedwhite$

With respect to iii) and iv), looser, but simpler, bounds may be obtained using $\LaplaceL_{\GM(\bold{F}, \boldsymbol{\theta})}(s; \Na)$ and $\LaplaceL_{\GM(\bold{F}, \boldsymbol{\theta})}(s; \Na)$, in (\ref{eq:lap-gm-lbd}) and (\ref{eq:lap-gm-ubd}) respectively. Further, Theorem 4 may be used in tandem with Corollary 3 to obtain both approximate and exact upper and lower bounds on the ergodic sensing efficiency. Overall, these results offer a highly expressive yet tractable means to analytically characterize the performance of parameter estimation in JCAS networks. In addition to capturing the effects of non-isotropic interference, blockages, and  varying antenna gains, our model further accounts for the structure of the excitation waveform over time and frequency.

Finally, we note that, although we have focused on bounds and approximations, the fact we obtain true upper and lower bounds on $\Pcr(\tau)$ allows us to immediately obtain analytical characterizations of the tightness of the bounds and approximations. For instance the approximation error with respect to $\SINRr$ and $\SINRgm$ may be upper bounded as
\begin{align}
	 \lvert \Pb(\SINRr &\ge \tau) - \Pb(\SINRgm \ge \tau) \rvert \le \nonumber
	\\
	&\Pcru\left(\tau; \LaplaceU_{\HM(\bold{F}, \boldsymbol{\theta})}(s; \NL), \LaplaceU_{\HM(\bold{F}, \boldsymbol{\theta})}(s; \NN) \right) - \Pcrl\left(\tau; \LaplaceL_{\AM(\bold{F}, \boldsymbol{\theta})}(s; \NL), \LaplaceL_{\AM(\bold{F}, \boldsymbol{\theta})}(s; \NN) \right).
\end{align}


\section{Communication Coverage Probability and Ergodic Efficiency}
\label{sec:pcc_analysis}
We now conclude the presentation of our analytical results by characterizing the communication coverage probability and ergodic communication efficiency. As there is little novel in these (aside from a usage of Theorem 2 in Appendix \ref{app:LST_bds}), we present the main results without proof.

{\bf Theorem 5: (Characterization of Communication Coverage Probability)} {\em The path loss Mellin Transform of the LoS interference process may be expressed with respect to the $G_L$ function defined in (\ref{eq:GL-func}) in Appendix \ref{app:int_mt},
\begin{align}
	\Mellin_{\lambdaL^0 \circ \gL^{-1}}(p; A, R_0) = 2 \pi \lambdaB \KL^{p-1} G_L\left(p-1, A; [R_0, \infty), \alphaL, \gammaL, 1, \beta\right),
\end{align}
where $\lambdaL^0(r; R_0) = 2 \pi \lambdaB r e^{-\beta r} \indicator\{r \ge R_0\}$.

Similarly, for the NLoS interference process, using the $G_N$ function defined in in (\ref{eq:GN-func}) in Appendix \ref{app:int_mt}, we have
\begin{align}
	\Mellin_{\lambdaN^0 \circ \gN^{-1}}(p; A, R_0) = 2 \pi \lambdaB \KN^{p-1} G_N\left(p-1, A; [R_0, \infty), \alphaN, \gammaN, 1, 0, \beta\right),
\end{align}
where $\lambdaN^0(r; R_0) = 2 \pi \lambdaB r ( 1- e^{-\beta r}) \indicator\{r \ge R_0\}$.

The LoS equivalent distribution of the serving base station (see \cite[Lemma 5]{Olson2022}) may be expressed as 
\begin{align}
	f_{R_0}(r) = 2 \pi \lambdaB \lambdaEQ^{*}(r)  e^{-2 \pi \lambdaB \LambdaEQ^{*}([0, r])},
\end{align}
where
\begin{align}
	&\lambdaEQ^{*}(r) = r e^{-\beta r} + \frac{\psi^{-1}(r)( \alpha r^{-1} + \gammaL)}{\gammaN \psi^{-1}(r) + \alpha} \psi^{-1}(r) \left( 1 - e^{-\beta \psi^{-1}(r)} \right),
\end{align}
and
\begin{align}
	 \LambdaEQ^{*}([0,  r]) 
 &= \beta^{-2} \left( \frac{\beta^2}{2} [\psi^{-1}(r)]^2 + e^{-\beta \psi^{-1}(r)} \left( \beta \psi^{-1}(r) + 1 \right) - e^{-\beta r}( \beta r + 1)  \right).
\end{align}
Finally, the Laplace Transform for the LoS/NLoS interference fading terms may be expressed as 
\begin{align}
	\Laplace_{F_a}(s) = \E_{B, Z_U}\left[ \left(1 + \frac{s B Z_U}{\Na} \right)^{-\Na} \right].
\end{align}

Then, letting $N_w \in \N$ and defining $\{W_i^L\}_{i \in [N_w + 1]}$ and $\{W_i^N\}_{i \in [N_w + 1]}$ as in Definition 3, the following hold

\begin{enumerate}[label=\roman*)]
	\item The communication coverage probability is lower bounded as $\Pb(\SINRc \ge \tau) \ge \Pccl\left(\tau; \Laplace_{F_L}, \Laplace_{F_N}\right)$ where
		\begin{align}
 &\Pccl(\tau; \Laplace_{F_L}, \Laplace_{F_N}) = \int_{\R_+} f_{R_0}(u) \exp\left({\frac{-\tau \nur}{\gL(u)}}\right) \nonumber
 \\
 &\exp\left(-\sum_{i = 1}^{N_w +1} \HL\left(\frac{\tau}{\gL(u)}, W^L_i; 1 -\Laplace_{F_L}, \Mellin_{\lambdaL^0 \circ \gL^{-1}}(\cdot; \cdot, u)\right) \right) \nonumber
\\
 &\exp\left(-\sum_{i = 1}^{N_w} \HL\left(\frac{\tau}{\gL(u)}, W^N_i; 1 -\Laplace_{F_N}, \Mellin_{\lambdaN^0 \circ \gN^{-1}}(\cdot; \cdot, u)\right)  + \frac{\tau}{\gL(u)} \E[F_N] \Mellin_{\lambdaN^0 \circ \gN^{-1}}(2; W^N_{N_w + 1}, u) \right) du.
\end{align}
\item The communication coverage probability is upper bounded as $\Pb(\SINRc \ge \tau) \le \Pccu\left(\tau; \Laplace_{F_L}, \Laplace_{F_N}\right)$ where
		\begin{align}
 \Pccu&(\tau; \Laplace_{F_L}, \Laplace_{F_N}) = \int_{\R_+} f_{R_0}(u) \exp\left({\frac{-\tau \nur}{\gL(u)}}\right) \nonumber
 \\
 &\exp\left(-\sum_{i = 1}^{N_w +1} \HU\left(\frac{\tau}{\gL(u)}, W^L_i; 1 -\Laplace_{F_L}, \Mellin_{\lambdaL^0 \circ \gL^{-1}}(\cdot; \cdot, u)\right) \right) \nonumber
\\
 &\exp\left(-\sum_{i = 1}^{N_w} \HU\left(\frac{\tau}{\gL(u)}, W^N_i; 1 -\Laplace_{F_N}, \Mellin_{\lambdaN^0 \circ \gN^{-1}}(\cdot; \cdot, u)\right) \right) du.
\end{align}
\end{enumerate}}

In a similar manner to Corollary 3 we bound the ergodic communication efficiency as follows

{\bf Corollary 5: (Bounds on the Ergodic Communication Efficiency)} {\em The ergodic communication efficiency may be bounded as follow.  
\begin{align}
	\Ts \E\left[ \Rc^{(m,n)} \right] =  \E\left[\log_2(1 + \SINRc)\right] \ge \frac{1}{\ln(2)}\int_{\R_+} \Pcrl \left(x; \Laplace_{F_L}, \Laplace_{F_N} \right) \left(1 + \tau\right)^{-1} d\tau,
\end{align}
and
\begin{align}
	\Ts \E\left[ \Rc^{(m,n)} \right] =  \E\left[\log_2(1 + \SINRc)\right] \le \frac{1}{\ln(2)}\int_{\R_+} \Pcru \left(x; \Laplace_{F_L}, \Laplace_{F_N} \right) \left(1 + \tau\right)^{-1} d\tau.
\end{align}}

\section{Numerical Analysis}
\label{sec:num_res}
Leveraging these analytical results, we now present numerical analysis of the JCAS network and discuss the insights they provide from a system design perspective. Following, \cite{Wild2021} we consider a sensing scenario in which BSs in the network are primarily used for sensing vehicles in an urban environment.

Unless otherwise stated, the network parameters are set as follows. We consider the center frequency to be $f_c = 75$ GHz. Using the Friis transmission model the pathloss intercept for the LoS case is set to $\KL = \left(\frac{c_0}{4 \pi f_c}\right)^2$, and the NLoS intercept is set to be 15 dB lower. Based on \cite{Rappaport2014}, we set the LoS path loss exponent to $\alphaL = 2$ and the NLoS path loss exponent to $\alphaN = 3.2$. Similarly, from \cite{ITUR2013} the LoS absorption coefficient is set to $\gammaL = 5\text{e}^{-6}$ and the NLoS absorption coefficient to $\gammaN = 5\text{e}^{-3}$. The beam widths are set to $\thetabt = 5$ degrees at the BS transmitter, $\thetabr = 5$ degrees at the BS receiver, and $\thetaur = 30$ degrees at the UE. The front to back ratios are $\xibt = -35$ dB at the BS transmitter, $\xibr = -20$ dB at the BS receiver, and $\xiur = -15$ dB at the UE. We set main lobe gain to a physically realizable value given the beam width and front to back ratios following \cite{Petrov2017}. This results in antenna gains of $\Gbt = 31$ dB, $\Gbr = 19.8$ dB, and $\Gur = 13.2$ dB. Thus, assuming a transmit power of $P_\text{t} =15$ dBm and noise power of $P_\text{n} = -123.2$ dBm, the normalized SNR for communication case is $\nuc^{-1} = 182.4$ dB and $\nur^{-1} = 189.02$ dB. Finally, we assume that the LoS interference fading terms are of the order $\NL = 3$, and the NLoS terms are $\NN = 2$.

Regarding the parametrization of the  waveform and the composition of $\Sr$, we follow \cite{Wild2021} and \cite{dahlman2021}. Using numerology 3 in the 5G NR specification, we set the subcarrier spacing to $\Delta f = 120$ kHz and the guard interval to $T_g = 570$ ns. Thus, the constants $k_1$ and $k_2$ from Theorem 1 are $k_1 = 8\text{e}^{-4}$ and $k_2 = 8.903\text{e}^{-3}$. Moreover, $\Sr$ is composed in a manner following \cite{Wild2021}. Under the objectives that our range resolution is $\Delta r  = 1$ m, our velocity resolution is $\Delta v = 1.33$ m/s, our maximum unambiguous range is $r_\text{max} = 300$ m, and our max velocity is $v_\text{max} = 200$ km/h we use one in every $\Delta N_\text{s} =3$  symbols and one in every $\Delta N_\text{c} =14$ subcarriers. In total, the excitation signal covers $N_\text{s} = 264$  symbols and uses the total bandwidth, corresponding to $N_\text{c} = 3168$ subcarriers. We set the prior covariance for the range/velocity estimates to the identity matrix.

Finally, we set the blockage density to $\beta^{-1} = 140$ (which corresponds to an urban environment) and effective cell radius $r_c = \sqrt{1/(\pi \lambdaB)} = 100$ m. The key network parameters are summarized in the following table.
\begin{center}
	\begin{tabular}{ |M{2cm}|M{2.75cm}||M{2cm}|M{2.75cm}||M{2cm}|M{2.75cm}|}
	\hline
	 {\bf Parameter } &{\bf Value} &{\bf Parameter } &{\bf Value} &{\bf Parameter } &{\bf Value} \\
	 \hline \hline
	 $\boldsymbol{f_c}$ &$75$ GHz  &$\boldsymbol{\Gbr}$ &$19.8$ dB &$\boldsymbol{\Delta r}$  &$1$ m\\
	 \hline
	 $\boldsymbol{\KL}$ &$-75.96$ dB  &$\boldsymbol{\xibr}$ &$-20$ dB &$\boldsymbol{\Delta v}$ &$1.33$ m/s\\
	 \hline
	 $\boldsymbol{\KN}$ &$-90.96$ dB &$\boldsymbol{\thetabr}$ &$5$ deg. &$\boldsymbol{r_\text{max}}$ &$300$ m\\
	 \hline
	 $\boldsymbol{\alphaL}$ &$2$  &$\boldsymbol{\Gur}$ &$13.2$ dB  &$\boldsymbol{v_\text{max}}$ &$200$ km/h\\
	 \hline
	 $\boldsymbol{\alphaN}$ &$3.2$ &$\boldsymbol{\xiur}$ &$-15$ dB  &$\boldsymbol{N_\text{s}}$ &$264$\\
	 \hline
	 $\boldsymbol{\gammaL}$ &$5e^{-6}$ &$\boldsymbol{\thetaur}$ &$30$ deg. &$\boldsymbol{N_\text{c}}$ &$3168$\\
	 \hline
	 $\boldsymbol{\gammaN}$ &$5e^{-3}$ &$\boldsymbol{\NL}$ &$3$  &$\boldsymbol{\Delta N_\text{s}}$ &$3$\\
	 \hline
	 $\boldsymbol{P_\text{t}}$ &$15$ dBm  &$\boldsymbol{\NN}$ &$2$  &$\boldsymbol{\Delta N_\text{c}}$ &$14$\\
	 \hline
	 $\boldsymbol{P_\text{n}}$ &$123.2$ dBm &$\boldsymbol{\Delta f}$ &$120$ kHz &$\boldsymbol{Q}$ &$I$\\
	 \hline
	 $\boldsymbol{\Gbt}$ &$31$ dB &$\boldsymbol{T_g}$ &$570$ ns &$\boldsymbol{\beta^{-1}}$ &$140$\\
	 \hline
	 $\boldsymbol{\xibt}$ &$-35$ dB &$\boldsymbol{k_1}$ &$8e^{-4}$   &$\boldsymbol{r_c}$ &$100$ m\\
	 \hline
	 $\boldsymbol{\thetabt}$ &$5$ deg. &$\boldsymbol{k_2}$ &$8.903e^{-3}$ &$\cdot$&$\cdot$\\
	 \hline
	 \hline 
	\end{tabular}
	\end{center}

\subsection{Sensing and Communication Coverage Probability}
\begin{figure}[bt]
\includegraphics[width=1\linewidth]{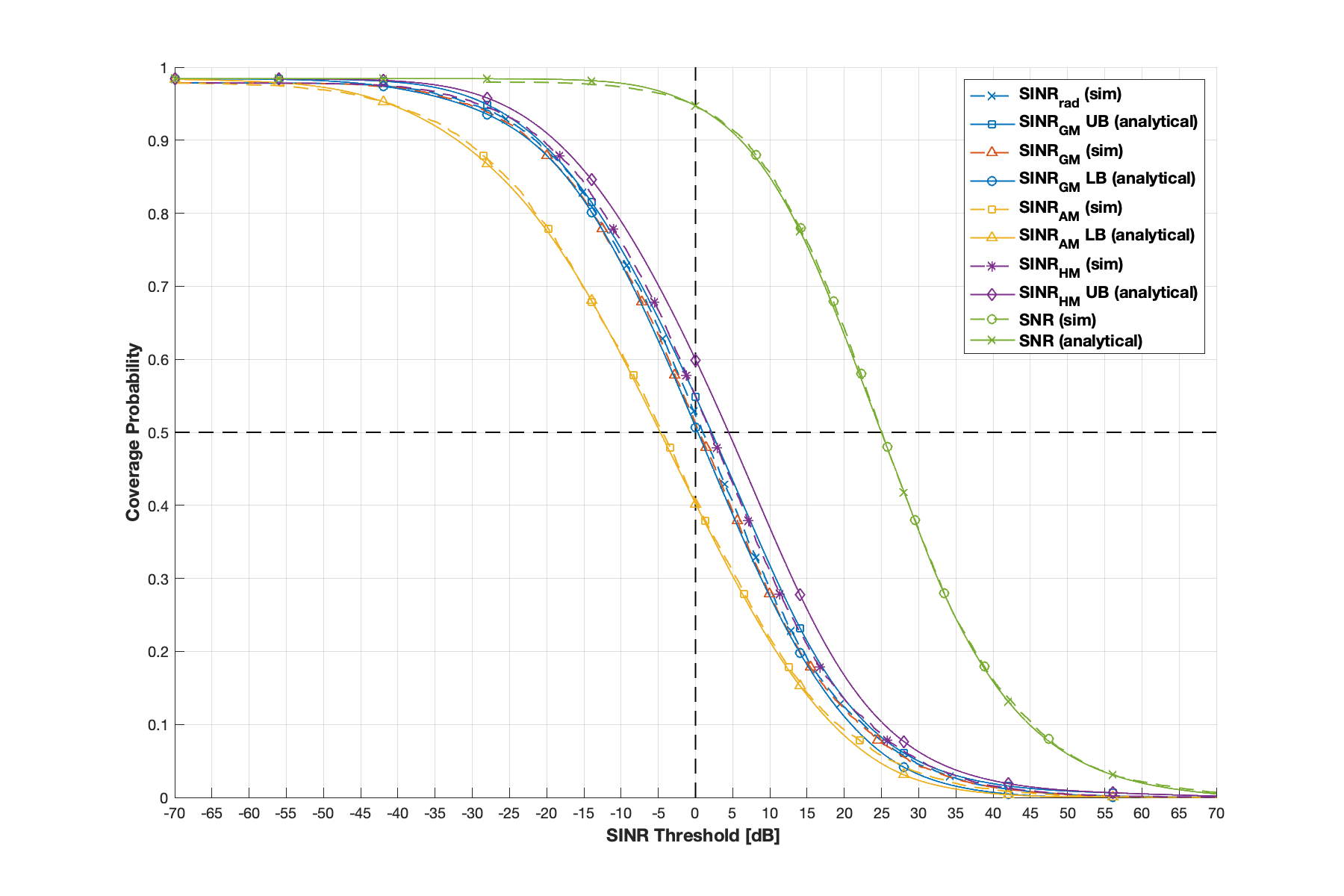}
			\centering
			\caption{This figure depicts analytical upper and lower bounds and simulated CCDFs for various radar  SINR models. These indicate that a) the analytical bounds are tight in most cases, and b) the approximate radar SINR model, $\SINRgm$, is a close fit to the true model, $\SINRr$.}
\label{fig:rad_cov_gam}
\end{figure}

\begin{figure}[bt]
\includegraphics[width=1\linewidth]{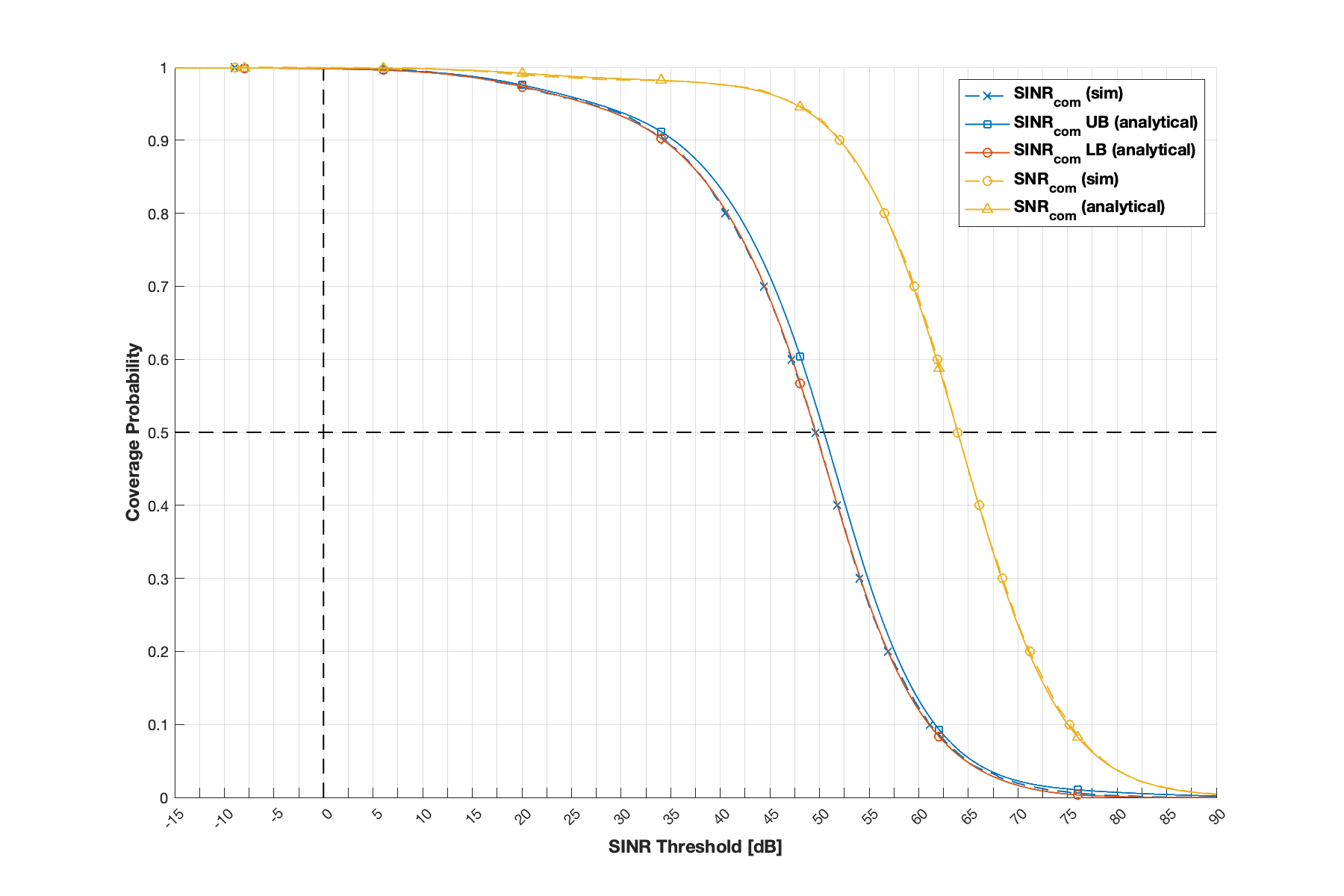}
			\centering
			\caption{This figure depicts analytical upper and lower bounds and simulated CCDFs for the communication SINR models. Like the radar case, these indicate that the analytical bounds are tight. }
\label{fig:comm_cov_gam}
\end{figure}

Figs. \ref{fig:rad_cov_gam} and \ref{fig:comm_cov_gam} depict the sensing and communication coverage probabilities for the JCAS network detailed above. For the sensing case both analytical bounds and simulated CCDFs are depicted for $\SINRam$, $\SINRgm$, $\SINRhm$, and the radar SNR. For the sake of comparison, simulated curves are depicted for $\SINRr$. For the communication case, analytical bounds and simulated curves are depicted for $\SINRc$. The analytical bounds were obtained using Theorems 4 and 5, and the simulated curves were obtained using Monte Carlo simulations of the JCAS network. An immediate takeaway is that the analytical upper and lower bounds are generally tight in all cases.

Figs. \ref{fig:rad_cov_gam} indicates that {\bf $\SINRgm$ is a good approximation of $\SINRr$}. For the aforementioned parametrization of the network, their respective CCDFs are empirically quite close. The usage of $\SINRgm$ is further motivated by the upper and lower bounding models $\SINRhm$ and $\SINRam$ from Proposition 2. The upper bounding model, $\SINRhm$, lies above the CCDFs for $\SINRgm$ and $\SINRr$ by a few dB. The closeness of $\SINRgm$ as an approximation for $\SINRr$ is further underscored by the analytical upper bound. The lower bounding model differs by 5 to 10 dB for low SINR thresholds, $\tau$. It is expected to be somewhat looser, as its derivation involve two applications of Jensen's Inequality with respect to $\SINRr$ in comparison to only one for $\SINRhm$.

{\bf These figures also demonstrate the utility of Theorem 2 in obtaining tractable bounds on the distributions of SINR models}, when their distributions may be expressed in terms of the Laplace Transform of their interference process. In both figures, matching upper and lower bounds are shown for $\SINRgm$ and $\SINRc$. These bounds are close -- especially for the case of $\SINRc$ -- and differ by only a few dB in the sensing case and less than a dB in the communication case. 

From a system design perspective, these figures demonstrate that {\bf interference and blockages play a greater role in the link quality of a typical sensing target than that of a typical communication user}. While the overall average SINR is lower in the sensing case due to the decreased path loss, this is made up for in part by the distribution of the excitation signal over time and frequency (which increases the processing gain analog, $G$). Looking at relative features, then, we see a prominent downward shift in the sensing SINR curves due to the probability that a typical SO may have no LoS BSs. Moreover, in comparison to the communication case, we see that the gap between the SNR and SINR is slightly wider. This is partially explained by the fact that the BS serving the typical SO is also the radar receiver due to the considered mono-static configuration, which leaves open the possibility of interferers being closer than the sensing target.
\subsection{Ergodic Estimation and Communication Efficiency with Varying Network Density}
\begin{figure}[bt]
\includegraphics[width=1\linewidth]{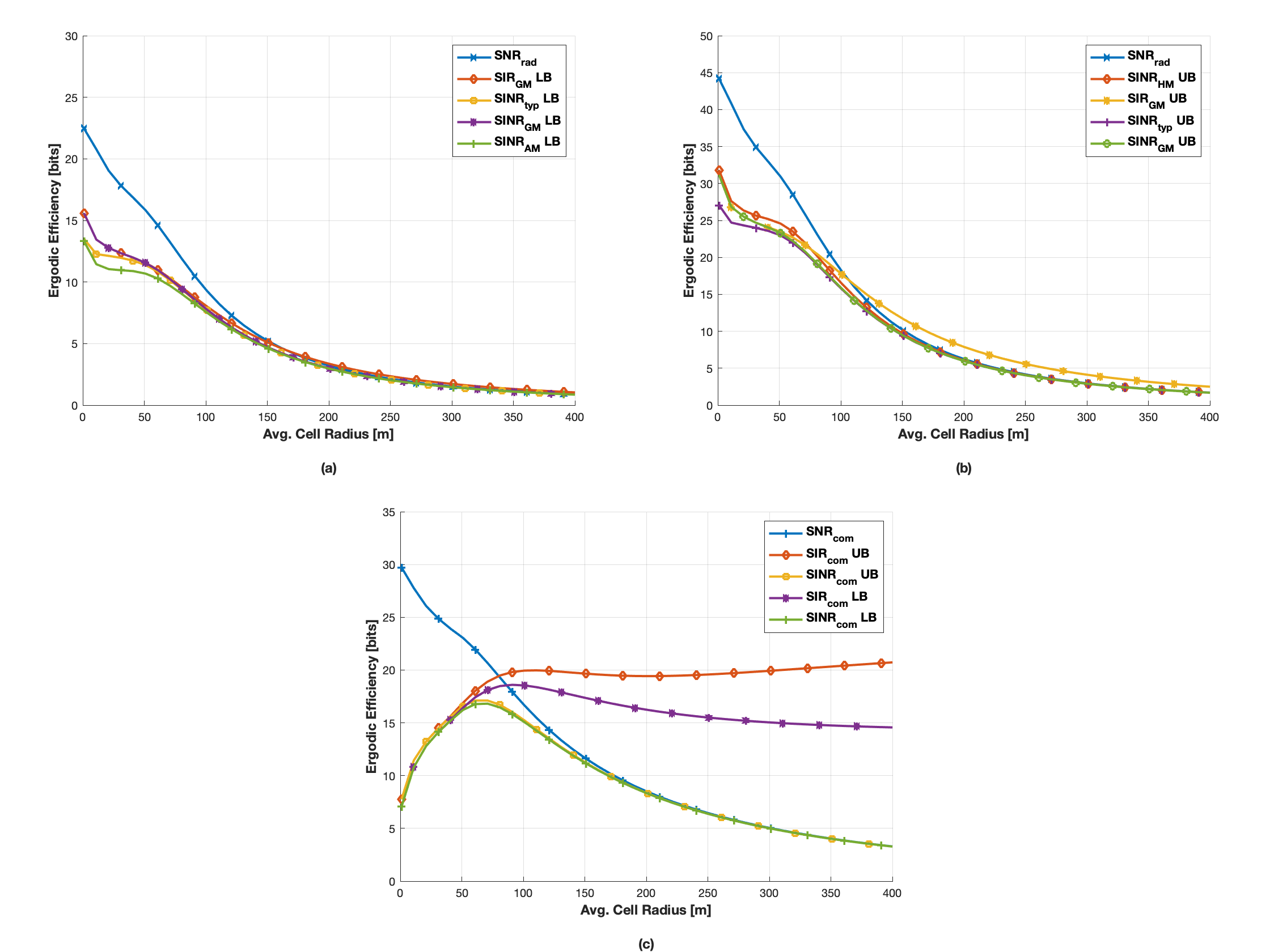}
			\centering
			\caption{This figure depicts lower (a) and upper (b) bounds for the ergodic estimation efficiency of the JCAS network with respect to various SINR models as well as bounds on the ergodic communication efficiency (c) in the high blockage regime with $\beta^{-1} = 360.67$.}
\label{fig:rc_low_block}
\end{figure}

\begin{figure}[bt]
\includegraphics[width=1\linewidth]{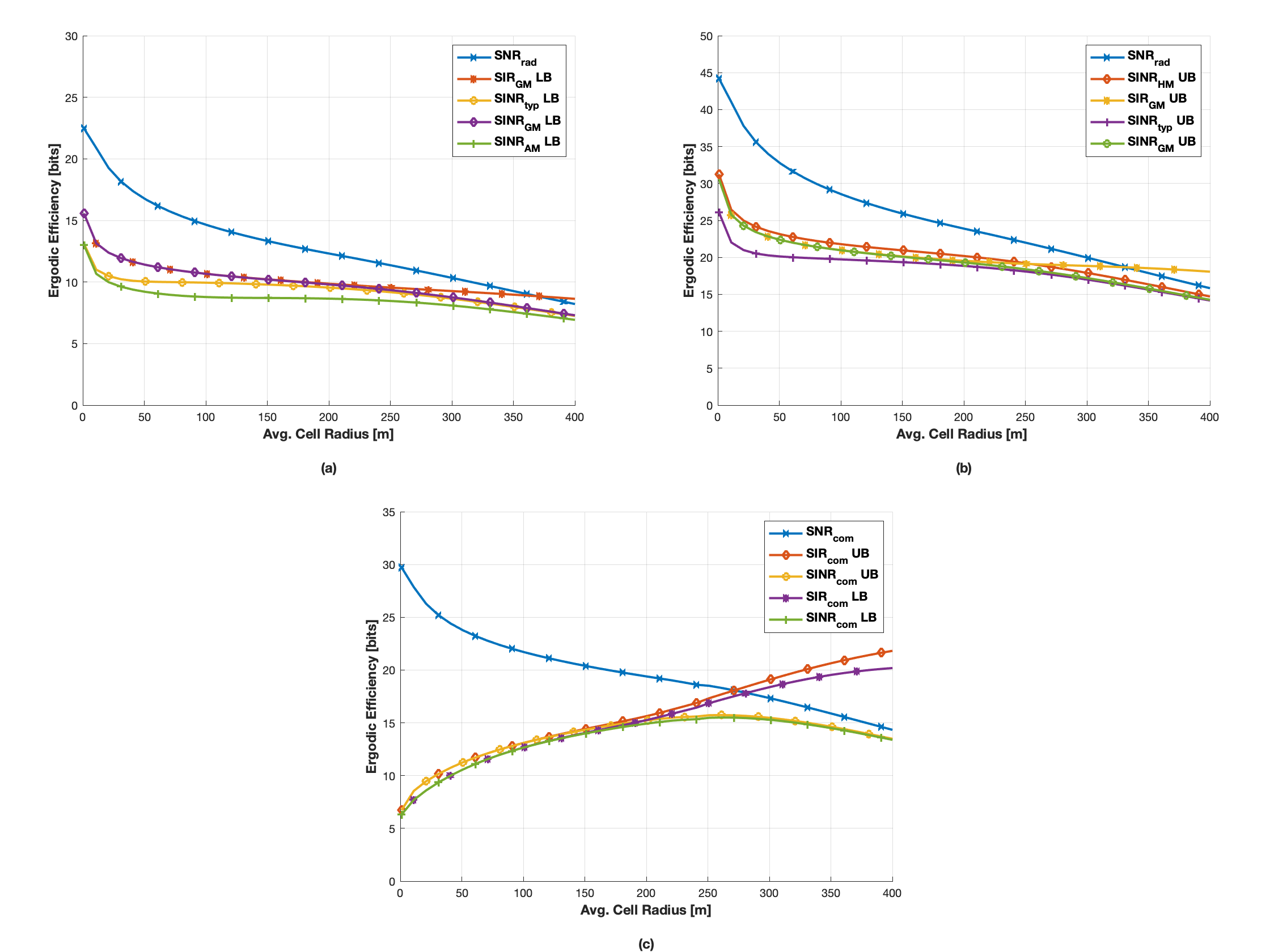}
			\centering
			\caption{This figure depicts lower (a) and upper (b) bounds for the ergodic estimation efficiency of the JCAS network with respect to various SINR models as well as bounds on the ergodic communication efficiency (c) in the low blockage regime with $\beta^{-1} = 72.13$.}
\label{fig:rc_high_block}
\end{figure}

\begin{figure}[bt]
\includegraphics[width=1\linewidth]{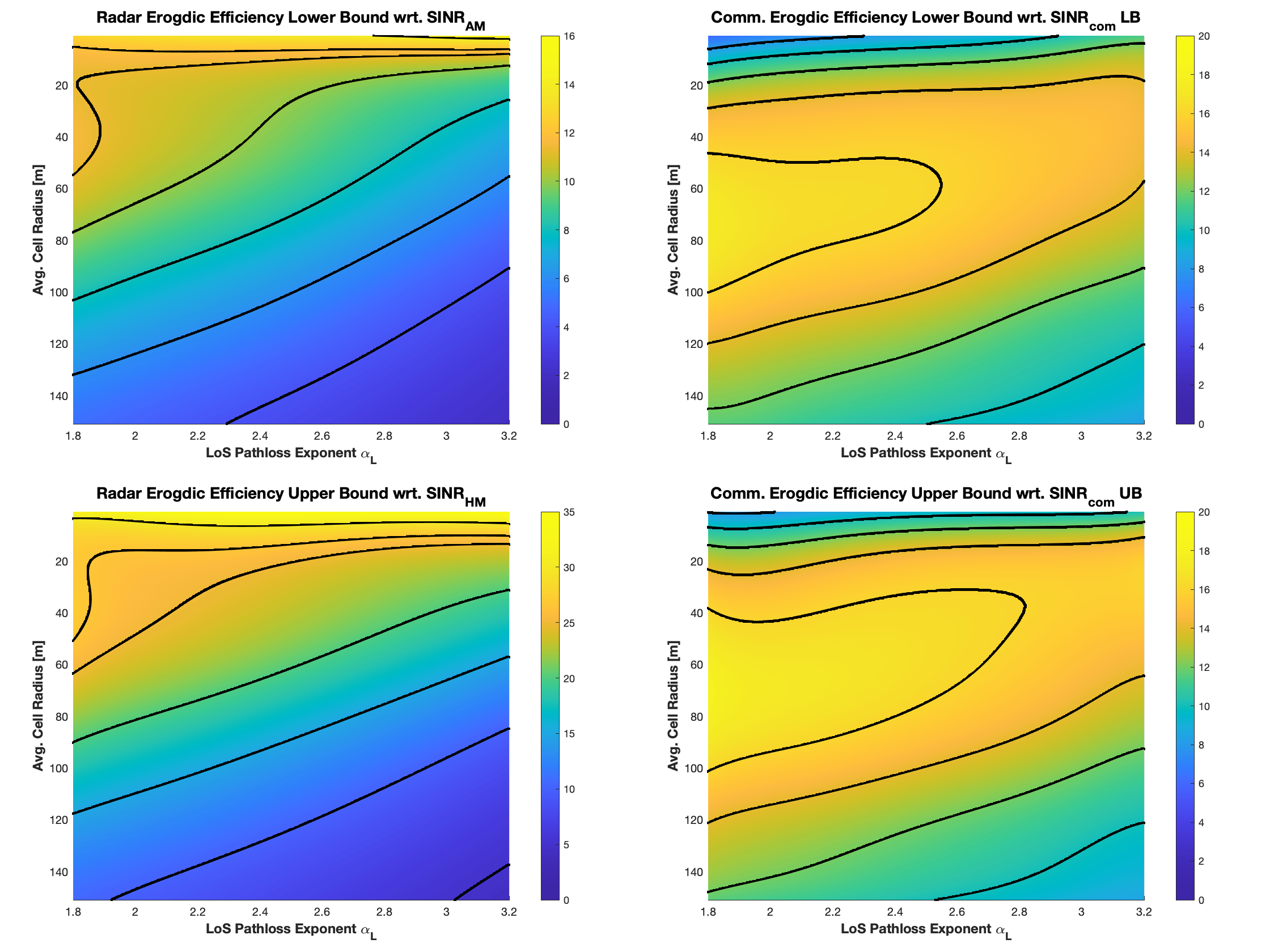}
			\centering
			\caption{This figure depicts how the sensing and communication ergodic efficiencies vary jointly with the network density and the LoS pathloss exponent, $\alphaL$, in the high blockage regime. Lower bounds for the ergodic efficiencies are depicted for sensing with respect to the lower bounding model, $\SINRam$, (a) and communications (b). Similarly, upper bounds for the ergodic efficiencies are depicted for sensing with respect to the upper bounding model, $\SINRhm$, (c) and communications (d). A few representative level sets for these surfaces are shown as block lines to improve clarity.}
\label{fig:pl_sweep}
\end{figure}

In Figs. \ref{fig:rc_low_block} and \ref{fig:rc_high_block} we consider the impact of base station and blockage density on the ergodic sensing  and communication efficiency of the JCAS network. To this end, we sweep the average cell radius at different levels of blockage density -- as captured by $\beta$. We consider a low blockage regime $\beta^{-1} = 360.67$ and a high blockage regime, $\beta^{-1} = 72.13$. These are set such that the median LoS probability occurs at $r = 50$ and $r = 250$ meters, respectively. Note that we also adjust the maximum unambiguous range for sensing to be $3 r_c$ to track with varying network density. This changes the structure of $\Sr$.

The curves depicted in the figures were obtained using Corollaries 3 and 5. For the sake of discussion, we have included new radar SINR model for the sensing case, $\SINRt$. This model is simply the SINR of the radar return at an arbitrary resource element, and is meant to highlight the impact of the distribution of the excitation signal over time and frequency. Formally, it is defined as
\begin{align}
\SINRt = \indicator\{\PhiL(\R^2) > 0\} \frac{ \kc  \gLr(\norm{X_0})  }{ \sum_{X_k \in \PhitB \setminus \{X_0\}} \lvert H^k\rvert^2 B Z_B^k L(\norm{X_k - X_0}) + \nur}.
\end{align}	
Note that bounds on its CCDF may be obtained  using the bounds on $\SINRgm$ from Theorem 4 with $\Sr$ having support over a single resource element.

 In each of the figures, the bounds for the radar case depict the ergodic efficiency with respect to the approximate model, $\SINRgm$, the typical resource element, $\SINRt$, the SIR with respect to the approximate model, $\SIRgm$, and the SNR, $\SNRr$. The lower bounds additionally include the sensing efficiency with respect to the lower bounding SINR model, $\SINRam$, and the upper bounds include the estimation efficiency with respect to the upper bounding SINR model, $\SINRhm$. The communication ergodic efficiency is depicted with respect to the SINR $\SINRc$, SIR, $\SIRc$, and SNR, $\SNRc$. Upper and lower bounds are obtained using the corresponding bounds in Theorems 4 and 5 along with Corollaries 3 and 5. While the upper and lower bounds on ergodic sensing efficiency differ by around a factor of 2, they exhibit the same trends in all cases.
 
 These figures reveal several insights into the performance trends of communication and sensing in JCAS networks. Blockage density plays a significant role in the performance of both functions, though its impact on sensing is more pronounced. Generally, increasing blockage density improves JCAS network performance at high to moderate base station densities, and degrades performance at lower base station densities. The improvement is due to the reduction in overall interference power, while the reduction is due to the degradation of the desired signal power resulting from the scarcity of LoS base stations. This effect is compounded on sensing, as it requires a LoS base station, while communication can leverage strong NLoS base stations. Hence, the faster decreasing of ergodic sensing efficiency with increasing cell radius at moderate and high blockage densities is due to increasing probability that the target is invisible to the network (i.e. the probability that $R_0 = \infty$).
 
 More interesting trends are observed with respect to the variation of the performance of communication and sensing with respect to base station density. These trends are similar at different blockage densities, so we focus primarily on Fig. \ref{fig:rc_high_block}. At moderate to high base station densities, in the range of $r_c \le 75$ m, the sensing efficiency is relatively insensitive or increases to changes in base station density. 
 While the strength of the desired signal improves, as indicated by the sensing efficiency with respect to $\SNRr$, these improvements are mostly offset by the increased interference power. 
 Interestingly, this is not the case for communication efficiency, which decreases with increasing base station density in this range -- a well known phenomena in the cellular networks \cite{AlAmmouri2019}.

 To provide further insight into these trends, we investigate the variation of communication and sensing ergodic efficiency with respect to both the LoS pathloss exponent, $\alphaL$, and network density in Fig. \ref{fig:pl_sweep}. Note that the upper and lower bounds for the sensing case are computed with respect to $\SINRhm$ and $\SINRam$, respectively, as these models provide true upper and lower bounds. This figure confirms the trends observed in Figs. \ref{fig:rc_low_block} and \ref{fig:rc_high_block}: over the range of pathloss exponents considered, {\bf sensing ergodic efficiency is non-decreasing with increasing network density, while communication efficiency is maximized for some value of $r_c$ and then decreases thereafter}.
  We hypothesize that the differences between the densification trends for communication and sensing are principally driven by the additional factor of two in the pathloss exponent and absorption factor for the two way radar path loss model. While the additional distance induced by the reflection reduces overall radar pathloss, the derivative of the radar path loss is steeper due to the additional factor of two in the path loss exponent and absorption factor. Viewing the increase in base station density as  decreasing the distances of all links, {\bf the marginal improvement on the desired signal, which is based on the radar path loss model, outstrips the marginal increase in interference power, which is based on either the LoS or NLoS path loss models}. Hence, we observe the aforementioned insensitivity of ergodic sensing efficiency at moderate base station densities, and the increase at high densities. {\bf Overall, these results highlight a tension between optimal regimes of base station density for communication and sensing: the sensing function performs best at high base station densities, while the communication performance is optimized at moderate base station densities.}
 
 Finally, we comment on a few remaining features of interest. First, note that the sensing efficiency with respect to $\SINRgm$ and $\SINRhm$ dominates the sensing efficiency with respect to $\SINRt$. While the this is not the case for the efficiency with respect to $\SINRam$, this is likely due to the model being more pessimistic, as mentioned previously. This indicates that {\bf the distribution of the excitation signal over time likely reduces the impact of worst case interference scenarios in which a BS in the main lobe of the receive beam is also pointing its transmit beam in the direction of the serving BS. }Additionally, at low base station densities, the ergodic efficiency with respect to the $\SIRgm$ tracks more closely with $\SNRr$ than does the ergodic efficiency of $\SIRc$ with respect to $\SNRc$. This indicates that, even at low base station densities, {\bf interference remains a more prominent feature on the performance of the sensing function than for communications.}
 
\section{Conclusion}
\label{sec:conclusion}
We have presented a novel analytical framework for characterizing the coverage probability and ergodic rate of JCAS in cellular networks. Using a characterization of parameter estimation based on mutual information, we extended the notion of coverage probability to the radar setting, defining it as the probability that the rate of information extracted about the parameters of interest associated with a typical sensing target being tracked by the network exceeds some threshold. Similarly, we define ergodic sensing rate as the expected value of this quantity. Focusing on the setting of doppler and delay estimation, we established upper and lower bounds on the sensing rate in terms of an aggregate SINR induced by taking a weighted average of per-resource element SINRs experienced by the radar receiver over the excitation signal. Using this model, we developed a stochastic geometry framework with which to characterize the sensing and communication coverage probabilities and ergodic capacities in a mmWave JCAS network employing a shared multi-carrier waveform and directional beamforming. As a first step in our analysis, we derived a generic method for obtaining closed form upper and lower bounds on the Laplace Transform of a shot noise process. We leveraged this result to establish bounds on the CCDFs of a series of bounding and approximate sensing SINR models and the communication SINR model. We further established expressions for the ergodic sensing and communication efficiencies as corollaries with respect to these bounds. Over the course of our analysis, we developed some noteworthy, independent results: an analog to H{\"o}lder's inequality in the setting of harmonic, rather than geometric, means of functions in a measure space, and a result linking the Laplace Transform of a non-negative random variable to its Mellin Transform via an inverse Mellin Transform. Using our bounds for JCAS coverage and rate, we investigated performance trends in a numerical case study. Among several insights, our numerical analysis indicates the possibility that network densification improves sensing performance -- in contrast to the communication function. 

\section{Future Work}
\label{sec:future-work}
Our work is an initial foray into the analysis of JCAS networks, and could provide a foundation for an array of interesting potential extensions. One such extension would be to generalize our notion of sensing performance to account for both parameter estimation and detection. This would necessarily require a richer metric as now one would want to quantify the utility of sensing measurements in enabling the BSs to revise their beliefs about both the number of SOs and their associated parameters of interest. An additional challenge also arises from need to address the impact of false detections.  One approach to tackle these issues would be to leverage formulations of multi-target tracking and detection based on point processes -- also called random finite sets in the related literature \cite{Mahler2013}. This allows for interesting connections to be drawn with multi-user communication and thus for corresponding information theoretic performance measures to be obtained \cite{Clark2021}. Such an approach would fit nicely within our framework. 

Another avenue of interest would be to consider the performance impacts inherent in moving from monostatic to multistatic sensing. This would enable insights to be drawn regarding the important tradeoff between ``competition'' -- BSs acting as independent sensors -- and ``cooperation'' -- BSs acting in tandem in some manner \cite{Zhang2021}. The consideration of multistatic sensing would further facilitate the analysis of more advanced data fusion policies, where tracks are formed using measurements from multiple BSs. There are variety of ways in which multistatic sensing may be implemented in a JCAS network, each with their own impacts on our modeling framework. At a minimum, the incorporation of multistatic  sensing would require a more nuanced sensing channel model accounting for the difference among LoS and NLoS paths in a spatially consistent manner and of course the geometry of cooperating BSs. With this model in hand, one could then obtain the associated BCRB \cite{VanTrees2006} and exploit our notion of sensing coverage and rate. Multistatic sensing may further be implemented either with or without full duplex transceivers. When full duplex transceivers  are not used, one must consider which BSs are to serve as receivers and which are to serve as transmitters. This opens the door to complex tradeoffs between sensing and communication coverage.

While we have focused on notions of JCAS performance amenable to a snapshot model of the network, tradeoffs in the temporal domain are certainly an important area of interest. A complete treatment of these issues is perhaps better served by framework based on queueing theory, \cite{Husheng22}. However, due to the non-orthogonal implementation of communication and sensing in JCAS networks, geometry plays an important role in these issues, and thus stochastic geometry may be able to provide some insights. One such way to go about this would be to augment the ergodic rate expression by a potentially random scheduling gain factor -- similar to \cite{Ohto2017} -- to obtain a metric corresponding to average throughput. A key challenge in this area would be to address the different time scales inherent in communication and sensing in a rigorous manner. As we have shown, the performance of sensing corresponds to properties of the received waveform over multiple communication time slots.

Finally, we conclude noting that the above list of extensions is by no means exhaustive: one could further build upon our model by incorporating various interference mitigation techniques; by considering different parameters of interest -- such as angular resolution; by performing meta-distribution analysis for communication and sensing; or by applying MMSE bounds other than the BCRB in our formulation of sensing coverage and rate.

\pagebreak

\appendices

\section{Sensing Signal Model and Bayesian Fisher Information}
\label{app:sens-sig-mod}

\subsection{Sensing Signal Model}

The usage of multi-carrrier  waveforms for radar (most commonly OFDM)  has gained interest over the past decade \cite{Wild2021}, \cite{Braun2014ofdm}, \cite{Strum2011}. Such waveforms are an attractive option for JCAS networks in that they provide excellent communication capability while providing sufficient flexibility for sensing purposes \cite{Graff2021}. Moreover, the usage of a  waveform allows for parameter estimation to be performed at the symbol level in the frequency domain, rather than directly through the baseband waveform \cite{Strum2011}. This allows for some of the benefits of multi-carrier methods for communication, such as reduction of ISI, to carry over for sensing. Motivated by \cite{Strum2011}, we consider sensing to take place in the frequency domain (that is, estimation is performed using the output of an  receiver) as this has the added benefit of facilitating the joint analysis of sensing and communication. 

A JCAS BS performing monostatic sensing of SOs and downlink communication to UEs may thus use the avaialable resource elements of the  waveform to conduct both functions. Often, this may be done in tandem -- for instance in the beamforming based framework we consider an SO and UE in the same beam may be "served" simultaneously. However, depending on the spatial distribution of the UEs and SOs, the queue states of the UEs, and level of certainty the network has about the parameters of interest of the SOs, simultaneous operation may not always be possible. Hence, over some number of symbols, say $N_{\rm s}$, a subset of resource elements will be dedicated to a particular UE or SO. In light of these considerations, for a mulitcarrier waveform with $N_{\rm c}$ sub-carriers we denote the resource elements used for sensing and communications as $\Sr \in \{0, 1\}^{N_\text{s} \times N_\text{c}}$ and $\Sc \in \{0, 1\}^{N_\text{s} \times N_\text{c}}$, respectively. In general, these objects would arise from some underlying resource allocation processes -- which is not the focus of our present work. Rather, for the purposes of our analysis we treat these allocation matrices as fixed and focus instead on the quantifying the utility of the resulting allocations for communication and sensing in terms of the coverage and rate.

In particular, our sensing rate metric is expressed in terms of the Fisher Information Matrix (FIM) of the underlying parameter estimation problem. We characterize this now in the setting wherein the  waveform is used to sense a single SO treating interference from the JCAS network as noise. Although interference from the network would in theory be informative with respect to the SO's parameters of interest due to secondary reflections, the exploitation of these features would require knowledge of the environment geometry and relative orientation of the interfering BSs. Moreover, the additional information provided by the exploitation of this phenomena is limited unless the signal structure is also known by the receiver -- and thus some degree of cooperation is required. This would further increase the complexity of the theoretically optimal estimator. In light of these considerations, and noting similar observations by others, we the focus on the more practical setting in which JCAS radar receivers do not exploit interference and hence treat it as noise \cite{Griffiths2015}. We note further, that this modeling assumption would only serve to reduce the sensing mutual information.

The setting we consider is as follows.

\begin{enumerate}
	\item  The parameters of interest for the SOs are taken to be the doppler and delay, and hence $\Theta = (\tau, f_D)$. The target is modeled as a point source and characterized by its relative range, $r$, and relative velocity $v_\text{rel}$. Letting $c_0$ denotes the speed of light and $f_\text{c}$ denote the center frequency, these translate to inducing a delay, $\nu  = 2r/c_0$, and Doppler shift, $f_\text{D} = 2 v_\text{rel} f_\text{c} /c_0$ to the transmitted waveform. We assume that $f_\text{c} \gg B$, which implies that $f_D$ is constant over the sub-carriers. 
	\item The waveform consists of $N_\text{c}$ subcarriers, and the maximum burst duration of the radar excitation signal is $N_\text{s}$  symbols. A subset, $\Sr \in \{0, 1\}^{N_\text{s} \times N_\text{c}}$, of these resource elements are used for sensing.
	\item The total length of each  symbol is $\Tofdm = \Ts + T_\text{g}$ where $\Ts$ denotes the symbol duration and $T_\text{g}$ the guard interval, yielding a sub-carrier spacing of $\Delta f = 1/\Ts$. It is assumed $\nu < T_\text{g}$ and $f_\text{D} \ll \Delta f$, which simplifies the signal model and very commonly holds in practice \cite{Wild2021}, \cite{Braun2014ofdm}.
	\item The waveform may be shared by the communication function. Hence, arbitrary message bearing symbols are transmitted on the scheduled resource elements. Let the frequency domain symbols be denoted as $X_{m,n}$ for $(m,n) \in \Sr$. 
\end{enumerate}

Under these assumptions, the frequency domain representation of the  radar return for a single target without clutter is given by the matrix $\bold{F} \in \C^{N_s \times N_c}$ with support over $\Sr$ (following \cite{Braun2014ofdm} \cite{Strum2011}) 
\begin{align}
&(\bold{F})_{m,n} = e^{-j 2 \pi \Delta f \tau n} e^{j 2 \pi \Tofdm f_D m} + Z_{m,n}. &\bold{Z} \sim \mathcal{CN}(0, \text{diag}\{\lvert X_{m,n} \rvert^{-2} \SINR_{m,n}^{-1}\}_{(m,n) \in \Sr}).
\label{eq:ofdm_radar_est_prob}
\end{align}
The Fisher Information Matrix corresponding to this signal model may be expressed as \cite{Gaudio2019}
\begin{align}
\begin{split}
	\boldsymbol{\mathcal{J}} &=  \sum_{(m,n) \in \Sr} 8 \pi^2 |X_{m,n}|^2 \SINR_{m,n}
	\begin{bmatrix}
		&\Delta f^2 \left( \frac{2}{C_0}\right)^2 n^2 &-\Tofdm \frac{2 f_c}{c_0} \Delta f \frac{2}{C_0} m n \\ &-\Tofdm \frac{2 f_c}{c_0} \Delta f \frac{2}{C_0} m n &\Tofdm^2 \left( \frac{2 f_c}{c_0}\right)^2 m^2
	\end{bmatrix},
	\end{split}
	\label{eq:ofdm-fim}
\end{align}
which does not depend one true the value of $\Theta$, and hence we take $\boldsymbol{\mathcal{J}} = \boldsymbol{\mathcal{J}}(\Theta)$ from our earlier notation.

The reader may note that our signal model does not explicitly account for either the presence of other targets or exogenous reflections -- which would jointly contribute to an additional clutter term in the signal model. We note however, that these may be accounted for by scaling the so called "clean" FIM given above by a scale factor between $0$ and $1$, \cite{Payan2019}. Thus, although an important phenomena in practical sensing problems, this would have no impact on our analytical results other than to attenuate the $G$ factor stated in (\ref{eq:G-def}) in Theorem 1.

\section{Comments on the Bayesian Cram{\'e}r Rao Lower Bound}
\label{app:bcrb}
Our formulation of the approximate lower bound for the sensing MI which we use to define our notion of sensing rate in (\ref{eq:R_est_formula1}) relies on exchanging the MMSE covariance with the Bayesian FIM. Such an exchange is only meaningful when the Bayesian Cram{\'e}r Rao Lower Bound (BCRB) holds  
\begin{align*}
	\mathbf{R}_{\rm MMSE}^{-1} \le \E[\boldsymbol{\mathcal{J}}(\Theta)] + \E\left[ \nabla_{\theta}\log(p(\Theta)) \nabla_{\theta}\log(p(\Theta))^{\rm T} \right].
\end{align*}
\ifdefined\proveBCRB

Here we summarize the conditions necessary to establish this inequality and provide a proof of the Bayesian Cram{\'e}r Rao Bound based (BCRB) on the Hammersley-Chapman-Robbins (HCR) bound. The conditions under which the bound holds are provide a different perspective on Van Tree's original formulation \cite{van2013detection}. 
\else
Note that there exist priors for which this bound is attainable \cite{Fau2021}. Here we summarize the conditions necessary to establish this inequality using the Hammersley-Chapman-Robbins Bound.
\fi 

{\bf Proposition 4: Bayesian Cram{\'e}r Rao Lower Bound.} {\em Let $\Theta$ and $Y$ be random variables defined on a common probability space with $\Theta$ taking values in $\R^{n}$ and $Y$ taking values in an arbitrary space $\mathcal{Y}$. Denote their joint distribution as $\Pb_{Y,\Theta}$. Let the following sufficient conditions hold.
\begin{enumerate}[label=\roman*)]
	\item For some common denominating measure $\mu$ define
	{\em \begin{align}
		\frac{\partial \Pb_{Y\vert \Theta}}{\partial \mu} = f(y, \theta).
	\end{align}}
	Let $\nabla_{\theta} f(y, \theta)$ exist $\Pb_{Y, \Theta}$-almost surely.
	\item Let $\Pb_{\Theta}$ admit a density with respect the Lebesgue measure over $\R^n$, $\ell^{n}$,
	{\em \begin{align}
		\frac{\partial \Pb_{\Theta}}{\partial \ell^{n}} = p(\theta).
	\end{align}}
	Moreover, let $\nabla_{\theta}p(\theta)$ exist $\Pb_{\Theta}$-almost surely.
	\item Let the support of $\Pb_{Y, \Theta}$ be such that
	{\em \begin{align}
	\text{dim}(\text{span}(\{ \mathbf{v}\mathbf{v}^{\rm T}: \exists \delta > 0 ~ s.t. ~ S_{(0, \delta \mathbf{v}) } \Pb_{Y, \Theta} \ll \Pb_{Y, \Theta}\})) = \frac{n(n+1)}{2}.
		\end{align}}
	Where $0$ is the zero element in $\mathcal{Y}$.
	\item Finally assume that the following limit and expectation may be interchanged
	{\em \begin{align}
		\lim_{\delta \to 0} \E_{Y, \Theta}&\left[\left( \frac{f(Y, \Theta + \delta \mathbf{v})p(\Theta + \delta \mathbf{v}) + f(Y, \Theta)p(\Theta)}{\delta f(Y,\Theta)p(\Theta)}\right)^2 \right] = \nonumber
		\\
		& \E_{Y, \Theta}\left[\lim_{\delta \to 0}\left( \frac{f(Y, \Theta + \delta \mathbf{v})p(\Theta + \delta \mathbf{v}) + f(Y, \Theta)p(\Theta)}{\delta f(Y,\Theta)p(\Theta)}\right)^2 \right].
	\end{align}}
	As a consequence of the Dominated Convergence Theorem, a sufficient condition for this is if the gradients are bounded by another random variable with finite $L_1$-norm.

\end{enumerate}

	Now, let $\hat{\theta}: \mathcal{Y} \rightarrow \R^{n}$ denote an arbitrary estimator that is unbiased in expectation. That is 
	{\em \begin{align}
     	\E_{Y, \Theta}[ \Theta - \hat{\theta}(Y)] = 0. 	
     \end{align}}
     Let the error covariance of $\hat{\theta}$ be denoted as $\mathbf{R} = \E[(\Theta - \hat{\theta}(Y))(\Theta - \hat{\theta}(Y))^{\rm T}]$. Then, the following inequality holds
     \begin{align}
     	\mathbf{R}^{-1} \le 	\E[\boldsymbol{\mathcal{J}}(\Theta)] + \E\left[ \nabla_{\theta}\log(p(\Theta)) \nabla_{\theta}\log(p(\Theta))^{\rm T} \right],
     \end{align}
     where $\boldsymbol{\mathcal{J}}(\Theta)$ is the FIM
     \begin{align}
     	\boldsymbol{\mathcal{J}}(\Theta) = \E_Y\left[ \nabla_{\theta}\log(f(Y, \Theta)) \nabla_{\theta}\log(f(Y, \Theta))^{\rm T} \big\vert \Theta\right].
     \end{align}}
     
    \ifdefined\proveBCRB
	{\em Proof:} The proof relies on the Hammersley-Chapman-Robbins (HCR) bound, which states that for any measures $P$ and $Q$ on some measures space $(\Omega, \mathcal{F})$ and measurable function $g: \Omega\rightarrow \R$ 
	\begin{align*}
		\chi^2(P \vert \vert Q) \ge \frac{\left(\E_P[g] - \E_Q[g] \right)^2}{\E_Q[(g - \E_Q[g])^2]},
	\end{align*}
	where $\chi^2(P \vert \vert Q)$ denotes the $\chi^2$-divergence.
	
	Let $\mathbf{v} \in \R^{n}$ such that $\exists ~\delta >0$ such that $S_{\delta \mathbf{v}}\Pb_{Y,\Theta} \ll \Pb_{Y, \Theta}$. Note that, for any measurable set, $A$, we have 
	\begin{align*}
		S_{(0, \delta \mathbf{v})}\Pb_{Y, \Theta}(A)	 &= \int_{\mathcal{Y} \times \R^n} \indicator\{(y, \theta) \in A + \delta \mathbf{v}\} f(y, \theta)p(\theta)\mu(dy) \ell^{n}(d\theta)
		\\
		&= \int_{\mathcal{Y} \times \R^n} \indicator\{(y, \tilde{\theta}) \in A\} f(y, \tilde{\theta} + \delta \mathbf{v})p(\tilde{\theta} + \delta \mathbf{v})\mu(dy) \ell^{n}(d\tilde{\theta}).
	\end{align*}
	Where the last equality follows from the change of variables $\tilde{\theta} = \theta - \delta \mathbf{v}$ and stationarity of $\ell^{n}$. Thus, we have 
	\begin{align*}
		\frac{\partial S_{(0, \delta \mathbf{v})} \Pb_{Y, \Theta}}{\partial \Pb_{Y, \Theta}} = f(y, \tilde{\theta} + \delta \mathbf{v})p(\tilde{\theta} + \delta \mathbf{v}).
	\end{align*}
	Thus, consider
	\begin{align*}
		\chi^{2}\left(S_{(0, \delta \mathbf{v})} \Pb_{Y, \Theta} \lvert \lvert \Pb_{Y, \Theta} \right) &= \E_{Y, \Theta} \left[\left(\frac{f(Y, \Theta + \delta \mathbf{v})p(\Theta + \delta \mathbf{v})}{f(Y, \Theta) p(\Theta)} - 1 \right)^2 \right]
		\\
		&= \E_{Y, \Theta} \left[\left(\frac{f(Y, \Theta + \delta \mathbf{v})p(\Theta + \delta \mathbf{v}) - f(Y, \Theta) p(\Theta)}{f(Y, \Theta) p(\Theta)}\right)^2 \right].
	\end{align*}
	Thus
	\begin{align*}
		\lim_{\delta \to 0} \frac{\chi^{2}\left(S_{(0, \delta \mathbf{v})} \Pb_{Y, \Theta} \lvert \lvert \Pb_{Y, \Theta} \right)}{\delta^2} &= \lim_{\delta \to 0} \E_{Y, \Theta} \left[\left(\frac{f(Y, \Theta + \delta \mathbf{v})p(\Theta + \delta \mathbf{v}) - f(Y, \Theta) p(\Theta)}{\delta \cdot f(Y, \Theta) p(\Theta)}\right)^2\right]
		\\&=  \E_{Y, \Theta} \left[\lim_{\delta \to 0} \left(\frac{f(Y, \Theta + \delta \mathbf{v})p(\Theta + \delta \mathbf{v}) - f(Y, \Theta) p(\Theta)}{\delta \cdot f(Y, \Theta) p(\Theta)}\right)^2 \right]
		\\&=  \E_{Y, \Theta} \left[\mathbf{v}^{\rm T} \left( \nabla_{\theta}\log(f(Y, \Theta) p(\Theta) \nabla_{\theta}\log(f(Y, \Theta) p(\Theta))^{T} \right) \mathbf{v} \right]
		\\&= \mathbf{v}^{\rm T}\left(\E_\Theta[\boldsymbol{\mathcal{J}}(\Theta)] + \E_\Theta\left[ \nabla_{\theta}\log(p(\Theta) \nabla_{\theta}\log(p(\Theta)^{T}\right]\right)\mathbf{v}.
	\end{align*}
	
	Now, let $g(y, \theta) = \mathbf{w}^{\rm T}(\theta - \hat{\theta}(y))$ for some $\mathbf{w} \in \R^{n}$. The following relations hold.
	\begin{align*}
		\E_{(Y,\Theta) \sim S_{(0, \delta \mathbf{v})} \Pb_{Y, \Theta}} [ g(Y, \Theta)] &= \mathbf{w}^{\rm T}\left(\int_{\mathcal{Y} \times \R^{n}} (\theta - \hat{\theta}(y))	f(y, {\theta} + \delta \mathbf{v})p({\theta} + \delta \mathbf{v})\mu(dy) \ell^{n}(d\theta)  \right)
		\\
		&= \delta \mathbf{w}^{\rm T} \mathbf{v}.
	\end{align*}
	Similarly,
	\begin{align*}
		\E_{(Y,\Theta) \sim \Pb_{Y, \Theta}} [ g(Y, \Theta)] &= \mathbf{0},
	\end{align*}
	and
	\begin{align*}
		\E_{(Y,\Theta) \sim \Pb_{Y, \Theta}}[ g(Y, \Theta)^2] &= \mathbf{w}^{\rm T}\mathbf{R} \mathbf{w}.
	\end{align*}
	Therefore, applying the HCR bound we have
	\begin{align*}
		\chi^{2}\left(S_{(0, \delta \mathbf{v})} \Pb_{Y, \Theta} \lvert \lvert \Pb_{Y, \Theta} \right) &\ge \delta^2 \sup_{\mathbf{w} \ne 0} \frac{\mathbf{w}^{\rm T} \left( \mathbf{v} \mathbf{v}^{\rm T} \right) \mathbf{w}}{\mathbf{w}^{\rm T}\mathbf{R} \mathbf{w}}
		\\
		&= \delta^2 \mathbf{v}^{\rm T}\mathbf{R}^{-1} \mathbf{v}.
	\end{align*}
	Thus
	\begin{align*}
		\lim_{\delta \to 0} \frac{\chi^{2}\left(S_{(0, \delta \mathbf{v})} \Pb_{Y, \Theta} \lvert \lvert \Pb_{Y, \Theta} \right)}{\delta^2} \ge \mathbf{v}^{\rm T}\mathbf{R}^{-1} \mathbf{v}.
	\end{align*}
	This implies that
	\begin{align*}
		\mathbf{v}^{\rm T}\mathbf{R}^{-1} \mathbf{v} \le \mathbf{v}^{\rm T}\left(\E_\Theta[\boldsymbol{\mathcal{J}}(\Theta)] + \E_\Theta\left[ \nabla_{\theta}\log(p(\Theta) \nabla_{\theta}\log(p(\Theta)^{T}\right]\right)\mathbf{v}.
	\end{align*}
	Therefore, by assumption (iii), the proposition follows. $\qedwhite$
	\fi
	
	The conditions for which the BCRB holds are simply generalizations of the conditions for which the CRB holds to the Bayesian setting. For the signal model considered in the previous section, this is indeed the case under reasonable assumptions on the heretofore unspecified prior. 
	
	First, note that as we apply the BCRB to the MMSE estimator unbiasedness trivially holds. Moreover, since our received signal is conditionally Gaussian the differentiability and support conditions on $f(y, \theta)$ are satisfied. Thus, to apply the BCRB to our reference prior must satisfy the differentiability condition and the support condition:
	\begin{align}
	\text{dim}(\text{span}(\{ \mathbf{v} \mathbf{v}^{\rm T}: \exists \delta > 0 ~ s.t. ~ S_{(\delta \mathbf{v}) } \Pb_{\Theta} \ll \Pb_{\Theta}\})) = \frac{n(n+1)}{2}.
		\end{align}
		In the setting of doppler and delay estimation, which we consider, $\Theta$ may only take values in $\R_+ \times \R$. Nonetheless, if $p(\theta)$ is such that it is non-zero over $\R_+ \times \R$, and differentiable with bounded gradients, the above support, differentiability, and regularity conditions are satisfied. Hence, the BCRB is applicable.

\section{Proofs of Proposition 1 and Theorem 1}
\label{app:ofdm_rad}
\subsection{Proof of Proposition 1}
Under the conditions listed in the statement of the proposition, $I(Y; \Theta)$ may be expressed as
\begin{align*}
	I(Y; \Theta) = h(\Theta) - h(\Theta \vert Y).
\end{align*}
Using that fact that the Gaussian distribution maximizes differential entropy, subject to a covariance constraint, we have
\begin{align*}
	h(\Theta \vert Y) &\le \frac{n}{2} \log(2 \pi e) + \frac{1}{2} \E_{Y}\left[\log(\lvert \text{Cov}(\Theta \vert Y =y) \rvert)  \right]
	\\
	&\le \frac{n}{2} \log(2 \pi e) + \frac{1}{2} \log(\lvert \E_{Y}\left[\text{Cov}(\Theta \vert Y =y) \right] \rvert)
	\\
	&= \frac{n}{2} \log(2 \pi e) + \frac{1}{2} \log(\lvert \mathbf{R}_{\rm MMSE}\rvert).
\end{align*}
Where the second inequality follows from Jensen's Inequality.

Moreover, note that by a simple change of variables we have
\begin{align*}
	h(\Theta) = \frac{1}{2}\log(\lvert \mathbf{Q} \rvert) + h(\mathbf{Q}^{\frac{-1}{2}} \Theta).
\end{align*}
Therefore, we have
\begin{align*}
	I(Y; \Theta) \ge  \frac{1}{2} \log\left( \left \lvert \mathbf{Q}^{\frac{1}{2}} \mathbf{R}_{\rm MMSE}^{-1} \mathbf{Q}^{\frac{1}{2}} \right \rvert \right) - \frac{n}{2} \log(2 \pi e) + h(\mathbf{Q}^{\frac{-1}{2}} \Theta) . &&\qedwhite
\end{align*} 

\subsection{Proof of Theorem 1}
 The following inequalities hold.
\begin{align*}
 1 + {\rm Tr}\left( \bold{Q}^{\frac{1}{2}}\boldsymbol{\mathcal{J}}\bold{Q}^{\frac{1}{2}}\right)\le \left\lvert \bold{I} + \bold{Q}^{\frac{1}{2}}\boldsymbol{\mathcal{J}}\bold{Q}^{\frac{1}{2}} \right\rvert \le \frac{1}{4} {\rm Tr}\left( \bold{I} +\bold{Q}^{\frac{1}{2}}\boldsymbol{\mathcal{J}}\bold{Q}^{\frac{1}{2}}\right)^2.
\end{align*}
The lower bound follows from the positive semi-definiteness of $\bold{Q}^{\frac{1}{2}}\boldsymbol{\mathcal{J}}\bold{Q}^{\frac{1}{2}}$, and the upper bound follows from the AM-GM inequality. 

Recall that the FIM, $\boldsymbol{\mathcal{J}}$, for the measurement model stated in (\ref{eq:ofdm_radar_est_prob}) is given as (\ref{eq:ofdm-fim}). The proof follows noting that $G \cdot \SINRr = {\rm Tr}\left( \bold{Q}^{\frac{1}{2}}\boldsymbol{\mathcal{J}}\bold{Q}^{\frac{1}{2}}\right)$. $\qedwhite$

\section{Proof of Proposition 2}
\label{app:sinr_aprx}
This proposition inherently relies on the fact that $\SINRr$ is a function of a certain type of empirical mean of the interference over $\Srt$ (the harmonic mean specifically). One can then obtain upper and lower bounds using inequalities for different types of means, and a H{\"o}lder-like in Inequality for obtaining lower bounds on a mean of measures using the measure of the means. 

Hence, before proving this proposition, we first summarize some key inequalities. Of note, we prove a new H{\"o}lder-like inequality for measures of harmonic means.

\subsection{Summary of Intermediate Lemmas}

To prove Proposition 2, we shall exploit the following extension of H{\"o}lder's Inequality, which allows one to bound the geometric mean of ensembles of measures of functions.

\textbf{Lemma A: (H{\"o}lder's Inequality for Geometric Means of Sets of Functions)} {\em Let $p$ be a PMF on $[n]$ for some $n \in \N$, and let $(\G, \mathcal{G}, \mu)$ be a measure space. Let $\{f_k\}_{k \in [n]}$ be a sequence measurable functions from $\G$ to $\C$. Then
\begin{align}
	\norm{\prod_{k = 1}^n f_k^{p_k}}_1 \le \prod_{k = 1}^{n} \norm{f_k}_{1}^{p_k}.
\end{align}
Or, alternatively
\begin{align}
	\norm{\GM(\{f_k\}_{k \in [n]}, p)}_1 \le \GM(\{\norm{f_k}_{1}\}_{k \in [n]}, p).
\end{align}
Note that the above norms are taken with respect to $\mu$.}

We now generalize H{\"o}lder's Inequality to the setting of Harmonic means. 

\textbf{Lemma B: (A H{\"o}lder-Like Inequality for Harmonic Means of Sets of Functions)} {\em Let $p$ be a PMF on $[n]$ for some $n \in \N$, and let $(\G, \mathcal{G}, \mu)$ be a measure space. Let $\{f_k\}_{k \in [n]}$ be a sequence measurable functions from $\G$ to $\R_+$. Then
\begin{align}
	\norm{\left(\sum_{k = 1}^n p_k f_k^{-1}\right)^{-1}}_1 \le \left(\sum_{k = 1}^{n} p_k\norm{f_k}_{1}^{-1}\right)^{-1}.
\end{align}
Or, alternatively
\begin{align}
	\norm{\HM(\{f_k\}_{k \in [n]}, p)}_1 \le \HM(\{\norm{f_k}_{1}\}_{k \in [n]}, p).
\end{align}
Note that the above norms are taken with respect to $\mu$.

Moreover, equality is obtained iff $\{f_k\}_{k \in [n]}$ are linearly dependent. }

{\em Proof}: See Appendix \ref{app:tech_res}.

\subsection{Proof of Proposition 2}
Note that the SINR models differ only in the interference terms. Thus, it suffices to establish bounds on the these terms only. For conciseness, let
\begin{align*}
	&I_{t,n} = \sum_{R_k \in \PiBt^0} L(R_k) Z_B^k F_{t,n}^k + \nur,
\end{align*}
and defeine $\mathbf{I} = \{I_{t,n}\}_{(t,n) \in \Srt}$. By the AM-GM-HM Inequality, we have that
\begin{align}
	\HM(\mathbf{I} + \nur, \boldsymbol{\theta}) \le \GM(\mathbf{I} + \nur, \boldsymbol{\theta}) \le \AM(\mathbf{I} + \nur, \boldsymbol{\theta}).
\end{align}
We now show that we may lower bound $\HM(\mathbf{I} + \nur, \boldsymbol{\theta})$ and $\GM(\mathbf{I} + \nur, \boldsymbol{\theta})$ using the H{\"o}lder-like Inequalities in Lemmas A and B. $\GM(\mathbf{I} + \nur, \boldsymbol{\theta})$ may be expressed as
\begin{align}
	\GM(\mathbf{I} + \nur, \boldsymbol{\theta})  = \prod_{(t,n) \in \Srt} \left(\sum_{R_k \in \PiBt^0} L(R_k) Z_B^k F_{t,n}^k+ \nur \right)^{\theta_{t,n}}.
	\label{eq:GM_rel_1_prop1_proof}
\end{align}
Note that the fading and beam alignment terms inside the sum may be interpreted as marks associated with $R_k \in \PiBt^0$, which is itself a marked process. More precisely, we shall let
\begin{align}
	\PiBt' = \{(R_k, \mathbf{F}_k, Z_B^k, M_k): R_k \in \PiB^0\}
\end{align}
denote the corresponding marked process. Since $\PiBt'$ is defined with respect to a Palm process which is itself defined with respect to the stationary process $\PhiB$, by \cite[Prop-7.1.8]{baccelliSG} we may conclude the existence of the random fields $\mathbf{F}(r)$, $Z_B(r)$, and $M(r)$ over the index set $\R_+$ such that $\mathbf{F}(R_k) = \mathbf{F}_k$, $Z_b(R_k) = Z_B^k$, and $M(R_k) = M_k$ almost surely. Then, we may express (\ref{eq:GM_rel_1_prop1_proof}) as
\begin{align}
	\GM(\mathbf{I} + \nur, \boldsymbol{\theta}) = \prod_{(t,n) \in \Srt} \left( \int_{\R_+} L(r, M(r)) F_{t,n}(r) Z_B(r) \PiB^0(dr) + \nur \right)^{\theta_{t,n}}.
	\label{eq:GM_rel_2_prop1_proof}
\end{align}
We argue Lemma A (H{\"o}lder's Inequality) may be applied to (\ref{eq:GM_rel_2_prop1_proof}). This requires us to absorb the $\nur$ term into the above measure, which we shall do as follows. First, pick some $a \in \R_+$. Since $\PiB^0$ is a simple point process, we have that $\PiB(\{a\}) = 0$ almost surely. Now, we may construct new random fields from $\mathbf{F}(r)$ and $Z_B(r)$, $\mathbf{\tilde{F}}(r)$ and $\tilde{Z}_B(r)$ respectively. Let $\mathbf{\tilde{F}}(r) = \mathbf{F}(r)$, and $\tilde{Z}_B(r) = Z_B(r)$ for all $r \in \R_+ \setminus{\{a\}}$, and $\mathbf{\tilde{F}}(a) = \mathbf{0}$ and $\tilde{Z}_B(a) = 0$. Then we may conclude that
\begin{align*}
	&\GM(\mathbf{I} + \nur, \boldsymbol{\theta})
	\\
	&= \prod_{(t,n) \in \Srt} \left( \int_{\R_+} L(r, M(r)) F_{t,n}(r) Z_B(r) \PiB^0(dr) + \nur \right)^{\theta_{t,n}}
	\\
	&= \prod_{(t,n) \in \Srt} \left( \int_{\R_+}  \left(L(r, M(r)) F_{t,n}(r) Z_B(r)+ \indicator\{r =a\} \nur \right)(\PiB^0 + \delta_a)(dr) \right)^{\theta_{t,n}}
\end{align*}
almost surely. Hence applying Lemma A (H{\"o}lder's Inequality), we have that
\begin{align*}
	&\GM(\mathbf{I} + \nur, \boldsymbol{\theta})
	\\
	&\ge    \int_{\R_+} \prod_{(t,n)\in \Srt} \left(\left(L(r, M(r)) F_{t,n}(r) Z_B(r) + \indicator\{r =a\} \nur \right) \right)^{\theta_{t,n}} (\PiB^0 + \delta_a)(dr)
	\\
	&= \int_{\R_+} \prod_{(t,n)\in \Srt} \left(L(r, M(r)) F_{t,n}(r) Z_B(r) \right)^{\theta_{t,n}} \PiB^0 (dr)	+ \nur
	\\
	&= \sum_{R_k \in \PiBt^0} L(R_k) \GM(\mathbf{F}, \boldsymbol{\theta}) + \nur.
\end{align*}
Using a similar argument and applying Lemma B, we have that
\begin{align*}
	&\HM(\mathbf{I} + \nur, \boldsymbol{\theta}) \ge \sum_{R_k \in \PiBt^0} L(R_k) \HM(\mathbf{F}, \boldsymbol{\theta}) + \nur
\end{align*}
Moreover, by linearity of the $AM(\cdot, \cdot)$ operator we have that
\begin{align*}
	&\AM(\mathbf{I} + \nur, \boldsymbol{\theta}) = \sum_{R_k \in \PiBt^0} L(R_k) \AM(\mathbf{F}, \boldsymbol{\theta}) + \nur
\end{align*}
Therefore, we may conclude the following inequalities almost surely
\begin{align}
	\sum_{R_k \in \PiBt^0} L(R_k) \HM(\mathbf{F}, \boldsymbol{\theta}) + \nur \le \HM(\mathbf{I} + \nur, \boldsymbol{\theta}) \le \sum_{R_k \in \PiBt^0} L(R_k) \AM(\mathbf{F}, \boldsymbol{\theta}) + \nur.
\end{align}
Additionally, by the AM-GM-HM Inequality, we have almost surely
\begin{align}
	\sum_{R_k \in \PiBt^0} L(R_k) \HM(\mathbf{F}, \boldsymbol{\theta}) + \nur \le \sum_{R_k \in \PiBt^0} L(R_k) \GM(\mathbf{F}, \boldsymbol{\theta}) + \nur \le \sum_{R_k \in \PiBt^0} L(R_k) \AM(\mathbf{F}, \boldsymbol{\theta}) + \nur.
\end{align}
These imply that
\begin{align*}
	&\SINRam \le \SINRr \le \SINRhm &\Pb-a.s.
\end{align*}
and
\begin{align*}
	&\SINRam \le \SINRgm \le \SINRhm &\Pb-a.s.
\end{align*}
 which imply the proposition. $\qedwhite$
 
 \section{Proof of Theorem 2}
 \label{app:LST_bds}
 \subsection{Statement of Theorem 2 and Discussion}
$\SINRgen$ and $\SINRc$ belong to a common class of SINR models in that 
\begin{enumerate}[label=\roman*)]
	\item the fading term on the desired signal is exponentially distributed,
	\item conditioned on the location of the serving base station, the interference is independent of the desired signal and follows a Poisson shot noise process,
	\item and the interference fading terms are IID (per each class of LoS/NLoS points).
\end{enumerate}
As a consequence of i) the distribution of an SINR model from this class may be characterized in terms of the Laplace Transform of its interference process. By ii) and iii), the conditional interference Laplace Transform admits an integral closed form expression in terms of Campbell's Formula. 

While a powerful framework, the usage of Campbell's Formula often leads to results expressed in terms of an incomputable integral. This is especially true in the case of $\SINRgen$. In light of this, we develop a general method for obtaining upper and lower bounds on the positive real branch of the Laplace Transform of a shot noise process (i.e. Poisson interference process). We consider the following generic interference process
%
\begin{align}
	I = \sum_{R_k \in \Phi} F_k g(R_k).
	\label{eq:gen_int_proc}
\end{align}
Where $\Phi$ is an arbitrary PPP with intensity measure $\Lambda$, $\{F_k\}$ are arbitrary fading terms, and $g$ is an arbitrary  path loss function.
Of particular interest is the Laplace Transform of $I$, which by Campbell's Formula may be expressed as
\begin{align}
	\Laplace_I(s) = \exp \left( - \int_{\R_+}\left(1 - \Laplace_F(s\cdot  g(r))\right) \Lambda(dr) \right).
	\label{eq:gen_int_LT}
\end{align}

We obtain upper and lower bounds on $\Laplace_{I}(s)$, for $s \in \R_+$, using the theorem to follow. The theorem requires some notation, which we define first before its statement.

Our general approach to obtaining upper and lower bounds on $\Laplace_I(s)$ for $s \in \R_+$ is to obtain lower and upper bounds on the integral term in (\ref{eq:gen_int_LT}). We obtain these bounds by approximating the intensity measure using lower and upper bounding atomic measures with a finite number of atoms.  These approximate measures are such that the moments of the path loss with respect to $\Lambda$ are matched over arbitrary intervals. To that end, we make use of the Mellin Transform of the path loss with respect to $\Lambda$, restricted to a set, $A \subseteq \R_+$, which we denote as \footnote{In many practical scenarios -- such as our case -- these may be obtained closed form.}
\begin{align}
	\Mellin_{(\Lambda \circ g^{-1})}(p; A) = \int_{A} g(r)^{p-1} \Lambda(dr).
\end{align}
Our bounds utilize the following two functionals, which have the interpretation as finite support approximations of $\Lambda$. The first, $\HL(s, A; 1 - \Laplace_F, \Mellin_{(\Lambda \circ g^{-1})})$, we define as 
\begin{align}
	\HL(s, A; 1 -\Laplace_F, \Mellin_{(\Lambda \circ g^{-1})}) = \Mellin_{(\Lambda \circ g^{-1})}(1; A)\left(p \left(1 - \Laplace_F (s x_1) \right) + (1-p)\left(1 - \Laplace_F(s x_2) \right)\right),
	\label{eq:H_LB_def}
\end{align}
\begin{align}
	\alpha &= \frac{\Mellin_{(\Lambda \circ g^{-1})}(1;A)\Mellin_{(\Lambda \circ g^{-1})}(4;A)-\Mellin_{(\Lambda \circ g^{-1})}(2;A)\Mellin_{(\Lambda \circ g^{-1})}(3;A)}{\Mellin_{(\Lambda \circ g^{-1})}(1;A)\Mellin_{(\Lambda \circ g^{-1})}(3;A)-\Mellin_{(\Lambda \circ g^{-1})}(2;A)^2}
	\\
	\beta &= \frac{\Mellin_{(\Lambda \circ g^{-1})}(3;A)^2-\Mellin_{(\Lambda \circ g^{-1})}(2;A)\Mellin_{(\Lambda \circ g^{-1})}(4;A)}{\Mellin_{(\Lambda \circ g^{-1})}(1;A)\Mellin_{(\Lambda \circ g^{-1})}(3;A)-\Mellin_{(\Lambda \circ g^{-1})}(2;A)^2}
	\\
	\gamma &= \sqrt{\alpha^2 +4\beta}
	\\
	x_1 &= \frac{\alpha - \gamma}{2}
	\\
	x_2 &= \frac{\alpha + \gamma}{2}
	\\
	p &= \frac{\Mellin_{(\Lambda \circ g^{-1})}(1;A)(\alpha + \gamma)-2\Mellin_{(\Lambda \circ g^{-1})}(2;A)}{2\gamma\Mellin_{(\Lambda \circ g^{-1})}(1;A)}.
\end{align}
Next, we define $\HU(s, A; 1 - \Laplace_F, \Mellin_{(\Lambda \circ g^{-1})})$ as
\begin{align}
\begin{split}
	&\HU(s, A; 1 - \Laplace_F, \Mellin_{(\Lambda \circ g^{-1})}) = 
	\\
	&\Mellin_{(\Lambda \circ g^{-1})}(1; A)\left((1 - p' - q') \left(1 - \Laplace_F (s z_1)\right) + p' \left( 1 -\Laplace_F (s z_2) \right) + q' \left(1 - \Laplace_F(s z_3)\right) \right),
	\end{split}
	\label{eq:H_UB_def}
\end{align}
\begin{align}
	z_1 &= \inf_{r \in A} g(r)
	\\
	z_3 &= \sup_{r \in A} g(r)
	\\
	\alpha &=  \frac{\Mellin_{(\Lambda \circ g^{-1})}(2;A)}{\Mellin_{(\Lambda \circ g^{-1})}(1;A)}  - z_1 
	\\
	\beta &= \frac{\Mellin_{(\Lambda \circ g^{-1})}(3;A) - 2z_1\Mellin_{(\Lambda \circ g^{-1})}(2;A)}{\Mellin_{(\Lambda \circ g^{-1})}(1;A)} - z_1^2
	\\
	\gamma &= \frac{\Mellin_{(\Lambda \circ g^{-1})}(4;A) - 3z_1\Mellin_{(\Lambda \circ g^{-1})}(3;A) - 3z_1^2\Mellin_{(\Lambda \circ g^{-1})}(2;A)}{\Mellin_{(\Lambda \circ g^{-1})}(1;A)} - z_1^3
	\\
	q' &= \frac{\alpha \gamma - \beta^2}{\alpha (z_3 - z_1)^3 - 2 \beta (z_3 - z_1)^2 + \gamma(z_3 - z_1)}
	\\
	p' &= \frac{\left(\alpha - q' z_3 \right)^2}{\beta - q' z_3^2}
	\\
	z_2 &= \frac{\beta - q' z_3^2}{\alpha - q' z_3}.
\end{align}
Using these functionals, we now state the theorem.

{\bf Theorem 2: (Bounds on the Laplace Transform of a Poisson shot noise  process)} {\em Let $\Phi$ be a PPP on $\R_+$ with intensity measure $\Lambda$, $\{F_k\}_{k \in \Phi}$ be a series of IID marks in $\R_+$ associated with $\Phi$ such that their first moment is finite, and $g: \R_+ \rightarrow \R_+$ be an arbitrary path loss function such that $\Lambda(g)$ is finite. Let $I$ denote the shot noise process as given in (\ref{eq:gen_int_proc}), and $\Laplace_I(s)$ denote its Laplace Transform.

Moreover, let $N_w \in \N$ and $\{d_i\}_{i \in [N_w]}$ be a partitioning of $\R_+$. Without loss of generality assume $d_i \le d_{i+1}$ and $d_0 = 0$. Then, for $s \in R_+$ the following hold
\begin{enumerate}[label=\roman*)]
	\item If $\Lambda(\R_+) = \infty$
	\begin{align}
		\Laplace_I(s) \ge \exp\left(-\sum_{i = 1}^{N_w} \HL\left(s, [d_{i-1}, d_i]; 1 -\Laplace_F, \Mellin_{(\Lambda \circ g^{-1})}\right) + s \E[F] \Mellin_{(\Lambda \circ g^{-1})}\left(s; [d_{N_w}, \infty)\right) \right),
	\end{align}
	and
	\begin{align}
		\Laplace_I(s) \le \exp\left(-\sum_{i = 1}^{N_w} \HU\left(s, [d_{i-1}, d_i]; 1 - \Laplace_F, \Mellin_{(\Lambda \circ g^{-1})}\right) \right).
	\end{align}
	\item Otherwise, if $\Lambda(\R_+) < \infty$,
	\begin{align}
		\Laplace_I(s) \ge \exp\left(-\sum_{i = 1}^{N_w} \HL\left(s, [d_{i-1}, d_i]; 1  -\Laplace_F, \Mellin_{(\Lambda \circ g^{-1})}\right) + \HL\left(s, [d_{N_w}, \infty); 1 -\Laplace_F, \Mellin_{(\Lambda \circ g^{-1})}\right)\right),
	\end{align}
	and
	\begin{align}
		\Laplace_I(s) \le \exp\left(-\sum_{i = 1}^{N_w} \HU\left(s, [d_{i-1}, d_i]; 1 -\Laplace_F, \Mellin_{(\Lambda \circ g^{-1})}\right) + \HU\left(s, [d_{N_w}, \infty); 1 -\Laplace_F, \Mellin_{(\Lambda \circ g^{-1})}\right)\right).
	\end{align}
\end{enumerate}}

{\em Proof:} See Appendix \ref{app_subsec:Thm2}.

In these expressions, $\HL\left(\cdot, [d_1, d_2]; (\cdot), \Mellin_{(\Lambda \circ g^{-1})}\right)$ and $\HU\left(\cdot, [d_1, d_2]; (\cdot), \Mellin_{(\Lambda \circ g^{-1})}\right)$ may be interpreted as finite support approximations of $\Lambda$ over the interval $[d_1, d_2]$. Hence the summation of these functionals over the partitions represents a finite order approximation of $\Lambda$ over $\R_+$. These functionals arise as generalizations of results in \cite{Eckberg77}, which enable us to conclude that they result in upper and lower bounds for the complementary Laplace Transform, $1 - \Laplace_F$, in the argument of of exponential function in (\ref{eq:gen_int_LT}).
 \subsection{Intermediate Lemmas}
 The proof of Theorem 2 follows from the following lemmas, the first of which is from \cite{Eckberg77}.
 
 {\bf Lemma C: (Bounds on the Laplace Transform of a Finite Measure)} {\em Let $\nu$ be a measure  over $\R_+$ such that $\nu(\R_+) < \infty$. Then for $s \in \R_+$, the following bounds hold
 \begin{align}
 	\int_{\R_+}e^{-sx} \nu(dx) \ge \HL\left(s, \R_+; e^{-s}, \Mellin_{\nu}\right),
 \end{align}
 and
  \begin{align}
 	\int_{\R_+}e^{-sx} \nu(dx) \le \HU\left(s, \R_+; e^{-s}, \Mellin_{\nu}\right).
 \end{align}}
 {\em Proof:} See \cite[Sec. 2.2.2]{Eckberg77}. 
 
 In the case where $\nu(\R_+) = \infty$ we require the following lemma
 
 {\bf Lemma D: (Bounds for the Complementary Laplace Transform of a Locally Finite, Globally Infinite Measure)} {\em Let $\nu$ be a locally finite measure over $\R_+$ such that $\nu(\R_+) = \infty$ and $\Mellin_{\nu}(2) <\infty$. Then, its Laplace Transform for $s \in \R_+$ may be bounded as
\begin{align}
	\int_{\R_+} (1 - e^{-sx}) \nu(d x) \le s \Mellin_{\nu}(2).
\end{align}}

{\em Proof:} Let $d \in \R_+$ and define $\nu^{(n)}$ as the measure induced by the restriction of $\nu$ to the interval $[0, nd]$ for $n \in \N$. Note that $\{\nu^{(n)}\}_{n \in \N}$ converges monotonically to $\nu$, and by the assumption that $\nu$ is locally finite, we have $\nu^{(n)}(\R_+) < \infty$. Now, by Jensen's Inequality, we have 
\begin{align*}
 	\int_{\R_+}(1 - e^{-sx}) \nu^{(n)}(d x) &\le \Mellin_{\nu^{(n)}}(1)\left(1 - \exp\left(-s \frac{\Mellin_{\nu^{(n)}}(2)}{\Mellin_{\nu^{(n)}}(1)} \right)\right)
 	\\
 	&\eqlabel{a} \Mellin_{\nu^{(n)}}(1)\left(\sum_{k = 1}^{\infty} \frac{(-1)^{k+1}(s)^k \Mellin_{\nu^{(n)}}(2)^{k}}{\Mellin_{\nu^{(n)}}(1)^{k}k!}\right)
 	\\
 	&= \sum_{k = 1}^{\infty} \frac{(-1)^{k+1}(s)^k \Mellin_{\nu^{(n)}}(2)^{k}}{\Mellin_{\nu^{(n)}}(1)^{k-1}k!}.
\end{align*}
Where (a) follows from the Taylor Series of $1 - e^{-x}$. Therefore, by the Monotone Convergence Theorem we have
\begin{align*}
 	\int_{\R_+} 1 - e^{-sx} \nu(d x) &\le \lim_{n \rightarrow \infty}\sum_{k = 1}^{\infty} \frac{(-1)^{k+1}(s)^k \Mellin_{\nu^{(n)}}(2)^{k}}{\Mellin_{\nu^{(n)}}(1)^{k-1}k!}
 	\\
 	&= s \Mellin_{\nu}(2).
\end{align*}
Where the expression for the limit follows from the assumption that $\Mellin_{\nu}(2) <\infty$ and $\Mellin_{\nu}(1) = \nu(\R_+)$. $\qedwhite$

 \subsection{Proof of Theorem 2}
 \label{app_subsec:Thm2}
 We first prove Part i). Consider
 \begin{align*}
 	\int_{\R_+} &\left(1 - \Laplace_{F}(s \cdot g(r)) \right) \Lambda(dr) = \sum_{i = 1}^{N_w} \int_{d_{i-1}}^{d_i} \left(1 - \Laplace_{F}(s \cdot g(r)) \right) \Lambda(dr) + \int_{d_{N_w}}^{\infty} \left(1 - \Laplace_{F}(s \cdot g(r)) \right) \Lambda(dr)
 	\\
 	&\eqlabel{a}\sum_{i = 1}^{N_w} \E_{F}\left[\int_{d_{i-1}}^{d_i} \left(1 - e^{-s F\cdot g(r)} \right) \Lambda(dr) + \int_{d_{N_w}}^{\infty} \left(1 - e^{-s F\cdot g(r)} \right)  \Lambda(dr)\right].
 	\\
 	&\eqlabel{b}\sum_{i = 1}^{N_w} \E_{F}\left[\int_{g([d_{i-1}, d_i])} \left(1 - e^{-s F\cdot x} \right) (\Lambda \circ g^{-1})(dx) + \int_{g([d_{N_w}, \infty))} \left(1 - e^{-s F\cdot x} \right)  (\Lambda \circ g^{-1})(dx)\right].
 \end{align*}
 Where the (a) follows from Fubini's Theorem, and (b) follows from a change of measure. Using that fact that $\Lambda$ is locally finite, we may apply Lemma C to obtain
  \begin{align*}
 	\int_{g([d_{i-1}, d_i])} \left(1 - e^{-sF \cdot x} \right) (\Lambda \circ g^{-1})(dx) \le \HL\left(s, [d_{i-1}, d_i]; 1 - e^{-sF}, \Mellin_{\Lambda \circ g^{-1}}\right),
 \end{align*}
 and
 \begin{align*}
 	\int_{g([d_{i-1}, d_i])} \left(1 - e^{-sF \cdot x} \right) (\Lambda \circ g^{-1})(dx) \ge \HU\left(s, [d_{i-1}, d_i]; 1 - e^{-sF}, \Mellin_{\Lambda \circ g^{-1}}\right).
 \end{align*}
 Moreover, by Lemma D and non-negativity of $1 - e^{-s g(r)}$ we have
 \begin{align*}
 	0 \le \int_{g([d_{N_w}, \infty))} \left(1 - e^{-sF \cdot x} \right) (\Lambda \circ g^{-1})(dx) \le s F \Mellin_{\Lambda \circ g^{-1}}(2, [d_{N_w}, \infty)).
 \end{align*}
 Hence, using the fact the $\HL$ and $\HU$ are linear in $\Laplace_F$ we may conclude that
  \begin{align*}
 	\int_{\R_+} &\left(1 - \Laplace_{F}(s \cdot g(r)) \right) \Lambda(dr) \le \sum_{i = 1}^{N_w} \HL\left(s, [d_{i-1}, d_i]; 1 - \Laplace_F(s), \Mellin_{\Lambda \circ g^{-1}}\right)  + s \E[F] \Mellin_{\Lambda \circ g^{-1}}(2, [d_{N_w}, \infty)),
 \end{align*}
 and
 \begin{align*}
 	\int_{\R_+} &\left(1 - \Laplace_{F}(s \cdot g(r)) \right) \Lambda(dr) \ge \sum_{i = 1}^{N_w} \HU\left(s, [d_{i-1}, d_i]; 1 - \Laplace_F(s), \Mellin_{\Lambda \circ g^{-1}}\right).
 \end{align*}
Therefore, part i) follows by using these bounds in the expression for $\Laplace_I(s)$ in (\ref{eq:gen_int_LT}).

Part ii) may be proved in a similar manner. However, in this case we may use Lemma A to obtain upper and lower bounds for the $[d_{N_w}, \infty)$ section by the assumption that $\Lambda(\R_+) < \infty$. $\qedwhite$

\section{Proof of the Exact Expression for the Palm Interference Process}
\label{app:palm_int}

\subsection{Proof of Lemma 2}
First recall that $\PhiB$ may be expressed as the superposition of the two independent processes $\PhiL$ and $\PhiN$. Since $X_0 \in \PhiL$, by assumption, we shall restrict our focus to this process for now. By Slivnyak's Theorem, for any $x \in \R^2$, we have that $\PhiL^{!x} \eqdist \PhiL$. Moreover, by definition of $X_0$, we have that 
\begin{align}
	\Pb(\PhiL^{!X_0} \in \cdot) = \Pb(\PhiL^{!x} \in \cdot \vert X_0 = x, \PhiL^{!x}(B(0, \norm{x}_2)) = 0 ).
\end{align}
Hence, we have that $\PhiL^{!X_0}$ is a PPP with intensity function
\begin{align}
	\lambda_{L}(x) = \lambdaB e^{-\beta \norm{x}_2} \indicator\{x \not\in B(0, R_0)\}.
\end{align}
Therefore, by the superposition theorem and independence of $\PhiL$ and $\PhiN$, $\PhiB^{!X_0}$ is a PPP with intensity function
\begin{align}
	\lambdaB(x; X_0) = \lambdaB \left(1 - e^{-\beta \norm{x}_2}\indicator\{x \in B(0, R_0)\}\right).
\end{align}
Where $B(x, r)$ denotes set of points in $\R^{2}$ within the circle of radius $r$ centered at $x$.

Now, consider
\begin{align*}
	\PsiB^0 &= \left\{ \norm{X_k - X_0}_2, \angle\{X_k - X_0\} : X_k \in \PhiB^{!X_0} \right\}
	\\
	&= \left\{ \norm{Y_k}_2, \angle\{Y_k\} : Y_k \in S_{X_0}\PhiB^{!X_0} \right\}
\end{align*}
Where $S_x$ is the shift by $x$ operator. By the Mapping Theorem, $\PsiB^0$ is a PPP with intensity measure (for some $\mathcal{A} \in \mathcal{B}(\R_+ \times [0, 2\pi))$, with $\mathcal{B}$ denoting Borel $\sigma$-algebra.)
\begin{align*}
	\E[\PsiB^0(\mathcal{A})] &= \int_{\R^2} \indicator\{(\norm{x}_2, \angle x \in \mathcal{A}\}\lambdaB(x + X_0; R_0)dx
	\\
	&= \lambdaB\int_{\R^2} \indicator\{\norm{x}_2, \angle x \in \mathcal{A}\} dx - \lambdaB\int_{\R^2} \indicator\{\norm{x}_2, \angle x \in \mathcal{A}\} \indicator\{x \in B(-X_0, R_0)\}  e^{-\beta \norm{x + X0}_2}dx.
\end{align*}

Now, we have that
\begin{align*}
	\lambdaB\int_{\R^2} \indicator\{\norm{x}_2 \angle x \in \mathcal{A}\} dx  =  \int_{\mathcal{A}} \lambda_B r dr d\theta,
\end{align*}
and
\begin{align*}
	&\lambdaB\int_{\R^2} \indicator\{\norm{x}_2, \angle x \in \mathcal{A}\} \indicator\{x \in B(-X_0, R_0)\} e^{-\beta \norm{x + X0}_2}dx.
	\\
	&\eqlabel{a} \lambdaB\int_{\mathcal{A}} \indicator\{(r, \theta) \in B((R_0, \theta_0 + \pi), R_0)\} e^{-\beta \sqrt{r^2 - 2rR_0cos(\theta - \theta_0 - \pi) + R_0^2}} r dr d\theta
	\\
	&\eqlabel{b} \lambdaB\int_{S_{(0, -\theta_0 - \pi)}\mathcal{A}}\indicator\{(r, \theta') \in B((R_0, 0), R_0)\} e^{-\beta \sqrt{r^2 - 2rR_0cos(\theta') + R_0^2}} r dr d\theta'
	\\
	&\eqlabel{c} \lambdaB \int_{S_{(0, -\theta_0 - \pi)}\mathcal{A}} \indicator\{r \le 2R_0\} \indicator\left\{\theta' \in \left[\pm \arccos\left(\frac{r}{2 R_0}\right)\right]\right\}e^{-\beta \sqrt{r^2 - 2rR_0cos(\theta') + R_0^2}} r d\theta' dr 
\end{align*}
Where (a) follows from changing from cartesian to polar coordinates, (b) follows from the change of variables $\theta' = \theta - \theta_0 - \pi$, and (c) follows from the fact that
\begin{align}
	\indicator\{(r, \theta') \in B((R_0, 0), R_0)\} \iff \indicator\{r \le 2R_0\} \indicator\left\{\theta' \in \left[\pm \arccos\left(\frac{r}{2R_0}\right)\right]\right\}.
\end{align}
 Putting these together, we have that $\PsiB^0$ admits the intensity function
\begin{align}
\begin{split}
	&\tilde{\lambdaB}^0(r, \theta + \theta_0 + \pi; R_0) = 
	\\
	&\lambdaB r\left(1 -  \indicator\{r \le 2R_0\} \indicator\left\{\theta \in \left[\pm \arccos\left(\frac{r}{2 R_0}\right)\right]\right\} e^{-\beta \sqrt{r^2 - 2rR_0cos(\theta) + R_0^2}} \right).
\end{split}
\end{align}
Moreover, by applying the Mapping Theorem again to $\PiB^0 = \{R_K : (R_k, \theta_k) \in \PsiB^0\}$, we have that $\PiB^0$ is a PPP with the intensity function specified int the theorm

Now, we may use $\PsiB^0$ to characterize $Z_B^k$. For $(R_k, \theta_k) \in \PsiB^0$, we shall assume that 
\begin{align}
	Z_B^k = 1 \iff  \theta_k \in [\theta_0 + \pi \pm \frac{\thetabr}{2}].
\end{align}
Note that this does not account for NLoS reflections being in the receive beam. Such a characterization requires one to account for the geometry induced by the set process of blockages $\XiB$, and is highly nontrivial (characterizing this is akin to modeling ray tracing). Hence, we maintain this simplifying assumption in keeping with assumptions used (usually implicitly) in prior mmWave/THz stochastic geometry work.

Consider the thinned version of $\PiB^0$, $\PiBa^0 = \{R_K : (R_k, \theta_k) \in \PsiB^0, \theta_k \in [\theta_0 + \pi \pm \frac{\thetabr}{2}]\}$. Applying the Mapping Theorem again with the further restriction on $\theta_k$ yields that $\PiBa^0$ is a PPP with intensity function
\begin{align*}
	&\lambdaBa^0(r; R_0) = 2\pi \lambdaB r\left(\frac{\thetabr}{2 \pi} - \frac{1}{\pi}J\left(R_k; R_0, \max\left\{\cos\left(\frac{\thetabr}{2}\right), \frac{R_k}{2 R_0}\right\} \right) \indicator\{R_k \le 2 R_0  )\}\right).
\end{align*}
Then using method similar to that of \cite[Corollary 2]{Olson2021}, one may interpret $\{Z_B^k\}_{k \in \PiB^0}$ as independent marks of $\PiB^0$ such that $\pBr(R_k) = \lambdaBa(R_k;R_0)/ \lambdaB(R_k;R_0)$.
$\qedwhite$

\section{Characterization of Bounds for the Intensity Measure of the Palm interference Process}
\label{app:palm_int_ub}

{\bf Intermediate Result: (Characterization of $\PiB^0$ Intensity measures)} {\em Using Lemma 2, $\lambdaBa(r; R_0)$ and $\lambdaBb(r; R_0)$ may be expressed as
\begin{align}
\lambdaBa^0(r;R_0) =
	\begin{cases}
		2 \pi \lambdaB r \left(\frac{\thetabr}{2 \pi} - \frac{1}{\pi}J\left(r; R_0, \cos\left(\frac{\thetabr}{2}\right)\right) \right) &r \in \left[0, 2 R_0 \cos\left( \frac{\thetabr}{2}\right) \right)
		\\
		2 \pi \lambdaB r \left(\frac{\thetabr}{2 \pi} - \frac{1}{\pi}J\left(r; R_0, \frac{r}{2R_0}\right) \right) &r \in \left[2 R_0 \cos\left( \frac{\thetabr}{2}\right), 2 R_0 \right)
		\\
		2 \pi \lambdaB r \frac{\thetabr}{2 \pi} &r \in [2 R_0, \infty )
	\end{cases}
\end{align}
\begin{align}
\lambdaBb^0(r;R_0) =
	\begin{cases}
		2 \pi \lambdaB r \left(1 - \frac{\thetabr}{2 \pi} - \frac{1}{\pi}J\left(r; R_0, \frac{r}{2 R_0}, \cos\left(\frac{\thetabr}{2}\right)\right) \right) &r \in \left[0, 2 R_0 \cos\left( \frac{\thetabr}{2}\right) \right)
		\\
		2 \pi \lambdaB r \left(1 - \frac{\thetabr}{2 \pi}\right) &r \in [2 R_0 \cos\left( \frac{\thetabr}{2}\right), \infty ).
	\end{cases}
\end{align}

Where we have used the notation $J(r; R_0, z_1, z_2) = J(r; R_0, z_1) - J(r; R_0, z_2)$.}

Hence, to characterize bounds for these expressions, we require upper and lower bounds on $J(r; R_0, z)$.

\subsection{Summary of Intermediate Lemmas}
We first establish the following lemmas, which we shall use to obtain the desired upper bounds on the above intensity functions.

\textbf{Lemma 3: (Bounds on $J(r; R_0, z)$)} {\em $J(r; R_0, z)$ admits the following lower bounds:
\begin{align}
	\arccos\left(z\right) e^{-\beta R_0} \le \arccos\left(z \right) \exp\left( -\beta \sqrt{r^2 + R_0^2 - 2rR_0 \sinc(\arccos(z)/\pi)} \right) \le J(r;R_0, z).
\end{align}
Additionally it may be upper bounded as 
\begin{align}
	 J(r;R_0, z) \le \arccos\left(z \right) e^{-\beta \lvert r - R_0\rvert}. 
\end{align}}

{\em Proof:} For conciseness, let $\theta_{max} = \arccos\left(z \right)$. Then
\begin{align*}
	J(r; R_0, z) &= \int_{z}^1(1 - u^2)^{-1/2} \exp\left(-\beta\sqrt{r^2 - 2rR_0u + R_0^2} \right)du
	\\
	&= \theta_{max} \int_{z}^1\theta_{max}^{-1}(1 - u^2)^{-1/2} \exp\left(-\beta\sqrt{r^2 - 2rR_0u + R_0^2} \right)du.
\end{align*}
Let $f(u) = \theta_{max}^{-1}(1 - u^2)^{-1}$. Note the $f(u)$ integrates to $1$ over the specified bounds of integration. Hence, we may interpret it as a probability measure. Then, letting $U \sim f(u)$ we have 
\begin{align*}
	J(r; R_0, z)
	&= \theta_{max} \E_{U}\left[ \exp\left(-\beta\sqrt{r^2 - 2rR_0u + R_0^2} \right)\right]
	\\
	&\gelabel{a} \theta_{max}  \exp\left(-\beta\sqrt{\E_{U}\left[r^2 - 2rR_0u + R_0^2\right]} \right)
	\\
	&= \arccos\left(z \right) \exp\left( -\beta \sqrt{r^2 + R_0^2 - \frac{2rR_0 \sin(\arccos(z))}{\arccos\left(z \right)}} \right)
	\\
	&\gelabel{b} \arccos\left(z \right) e^{-\beta R_0}.
\end{align*}
Where (a) follows from Jensen's Inequality and the fact that $e^{-\sqrt{x}}$ is convex, and (b) follows from maximizing the expression within the square root over $r$. 

Finally, the upper bounds may be obtained as follows:
\begin{align*}
	J(r; R_0, z) &= \int_{z}^1(1 - u^2)^{-1/2} \exp\left(-\beta\sqrt{r^2 - 2rR_0u + R_0^2} \right)du
	\\
	&\le \int_{z}^1(1 - u^2)^{-1/2} \exp\left(-\beta\sqrt{r^2 - 2rR_0 + R_0^2} \right)du.
	\\
	&= \arccos\left(z \right) e^{-\beta \lvert r - R_0\rvert}.
\end{align*}
$\qedwhite$

{\bf Lemma 4: (Finite Order Polynomial Bounds for $\arccos(z)$):} {\em Let $M_a \in \N$ and define
\begin{align}
	p(z, M_a) = \sum_{k = 0}^{M_a} \frac{z^{2k+1}\Gamma(k + 1/2)}{\Gamma(1/2) k! (1 + 2k)} + z^{2M_a + 3}\frac{\sqrt{2}}{\Gamma(1/2)}\left(\frac{\pi}{2} - \arctan(\sqrt{2 M_a + 3}) \right).
\end{align} 
Then for $z \in [0, 1]$,
\begin{align}
	\frac{\pi}{2} - p(z; M_a) \le \arccos(z) \le \frac{\pi}{2} - p(z; M_a) + z^{2M_a + 3}\frac{\sqrt{2}}{\Gamma(1/2)}\left(\frac{\pi}{2} - \arctan(\sqrt{2 M_a + 3}) \right).
\end{align}}

{\em Proof:} Note that $\arccos(z) = \pi/2 - \arcsin(z)$. Hence, we shall proceed by bounding $\arcsin(z)$. The taylor expansion of $\arcsin(z)$ is 
\begin{align*}
	\arcsin(z) &= \sum_{k = 0}^{\infty} \frac{z^{2k+1}\Gamma(k + 1/2)}{\Gamma(1/2) k! (1 + 2k)}
	\\
	&= \sum_{k = 0}^{M_a} \frac{z^{2k+1}\Gamma(k + 1/2)}{\Gamma(1/2) k! (1 + 2k)} +  \sum_{k = M_a + 1}^{\infty} \frac{z^{2k+1}\Gamma(k + 1/2)}{\Gamma(1/2) k! (1 + 2k)}
	\\
	&\lelabel{a} \sum_{k = 0}^{M_a} \frac{z^{2k+1}\Gamma(k + 1/2)}{\Gamma(1/2) k! (1 + 2k)} +  \sum_{k = M_a + 1}^{\infty} \frac{z^{2k+1}k^{-1/2}}{\Gamma(1/2) (1 + 2k)}
	\\
	&\lelabel{b} \sum_{k = 0}^{M_a} \frac{z^{2k+1}\Gamma(k + 1/2)}{\Gamma(1/2) k! (1 + 2k)} +  \frac{z^{2M_a+3}}{\Gamma(1/2)}\sum_{k = M_a + 1}^{\infty} \frac{k^{-1/2}}{ 1 + 2k}
	\\
	&\le \sum_{k = 0}^{M_a} \frac{z^{2k+1}\Gamma(k + 1/2)}{\Gamma(1/2) k! (1 + 2k)} +  \frac{z^{2M_a+3}}{\Gamma(1/2)}\int_{M_a + 1}^{\infty} \frac{(x-1)^{-1/2}}{ 1 + 2(x-1)}dx
	\\
	&=  \sum_{k = 0}^{M_a} \frac{z^{2k+1}\Gamma(k + 1/2)}{\Gamma(1/2) k! (1 + 2k)} +  \frac{z^{2M_a+3} \sqrt{2}}{\Gamma(1/2)} \left(\frac{\pi}{2} - \arctan(\sqrt{2 M_a + 3}) \right).
\end{align*}
Where (a) follows from the inequality $\Gamma(k + 1/2)/k! \le k^{-1/2}$, (b) follows from the fact that $z \in \text{dom}(\arcsin) = [0, 1]$. 
By a similar argument, we have 
\begin{align*}
	\arcsin(z) &= \sum_{k = 0}^{\infty} \frac{z^{2k+1}\Gamma(k + 1/2)}{\Gamma(1/2) k! (1 + 2k)}
	\\
	&\ge \sum_{k = 0}^{M_a} \frac{z^{2k+1}\Gamma(k + 1/2)}{\Gamma(1/2) k! (1 + 2k)}.
	\end{align*} $\qedwhite$

{\bf Lemma 5: (Concavity of $\sinc(\arccos(z)/\pi)$):} {\em The function $\sinc(\arccos(z)/\pi)$ is concave for $z \in [0, 1]$}

{\em Proof:} Follows from second order concavity conditions. $\qedwhite$

\subsection{Bounds on $J(r; R_0, z)$}
To obtain bounds on the aforementioned intensity measures, we require
\begin{enumerate}[label=\alph*)]
	\item A lower bound for $J\left(r; R_0, \frac{r}{2 R_0}\right)$ for $r \in [0, 2 R_0 \cos(\thetabr/2)]$
	\item A lower bound for $J\left(r; R_0, \frac{r}{2 R_0}\right)$ for $r \in (2 R_0 \cos(\thetabr/2), 2R_0]$
	\item A lower bound for $J\left(r; R_0, \cos\left(\frac{\thetabr}{2}\right)\right)$ for $r \in [0, 2 R_0 \cos(\thetabr/2)]$
	\item An upper bound for $J\left(r; R_0, \cos\left(\frac{\thetabr}{2}\right) \right)$ for $r \in [0, 2 R_0 \cos(\thetabr/2)]$.
\end{enumerate}

Note that we already have a sufficient bound for case d) from Lemma 3. It remains to characterize the lower bounds in cases a) through c). Each of these rely on piecewise linear bounds for the argument of the exponential function in the Jensen bound derived in Lemma 3. For conciseness, we shall exploit the function $h$ defined as follows:
\begin{align}
	&h\left(x; \{m_i, c_i\}_{i = 1}^k, \{r_i\}_{i = 1}^{k=1} \right) = \sum_{i = 1}^{k} (c_i + m_ix) \indicator\{r \in [r_{i-1}, r_i)\} &r_0 = -\infty, r_k = \infty.
\end{align}

 We begin with the bounds for $J\left(r; R_0, \frac{r}{2 R_0}\right)$.

{\bf Lemma 6: (Refined Lower bounds for $J\left(r; R_0, \frac{r}{2 R_0}\right)$):} {\em The following lower bound for $J\left(r; R_0, \frac{r}{2 R_0}\right)$ holds for $r \in [0, 2 R_0 \cos(\thetabr/2)]$: (For conciseness, let $r_M = 2 R_0 \cos(\thetabr/2)$)
\begin{itemize}
	\item Define
	\begin{align}
		&c_s^{\ell} = sinc(1/2)
		\\
		&m_s^{\ell} = \frac{\sinc\left( \frac{\thetabr}{2 \pi}\right) - \sinc(1/2)}{cos\left(\frac{\thetabr}{2}\right)} 
		\\
		&r_1^{\ell} = \frac{c_s^{\ell}}{1 - m_s^{\ell}}R_0
	\end{align}
	\item Let 
	\begin{align}
		f_2(r; R_0; c_s, m_s) = \sqrt{(1-m_s)r^2 + R_0^2 - 2 R_0 c_s r}
	\end{align}
	\item If $r_1^{\ell} \le r_M \iff cos\left( \frac{\thetabr}{2}\right) - \sinc\left( \frac{\thetabr}{2 \pi}\right) + \frac{1}{\pi} \ge 0$, let
	\begin{align}
		&k_{\ell}(\thetabr) = 2
		\\
		&m_1^{\ell} = \frac{f_2(r_1^{\ell}; R_0; c_s^{\ell}, m_s^{\ell}) - R_0}{r_1^{\ell}}
		\\
		&c_1^{\ell} = R_0
		\\
		&m_2^{\ell} = \frac{f_2(r_M; R_0; c_s^{\ell}, m_s^{\ell}) - f_2(r_1^{\ell}; R_0; c_s^{\ell}, m_s^{\ell})}{r_M - r_1^{\ell}}
		\\
		&c_2^{\ell} = f_2(r_1^{\ell}; R_0; c_s^{\ell}, m_s^{\ell}) - m_2^{\ell} r_1^{\ell}
	\end{align}
	\item Otherwise, let
	\begin{align}
		&k_l(\thetabr) = 1
		\\
		&m_1^{\ell} = \frac{f_2(r_M; R_0; c_s^{\ell}, m_s^{\ell}) - R_0}{r_M}
		\\
		&c_1^{\ell} = R_0
	\end{align}
\end{itemize}

Then for any $N_a \in \N$,
\begin{align}
	J\left(r; R_0, \frac{r}{2 R_0}\right) \ge \left(\frac{\pi}{2} - p\left(\frac{r}{2R_0}; N_a\right)\right)\exp\left(-\beta h\left(r; \{m_i^{\ell}, c_i^{\ell}\}_{i \in [k_l(\thetabr)]}, \{r_1^{\ell}\}_{i \in [k_l(\thetabr) - 1]}\right) \right).
\end{align}

Similarly, for $r \in (r_M, 2 R_0]$ the following lower bound holds
\begin{itemize}
	\item Define
	\begin{align}
		&m_s^u = \frac{1 - \sinc\left( \frac{\thetabr}{2 \pi}\right)}{1 - cos\left(\frac{\thetabr}{2}\right)} 
		\\
		&c_s^u = \sinc\left( \frac{\thetabr}{2 \pi}\right) - m_s^u \frac{r_M}{2 R_0}
		\\
		\\
		&r_1^u = \frac{c_s^u}{1 - m_s^u}R_0
	\end{align}
	\item If $r_1^u \le r_M \iff \frac{\thetabr}{2}\ge \frac{\pi}{3}$, let
	\begin{align}
		&k_u(\thetabr) = 1
		\\
		&m_1^u = \frac{f_2(2 R_0; R_0; c_s^u, m_s^u) - f_2(r_M; R_0; c_s^u, m_s^u)}{2R_0 - r_M}
		\\
		&c_1^u = f_2(r_M; R_0; c_s^u, m_s^u) - m_1^ur_M
	\end{align}
	\item Otherwise, let
	\begin{align}
		&k_u(\thetabr) = 2
		\\
		&m_1^u = \frac{f_2(r_1^u; R_0; c_s^u, m_s^u) - f_2(r_M; R_0; c_s^u, m_s^u)}{r_1^u - r_M}
		\\
		&c_1^u = f_2(r_M; R_0; c_s^u, m_s^u) - m_1^ur_M
		\\
		&m_2^u = \frac{f_2(2 R_0; R_0; c_s^u, m_s^u) - f_2(r_1^u; R_0; c_s^u, m_s^u)}{2R_0 - r_1^u}
		\\
		&c_2^u = f_2(r_1^u; R_0; c_s^u, m_s^u) - m_2^u r_1^u
	\end{align}
\end{itemize}
Then for any $M_a \in \N$,
\begin{align}
	J\left(r; R_0, \frac{r}{2 R_0}\right) \ge \left(\frac{\pi}{2} - p\left(\frac{r}{2R_0}; M_a\right)\right)\exp\left(-\beta h\left(r; \{m_i^u, c_i^u\}_{i \in [k_u(\thetabr)]}, \{r_1^u\}_{i \in [k_u(\thetabr) - 1]}\right) \right).
\end{align}}

{\em Proof:} Using the Jensen Bound from Lemma 3 and concavity of the sinc term (Lemma 5), one may conclude  that $f_2$ is a lower bound for the argument of the exponential term in the Jensen bound from Lemma 3. Moreover, $r_1^{\ell}$ and $r_1^u$ are inflection points of $f_2$. The different cases are simply when $r_1^{\ell}$ and $r_2^u$ are included in the range of interest. Note the $f_2$ is convex and apply Jensen's Inequality again to get the linearization parameters. Proposition 3 then follows, using Lemma 4 to handle the $\arccos$ term. $\qedwhite$

 Similar bounds may be obtained for $J\left(r; R_0, \cos \left(\frac{\thetabr}{2} \right)\right)$.

{\bf Lemma 7: (Refined Lower bounds for $J\left(r; R_0, \cos\left(\frac{\thetabr}{2}\right)\right)$):} {\em The following lower bound for \\ $J\left(r; R_0, \cos\left(\frac{\thetabr}{2}\right)\right)$ holds for $r \in [0, 2 R_0 \cos(\thetabr/2)]$: (For conciseness, let $r_M = 2 R_0 \cos(\thetabr/2)$)
\begin{itemize}
	\item Let 
	\begin{align}
		&f_1(r; R_0; \thetabr) = \sqrt{r^2 + R_0^2 - 2 R_0 \sinc\left(\frac{\thetabr}{2\pi} \right)r}
		\\
		&r_1 = R_0 \sinc\left(\frac{\thetabr}{2\pi} \right)
	\end{align}
	\item If $r_1 \le r_M \iff \sinc\left(\frac{\thetabr}{2\pi} \right)\le 2 \cos\left(\frac{\thetabr}{2} \right)$, let
	\begin{align}
		&k(\thetabr) = 2
		\\
		&m_1 = \frac{f_1(r_1; R_0; \thetabr) - R_0}{r_1}
		\\
		&c_1 = R_0
		\\
		&m_2 = \frac{f_1(r_M; R_0; \thetabr) - f_1(r_1; R_0; \thetabr)}{r_M - r_1}
		\\
		&c_2 = f_1(r_1; R_0; \thetabr) - m_2 r_1
	\end{align}
	\item Otherwise, let
	\begin{align}
		&k(\thetabr) = 1
		\\
		&m_1 = \frac{f_1(r_M; R_0; \thetabr) - R_0}{r_M}
		\\
		&c_1 = R_0
	\end{align}
\end{itemize}

Then,
\begin{align}
	J\left(r; R_0, \cos \left(\frac{\thetabr}{2} \right)\right) \ge \frac{\thetabr}{2}\exp\left(-\beta h\left(r; \{m_i, c_i\}_{i \in [k(\thetabr)]}, \{r_1\}_{i \in [k(\thetabr) - 1]}\right) \right).
\end{align}}

{\em Proof:} Note that $f_1$ is the argument of the exponential function in the Jensen bound from Lemma 3. One may determine that $r_1$ is the inflection point of $f_1$. The different cases are simply when $r_1$ is included in the range of interest. Noting that $f_1$ is convex, the linearization parameters follow from Jensen's Inequality. $\qedwhite$

\subsection{Proof of Lemma 8}
{\bf Lemma 8: (Bounds on the Intensity Measures of the Palm Interference Process)} {\em Let $k(\thetabr)$, $k^{\ell}(\thetabr)$, $k^u(\thetabr)$, $\{(r_i, c_i, m_i)\}_{i \in [k(\thetabr)]}$, $\{(r^{\ell}_i, c^{\ell}_i, m^{\ell}_i)\}_{i \in [k^{\ell}(\thetabr)]}$, $\{(r^u_i, c^u_i, m^u_i)\}_{i \in [k^u(\thetabr)]}$, and $\{\gamma_k\}_{k \in [M_a + 1]}$ (where $M_a \in \N$) be as defined in Appendix \ref{app:palm_int_ub}. Then
\begin{enumerate}[label=\roman*)]
	\item The following upper bounds hold: $\lambdaBa^0(r;R_0) \le \rhoBa(r;R_0)$ and $\lambdaBb^0(r;R_0) \le \rhoBb(r;R_0)$, where
	\begin{align}
	&\rhoBa(r;R_0) =  2 \lambdaB r \left(\frac{\thetabr}{2}   \right.
	-\frac{\thetabr}{2}\sum_{i = 1}^{k(\thetabr)} \indicator\left\{r \in \left[r_{i-1}, \min\left\{ r_i, 2 R_0\cos\left(\frac{\thetabr}{2}\right)\right\}\right]\right\} e^{-\beta(c_i + m_i r)} \nonumber
		\\
		&- \sum_{i = 1}^{k^u(\thetabr)}\frac{\pi}{2} \indicator\left\{r \in \left[\max\left\{ r_{i-1}^u, 2 R_0\cos\left(\frac{\thetabr}{2}\right)\right\}, \min\left\{ r_i^u, 2R_0\right\}\right]\right\} e^{-\beta(c_i^u + m_i^u r)} \nonumber
		\\
		&\left.+ \sum_{k = 0}^{M_a +1} \frac{\gamma_k}{(2 R_0)^{2k + 1}} \indicator\left\{r \in \left[\max\left\{ r_{i-1}^u, 2 R_0\cos\left(\frac{\thetabr}{2}\right)\right\}, \min\left\{ r_i^u, 2R_0\right\}\right]\right\} r^{2k + 1} e^{-\beta(c_i^u + m_i^u r)} \right),
		\label{eq:int-meas-ub1}
	\end{align}
	and
	\begin{align}
	&\rhoBb(r;R_0) = 2 \lambdaB r \left(\frac{2 \pi - \thetabr}{2}  \right.
		-\sum_{i = 1}^{k^{\ell}(\thetabr)} \frac{\pi}{2}\indicator\left\{r \in \left[r_{i-1}^{\ell}, \min\left\{ r_i^{\ell}, 2 R_0\cos\left(\frac{\thetabr}{2}\right)\right\}\right]\right\} e^{-\beta(c_i^{\ell} + m_i^{\ell} r)} \nonumber
		\\
		& + \sum_{k = 0}^{M_a +1} \frac{\gamma_k}{(2 R_0)^{2k + 1}} \indicator\left\{r \in \left[r_{i-1}^{\ell}, \min\left\{ r_i^{\ell}, 2 R_0\cos\left(\frac{\thetabr}{2}\right)\right\}\right]\right\} r^{2k + 1} e^{-\beta(c_i^{\ell} + m_i^{\ell} r)} \nonumber
		\\
		&+ \indicator\left\{r \in \left[0, \min\left\{ R_0, 2 R_0\cos\left(\frac{\thetabr}{2}\right)\right\}\right]\right\} \frac{\thetabr}{2} e^{-\beta(R_0 - r)} \nonumber
		\\
		&\left.+ \indicator\left\{r \in \left[R_0, 2 R_0\cos\left(\frac{\thetabr}{2}\right)\right]\right\} \frac{\thetabr}{2}e^{-\beta(r - R_0)} \right).
		\label{eq:int-meas-ub2}
\end{align}
\item The following lower bounds hold: $\lambdaBa^0(r;R_0) \ge \nuBa(r;R_0)$ and $\lambdaBb^0(r;R_0) \ge \nuBb(r;R_0)$, where
	\begin{align}
	&\nuBa(r;R_0) =  2 \lambdaB r \left(\frac{\thetabr}{2} -\indicator\left\{r \in \left[0, 2 R_0\cos\left(\frac{\thetabr}{2}\right)\right]\right\} \frac{\thetabr}{2}e^{-\beta \lvert R_0 - r \rvert}  \right. \nonumber
	\\
	&-\frac{\pi}{2} \indicator\left\{r \in \left[2 R_0\cos\left(\frac{\thetabr}{2}\right), 2R_0\right]\right\}  e^{-\beta \lvert R_0 - r \rvert} \nonumber
	\\
	&\left.+ \sum_{k = 0}^{M_a} \frac{\gamma_k}{(2 R_0)^{2k + 1}} \indicator\left\{r \in \left[2 R_0\cos\left(\frac{\thetabr}{2}\right), 2R_0\right]\right\}  r^{2k + 1} e^{-\beta \lvert R_0 - r \rvert}  \right),
	\label{eq:int-meas-lb1}
	\end{align}
	and
	\begin{align}
	&\nuBb(r;R_0) = 2 \lambdaB r \left(\frac{2 \pi - \thetabr}{2} -\frac{\pi}{2} \indicator\left\{r \in \left[0,2 R_0\cos\left(\frac{\thetabr}{2}\right)\right]\right\}  e^{-\beta \lvert R_0 - r \rvert} \right. \nonumber
	\\
	&+ \sum_{k = 0}^{M_a} \frac{\gamma_k}{(2 R_0)^{2k + 1}}  \indicator\left\{r \in \left[0,2 R_0\cos\left(\frac{\thetabr}{2}\right)\right]\right\} r^{2k + 1} e^{-\beta \lvert R_0 - r \rvert} \nonumber
	\\
	&\left.+\frac{\thetabr}{2}\sum_{i = 1}^{k(\thetabr)} \indicator\left\{r \in \left[r_{i-1}, \min\left\{ r_i, 2 R_0\cos\left(\frac{\thetabr}{2}\right)\right\}\right]\right\} e^{-\beta(c_i + m_i r)} \right).
	\label{eq:int-meas-lb2}
\end{align}
\end{enumerate}}

{\em Proof:}Using the intermediate result at the beginning of the appendix, the bounds in Lemma 8 follow from Lemmas 3, 6, and 7. These follow directly by selecting the appropriate lower bound in Lemma 6 or 7, or the upper bound in Lemma 3, depending on whether or not the $J$ function at hand is being added or subtracted. $\qedwhite$

\section{Characterization of the Radar Path Loss Mellin Transforms}
\label{app:int_mt}
\subsection{Statement of Pathloss Mellin Transforms}
The main utility of the bounds for the intensity functions of the Palm interference processes in Lemma 8 is that they enable us to obtain closed form expressions for the sectional path loss Mellin Transforms in terms of the   generalized incomplete gamma function
\begin{align}
	\Gamma(p, z_1, z_2) = \int_{z_1}^{z_2} x^{p-1}e^{-x}dx.
\end{align}
Note that this function is defined for complex $p, z_1, z_2$ by analytic continuation. 

For conciseness, we state the path loss Mellin Transforms in terms of the following functions

{\bf Definition: ($G_L$ function)} {\em Let $A$ and $B$ be compact intervals on $\R_+$, and let $\alpha, \gamma, m, n$ be arbitrary parameters and $p \in \C$. Moreover, define
\begin{align}
	&d_L = \inf\{ d \in A \cap B\} &d_U = \sup\{ d \in A \cap B\}.
\end{align}
Then the function $G_L$ is defined as 
\begin{align}
	G_L(p, A; B, \alpha, \gamma,  n, m) = \int_{\R_+} \indicator\{x \in A \cap B\} r^{n - p\alpha} e^{-(m + p\gamma)x}dx.
\end{align}
which evaluates to
\begin{align}
	G_L(p, A; B, \alpha, \gamma, n ,m) =
	\begin{cases}
		0 & \text{if } A \cap B = \emptyset
		\\
		(m +s\gamma)^{p\alpha -n  - 1} \Gamma(n + 1 - p \alpha, (m + p \gamma)d_L, (m + p \gamma)d_U) &\text{otherwise}.
	\end{cases}
	\label{eq:GL-func}
\end{align}}

{\bf Definition: ($G_N$ function)} {\em Let $A$ and $B$ be compact intervals on $\R_+$, and let $\alpha, \gamma, m_1, m_2, n$ be arbitrary parameters and $s \in \C$. Then the function $G_L$ is defined as 
\begin{align}
	G_N(p, A; B, \alpha, \gamma, n, m_1, m_2) = \int_{\R_+} \indicator\{x \in A \cap B\} r^{n - p\alpha} e^{-p\gamma x}\left(e^{-m_1 x} - e^{-m_2 x} \right)dx.
\end{align}
which evaluates to
\begin{align}
	G_N(p, A; B, \alpha, \gamma, n, m_1, m_2) = G_L(p, A; B, \alpha, \gamma, n, m_1) - G_L(p, A; B, \alpha, \gamma, n, m_2).
	\label{eq:GN-func}
\end{align}}

We summarize the path loss Mellin Transforms with respect to the Palm interference process in the following lemma.

{\bf Lemma 9: (Path Loss Mellin Transforms with Respect to the Bounding Radar Interference Process)} {\em Let $A \subseteq \R_+$ then,

\begin{enumerate}[label=\roman*)]
	\item The path loss Mellin Transforms with respect to $\rhoLa$ and $\rhoLb$
\begin{align}
	&\Mellin_{\rhoLa \circ \gL^{-1} }(p; A, R_0) = 2 \lambdaB \KL^{p-1}\left(\frac{\thetabr}{2} G_L\left(p-1, A; \R_+, \alphaL, \gammaL, 1, \beta\right) \right. \nonumber
		\\
		&-\frac{\thetabr}{2}\sum_{i = 1}^{k(\thetabr)} e^{-\beta c_i}G_L\left(p-1, A; \left[r_{i-1}, \min\left\{ r_i, 2 R_0\cos\left(\frac{\thetabr}{2}\right)\right\}\right], \alphaL, \gammaL, 1, \beta(m_i+1) \right) \nonumber
		\\ 
		 &-\sum_{i = 1}^{k^u(\thetabr)} \frac{\pi}{2} e^{-\beta c_i^u} G_L\left(p-1, A; \left[\max\left\{ r_{i-1}^u, 2 R_0\cos\left(\frac{\thetabr}{2}\right)\right\}, \min\left\{ r_i^u, 2R_0\right\}\right], \alphaL, \gammaL, 1, \beta(m_i^u+1) \right) \nonumber
		 \\
		  &+\sum_{k = 0}^{N_a +1} \frac{\gamma_k}{(2 R_0)^{2k + 1}} e^{-\beta c_i^u} \nonumber
		 \\
		 &\left.G_L\left(p-1, A; \left[\max\left\{ r_{i-1}^u, 2 R_0\cos\left(\frac{\thetabr}{2}\right)\right\}, \min\left\{ r_i^u, 2R_0\right\}\right], \alphaL, \gammaL, 2k +2, \beta(m_i^u+1) \right)\right),
		 \label{eq:mel-lb-los1}
\end{align}
and
\begin{align}
	&\Mellin_{\rhoLb \circ \gL^{-1} }(p; A, R_0) = 2 \lambdaB \KL^{p-1} \left(\frac{2 \pi - \thetabr}{2} G_L\left(p-1, A; \R_+, \alphaL, \gammaL, 1, \beta\right) \right. \nonumber
		\\
		 &-\sum_{i = 1}^{k^{\ell}(\thetabr)} \frac{\pi}{2} e^{-\beta c_i^{\ell}} _L\left(p-1, A; \left[r_{i-1}^{\ell}, \min\left\{ r_i^{\ell}, 2 R_0\cos\left(\frac{\thetabr}{2}\right)\right\}\right], \alphaL, \gammaL, 1, \beta(m_i^{\ell}+1) \right) \nonumber
		 \\
		 &+\sum_{k = 0}^{N_a +1} \frac{\gamma_k}{(2 R_0)^{2k + 1}} e^{-\beta c_i^{\ell}} G_L\left(p-1, A; \left[r_{i-1}^{\ell}, \min\left\{ r_i^{\ell}, 2 R_0\cos\left(\frac{\thetabr}{2}\right)\right\}\right], \alphaL, \gammaL, 2k +2, \beta(m_i^{\ell}+1) \right) \nonumber
		 \\
		 &+\frac{\thetabr}{2} e^{-\beta R_0}G_L\left(p-1, A; \left[0, \min\left\{ R_0, 2 R_0\cos\left(\frac{\thetabr}{2}\right)\right\}\right], \alphaL, \gammaL, 1, 0 \right) \nonumber
		 \\
		&\left.+\frac{\thetabr}{2} e^{\beta R_0}G_L\left(p-1, A; \left[R_0, 2 R_0\cos\left(\frac{\thetabr}{2}\right)\right], \alphaL, \gammaL, 1, 2\beta \right)\right).
		\label{eq:mel-lb-los2}
\end{align}
\item Similarly, the path loss Mellin Transforms with respect to $\nuLa$ and $\nuLb$ are:
\begin{align}
	&\Mellin_{\nuLa \circ \gL^{-1} }(p; A, R_0) = 2 \lambdaB \KL^{p-1} \left(\frac{\thetabr}{2} G_L\left(p-1, A; \R_+, \alphaL, \gammaL, 1, \beta\right) \right. \nonumber
		\\
		&-\frac{\thetabr}{2}e^{-\beta R_0}G_L\left(p-1, A; \left[0, \min\left\{ R_0, 2 R_0\cos\left(\frac{\thetabr}{2}\right)\right\}\right], \alphaL, \gammaL, 1, 0 \right) \nonumber
		\\
		&-\frac{\thetabr}{2}e^{\beta R_0}G_L\left(p-1, A; \left[R_0, 2 R_0\cos\left(\frac{\thetabr}{2}\right) \right], \alphaL, \gammaL, 1, 2\beta \right) \nonumber
		\\ 
		 &-\frac{\pi}{2} e^{-\beta R_0} G_L\left(p-1, A; \left[2 R_0\cos\left(\frac{\thetabr}{2}\right), R_0\right], \alphaL, \gammaL, 1, 0 \right) \nonumber
		 \\ 
		 &-\frac{\pi}{2} e^{\beta R_0} G_L\left(p-1, A; \left[\max\left\{ R_0, 2 R_0\cos\left(\frac{\thetabr}{2}\right)\right\}, 2R_0\right], \alphaL, \gammaL, 1, 2 \beta \right) \nonumber
		 \\
		  &+\sum_{k = 0}^{N_a} \frac{\gamma_k}{(2 R_0)^{2k + 1}} e^{-\beta R_0} G_L\left(p-1, A; \left[2 R_0\cos\left(\frac{\thetabr}{2}\right), R_0\right], \alphaL, \gammaL, 2k +2, 0\right) \nonumber
		 \\
		  &\left.+\sum_{k = 0}^{N_a} \frac{\gamma_k}{(2 R_0)^{2k + 1}} e^{\beta R_0} G_L\left(p-1, A; \left[\max\left\{ R_0, 2 R_0\cos\left(\frac{\thetabr}{2}\right)\right\}, 2R_0\right], \alphaL, \gammaL, 2k +2, 2 \beta \right)\right),
		  \label{eq:mel-ub-los1}
\end{align}
\begin{align}
	&\Mellin_{\nuLb \circ \gL^{-1} }(p; A, R_0) = 2 \lambdaB \KL^{p-1} \left(\frac{2 \pi - \thetabr}{2} G_L\left(p-1, A; \R_+, \alphaL, \gammaL, 1, \beta\right) \right. \nonumber
		\\
		&-\frac{\pi}{2}e^{-\beta R_0}G_L\left(p-1, A; \left[0, \min\left\{ R_0, 2 R_0\cos\left(\frac{\thetabr}{2}\right)\right\}\right], \alphaL, \gammaL, 1, 0 \right) \nonumber
		\\
		&-\frac{\pi}{2}e^{\beta R_0}G_L\left(p-1, A; \left[R_0, 2 R_0\cos\left(\frac{\thetabr}{2}\right) \right], \alphaL, \gammaL, 1, 2\beta \right) \nonumber
		 \\
		  &+\sum_{k = 0}^{N_a} \frac{\gamma_k}{(2 R_0)^{2k + 1}} e^{-\beta R_0} G_L\left(p-1, A; \left[0, \min\left\{ R_0, 2 R_0\cos\left(\frac{\thetabr}{2}\right)\right\}\right], \alphaL, \gammaL, 2k +2, 0\right) \nonumber
		 \\
		  &+\sum_{k = 0}^{N_a} \frac{\gamma_k}{(2 R_0)^{2k + 1}} e^{\beta R_0} G_L\left(p-1, A; \left[R_0, 2 R_0\cos\left(\frac{\thetabr}{2}\right) \right], \alphaL, \gammaL, 2k +2, 2 \beta \right) \nonumber
		  \\
		&\left.+\frac{\thetabr}{2}\sum_{i = 1}^{k(\thetabr)} e^{-\beta c_i}G_L\left(p-1, A; \left[r_{i-1}, \min\left\{ r_i, 2 R_0\cos\left(\frac{\thetabr}{2}\right)\right\}\right], \alphaL, \gammaL, 1, \beta(m_i+1) \right)\right).
		\label{eq:mel-ub-los2}
\end{align}
\end{enumerate} 

Similar expressions hold for the path loss Mellin Transforms of $\rhoNa$/$\rhoNb$ and $\nuNa$/$\nuNb$ using the $G_N$ function.}

{\em Proof:} Follows from Lemma 8 and the definitions of $G_L(\cdot)$ and $G_N(\cdot)$. $\qedwhite$

\section{Proof of Theorem 3}
\label{app:main_res}
\subsection{Proof of Theorem 3}
Using the notation developed in Sec. \ref{sec:palm_int_derivation} we may express the generic radar SINR model, $\SINRgen$ as
\begin{align}
	\SINRgen = \indicator\{\PhiL(\R^2) > 0\} \frac{ \kc  \gLr(R_0)  }{ \sum_{X_k \in \PiBt^0} F_k Z_B^k L(R_k) + \nur},
\end{align}
Hence, we have
\begin{align*}
	\Pb(\SINRgen \ge \tau) &= \Pb\left(\indicator\{R_0 < \infty\} \kc  \gLr(R_0) \ge \tau \sum_{R_k \in \PiBt^0} F_k Z_B^k L(R_k) +  \tau\nur \right)
		\\
	&= \Pb\left(\left.\kc  \gLr(R_0) \ge \tau \sum_{R_k \in \PiBt^0} F_k Z_B^k L(R_k)  + \tau\nur  \right\vert R_0 < \infty \right) \left(1 - \exp\left(\frac{-2 \pi \lambdaB}{\beta^2}\right) \right)
	\\
	&= \E_{R_0}\left[\Pb\left(\left.\kc  \gLr(R_0) \ge \tau \sum_{R_k \in \PiBt^0} F_k Z_B^k L(R_k)  + \tau\nur  \right\vert R_0 \right) \right]\left(1 - \exp\left(\frac{-2 \pi \lambdaB}{\beta^2}\right) \right).
\end{align*}
The term inside the expectation may be expressed as
\begin{align}
	\Pb\left(\kc  \gLr(R_0) \ge \tau \sum_{R_k \in \PiBt^0} \right.&\left. F_k Z_B^k L(R_k)  + \tau\nur  \bigg\vert R_0 \right)
	\\
&\eqlabel{a} \E_{\PiBt^0}\left[\exp\left(\frac{-\tau}{\gLr(R_0)} \sum_{R_k \in \PiBt^0} F_k Z_B^k L(R_k)  \right) \bigg\vert R_0\right]e^{\frac{-\tau \nur}{\gLr(R_0)}}.
\label{eq:thm3_cond_ccdf}
\end{align}
Where (a) follows from the fact that $\kc$ is exponentially distributed. 

Now, using the fact that the LoS/NLoS and receive antenna gain marks in $\PiBt^0$ are independent when conditioned on $R_0$, we have by the Independent Thinning Theorem that $\PiB^0$ may be expressed as the following superposition of independent PPPs
\begin{align}
	\PiB^0 = \PiLa^0 + \PiNa^0 + \PiLb^0 + \PiNb^0.
\end{align}
Where the above processes correspond to LoS/NLoS points inside/outside the receive beam, respectively. Define
\begin{align}
	&\Ila = \sum_{R_k \in \PiLa^0} F_{L,k} \gL(R_k), &\Ilb = \sum_{R_k \in \PiLb^0} F_{L,k} \xibr \gL(R_k),
	\\
	&\Ina = \sum_{R_k \in \PiNa^0} F_{N,k} \gN(R_k), &\Inb = \sum_{R_k \in \PiNb^0} F_{N,k} \xibr \gN(R_k).
\end{align}
Then,
\begin{align*}
	\E_{\PiBt^0}&\left[\exp\left(\frac{-\tau}{\gLr(R_0)} \sum_{R_k \in \PiBt^0} F_k Z_B^k L(R_k)  \right) \bigg\vert R_0 \right] = 
	\\
	&\Laplace_{\Ila}\left(\frac{\tau}{\gLr(R_0)} \bigg\vert R_0\right) \Laplace_{\Ilb}\left(\frac{\tau \xibr}{\gLr(R_0)} \bigg\vert R_0\right) \Laplace_{\Ina}\left(\frac{\tau}{\gLr(R_0)} \bigg\vert R_0\right) \Laplace_{\Inb}\left(\frac{\tau\xibr}{\gLr(R_0)} \bigg\vert R_0\right).
\end{align*}
Where we have used the notation $\Laplace_{I}(\cdot; R_0)$ to denote conditioning on $R_0$.

Consider $\Laplace_{\Ila}(s; R_0)$. By Campbells Formula we have
\begin{align*}
	\Laplace_{\Ila}(s \vert R_0) &= \exp\left( - \int_{\R_+} (1 - \Laplace_{F_L}(s\gL(r))) \lambdaLa^0(r, R_0)dr\right).
\end{align*}
Then, using the bounds on $\lambdaLa^0$ from Lemma 8, we have
\begin{align*}
	\Laplace_{\Ila}(s \vert R_0) \le \exp\left( - \int_{\R_+} (1 - \Laplace_{F_L}(s\gL(r))) \nuLa(r, R_0)dr\right),
\end{align*}
and
\begin{align*}
	\Laplace_{\Ila}(s \vert R_0) \ge \exp\left( - \int_{\R_+} (1 - \Laplace_{F_L}(s\gL(r))) \rhoLa(r, R_0)dr\right)
\end{align*}
Moreover, using the expressions for the path loss Mellin Transforms in Lemma 9 in addition to Theorem 2 yields further bounds
\begin{align*}
	 \exp\left( - \int_{\R_+} (1 - \Laplace_{F_L}(s\gL(r))) \nuLa(r, R_0)dr\right) \le \exp\left(-\sum_{i = 1}^{N_w +1} \HU\left(s, W^L_i; 1 -\Laplace_{F_L}, \Mellin_{\nuLa \circ \gL^{-1}}(\cdot; \cdot, R_0)\right) \right),
\end{align*}
and
\begin{align*}
	\exp\left( - \int_{\R_+} (1 - \Laplace_{F_L}(s\gL(r))) \rhoLa(r, R_0)dr\right) \ge \exp\left(-\sum_{i = 1}^{N_w +1} \HL\left(s, W^L_i; 1 -\Laplace_{F_L}, \Mellin_{\rhoLa \circ \gL^{-1}}(\cdot; \cdot, R_0)\right) \right).
\end{align*}

Bounds on $\Laplace_{\Ilb}(s \vert R_0)$, $\Laplace_{\Ina}(s \vert R_0)$, and $\Laplace_{\Inb}(s \vert R_0)$ follow in a similar manner. Therefore, using these bounds leads to upper and lower bounds on the conditional CCDF of $\SINRgen$ given $R_0$ in (\ref{eq:thm3_cond_ccdf}). Taking expectation with respect to the pdf of $R_0$ in Lemma 1 results in the bounds $\Pcrl\left(\tau; \Laplace_{F_L}, \Laplace_{F_N}\right)$ and $\Pcru\left(\tau; \Laplace_{F_L}, \Laplace_{F_N}\right)$ given in the statement of the theorem. $\qedwhite$

\section{Proofs of Results Pertaining to Interference Fading Terms}
\label{app:int_fad}
\subsection{Proof of Lemma 10}
Our expression for the Laplace Transform of $F$ requires the following transcendental function.

{\bf Definition: (Fox's H Function):} {\em Fox's $H$ function is defined as
\begin{align}
	H^{m,n}_{p,q}\left(x \bigg\vert \begin{matrix} (a_1, \alpha_1), \dots, (a_p, \alpha_p) \\ (b_1, \beta_1), \dots, (b_q, \beta_q) \end{matrix}\right) = \oint_L \frac{\prod_{i = 1}^m \Gamma(b_i - \beta_i s) \prod_{j = 1}^n \Gamma(1 - a_j + \alpha_j s)}{\prod_{j = n+1}^p  \Gamma(a_j + \alpha_j s) \prod_{i = m+1}^q \Gamma(1 - b_i + \beta_i s) }s^x ds,
\end{align}
where $m, n ,p ,q$ are non-negative integers such that $n \le p$ and $m \le q$, $\alpha_j, \beta_i \in \R_+$ for all $i,j$, and $\{(b_i, \beta_i)\}$ and $\{(a_i, \alpha_i)\}$ are such that the poles of $\Gamma(b_i - \beta_i s)$ do not coincide with the poles of $\Gamma(1 - a_j + \alpha_j s)$. Note that the above contour integral corresponds to the inverse Mellin Transform operator.}

Armed with this definition, we now prove Lemma 10.

{\em Proof:} Part i). Note that $\GM(\bold{F}, \boldsymbol{\theta})$ may be expressed as
\begin{align*}
	\GM(\bold{F}, \boldsymbol{\theta}) &= \prod_{(t,n) \in \Srt} \left(\lvert H_n \rvert^2 B_t \right)^{\theta_{t,n}}
	\\
	&= \prod_{n = 1}^{N} \left(\lvert H_n \rvert^2 \right)^{q_n} \prod_{t = 1}^{T} B_t^{w_t}
	\\
	&= \GM(\bold{H}, \bold{q}) \GM(\bold{B}, \bold{w})
\end{align*}
Hence, we may express $\Laplace_{\GM(\bold{F}, \boldsymbol{\theta})}(s)$ as
\begin{align*}
	\Laplace_{\GM(\bold{F}, \boldsymbol{\theta})}(s) &= \E[\exp(-s\GM(\bold{H}, \bold{q}) \GM(\bold{B}, \bold{w})] 
	\\
	&= \E_{\GM(\bold{H}, \bold{q})}[\Laplace_{\GM(\bold{B}, \bold{w})}(s\GM(\bold{H}, \bold{q}))].
\end{align*}
The Laplace Transform of $\GM(\bold{B}, \bold{w})$ may be expressed as 
\begin{align*}
	\Laplace_{\GM(\bold{B}, \bold{w})}(s) &= \E \left[ \exp\left( \prod_{t = 1}^{T}B_t^{w_{t}} \right) \right]
	\\
	&= \sum_{r = 0}^{T} \pB^{T - r} (1- \pB)^r \sum_{j = 1}^{T \choose r} \exp\left( -s \prod_{i \in T_j^r} \xibt^{w_{t_i}} \right).
\end{align*}
Using this expression, we have that
\begin{align}
	\Laplace_{\GM(\bold{F}, \boldsymbol{\theta})}(s) = \sum_{r = 0}^{T} \pB^{T - r} (1- \pB)^r \sum_{j = 1}^{T \choose r} \Laplace_{\GM(\bold{H}, \bold{q})}\left( s    \prod_{i \in T_j^r} \xibt^{w_{t_i}} \right).
	\label{eq:F_LT} 
\end{align}
Thus, it remains to characterize the Laplace Transform of $\GM(\bold{H}, \bold{q})$. To this end we exploit Proposition 3 and obtain the Laplace Transform of $\GM(\bold{H}, \bold{q})$ via its Mellin Transform. 

The Mellin Transform of $\GM(\bold{H}, \bold{q})$ is 
\begin{align*}
	\Mellin_{\GM(\bold{H}, \bold{q})}(p) &= \E\left[ \left(\prod_{n = 1}^{N} (\lvert H_n \rvert^2)^{q_{n}} \right)^{p-1}\right]
	\\
	&\eqlabel{a} \prod_{n \in \Srt} \E\left[ (\lvert H_n \rvert^2)^{q_{n}(p-1)} \right]
	\\
	&\eqlabel{b} \prod_{n \in \Srt} \frac{\Na^{-q_{n}(p-1)}}{\Gamma(\Na)} \Gamma(q_{n}(p-1) + \Na)
	\\
	&= \frac{\Na^{-p+1}}{\Gamma(\Na)^{N}} \prod_{n = 1}^{N}  \Gamma(q_{n}(p-1) + \Na).
\end{align*}
Where (a) follows from the independence of $\{H_n\}_{n \in \Srt}$ and (b) follows from the Mellin Transform of a Gamma random variable. Thus, we have
\begin{align*}
	\Laplace_{\GM(\bold{H}, \bold{q})}(s) &= \Mellin^{-1}\left\{ \frac{\Na^{p}}{\Gamma(\Na)^{N}} \prod_{n \in \Srt}  \Gamma(1 - (1 - \Na) - q_{n}p) \Gamma(p) \right\}
	\\
	&\frac{1}{\Gamma(\Na)^{N}} H^{1, N_c}_{N_c, 1}\left(\frac{s}{\Na} \Bigg\vert \begin{matrix} &(1 - N_a, q_{1}) \dots (1 - N_a, q_{N}) \\&(0,1)\end{matrix} \right).
\end{align*}
Hence, the the exact expression follows using this expression in (\ref{eq:F_LT}). 

Part ii). We may obtain the first and second moments of $\GM(\bold{H}, \bold{q})$ using $\Mellin_{\GM(\bold{H}, \bold{q})}(2)$ and $\Mellin_{\GM(\bold{H}, \bold{q})}(3)$. Whence, part ii) follows by fitting $\GM(\bold{H}, \bold{q})$ to a gamma distribution with parameters shape parameter $\alpha_0(\bold{q}, Na)$ and scale parameter $\beta_0(\bold{q}, Na)$ and exploiting the Laplace Transform of a gamma random variable. $\qedwhite$

\subsection{Proof of Lemma 11}
We now prove the upper and lower bounds for the approximate Laplace Transform of $\GM(\bold{F}, \boldsymbol{\theta})$. To this end we require the following intermediate Lemma.

\textbf{Lemma E: (Bound on Gamma Laplace Transform using Gamma Mixture)} {\em Let $\alpha >0$, $\gamma >0$, $p \in (0, 1)$, and define the functions $f_1:[0,1] \rightarrow \R_+$ and $f_2:[0,1] \rightarrow \R_+$ as
\begin{align}
	f_1(x) = (1 + \gamma p^x)^{-\alpha}
	\\
	f_2(x) = (1 + \gamma p)^{-\alpha x}.
\end{align}
Then for all $x \in [0, 1]$ $f_1(x) \le f_2(x)$.}

{\em Proof:} Note that $f_1(0) = (1 + \gamma)^{-\alpha} < 1 = f_2(0)$, and $f_1(1) = (1 + \gamma p)^{-\alpha} = f_2(1)$. Hence the Lemma follows from the fact that
\begin{align*}
	&\frac{\partial f_1(x)}{\partial x} = \frac{\alpha \gamma p^x \log(p^{-1})}{(1 + \gamma p^x)^{\alpha + 1}} > 0
	\\ 
	&\frac{\partial f_2(x)}{\partial x} = \frac{-\alpha \log(1 + \gamma p)}{(1 + \gamma p)^{\alpha x}} < 0.
\end{align*}
$\qedwhite$

We first establish the upper bound. Consider the approximate form for the the Laplace Transform of $GM(\bold{F}, \boldsymbol{\theta})$.
\begin{align*}
	\Laplace_{\GM(\bold{F}, \boldsymbol{\theta})}(s) &\approx \sum_{r = 0}^{T} \pB^{T - r} (1- \pB)^r \sum_{j = 1}^{T \choose r} \left(1 + \frac{s \prod_{i \in T_j^r} \xibt^{w_{t_i}}}{\beta_0(\bold{q}, N_a)} \right)^{-\alpha_0(\bold{q}, Na)}
	\\
	&\lelabel{a}  \sum_{r = 0}^{T} \pB^{T - r} (1- \pB)^r \sum_{j = 1}^{T \choose r} \left( 1 + \frac{s   \xibt}{\beta_0(\bold{q}, N_a)} \right)^{-\alpha_0(\bold{q}, Na){\sum_{i \in T_j^r}w_{t_i}}}
	\\
	&= \sum_{r = 0}^{T} \pB^{T - r} (1- \pB)^r \sum_{j = 1}^{T \choose r} \prod_{i \in T_j^r}\left( 1 + \frac{s   \xibt}{\beta_0(\bold{q}, N_a)} \right)^{-\alpha_0(\bold{q}, Na){w_{t_i}}}
	\\
	&= \prod_{t = 1}^{T} \left(\pB + (1- \pB)\left( 1 + \frac{s   \xibt}{\beta_0(\bold{q}, N_a)} \right)^{-\alpha_0(\bold{q}, Na){w_{t}}}\right)
	\\
	&= \LaplaceU_{\GM(\bold{F}, \boldsymbol{\theta})}(s).
\end{align*}

Where (a) follows from Lemma E. 

Regarding the lower bound, we have
\begin{align*}
	\Laplace_{\GM(\bold{F}, \boldsymbol{\theta})}(s) &\approx \sum_{r = 0}^{T} \pB^{T - r} (1- \pB)^r \sum_{j = 1}^{T \choose r} \left(1 + \frac{s \prod_{i \in T_j^r} \xibt^{w_{t_i}}}{\beta_0(\bold{q}, N_a)} \right)^{-\alpha_0(\bold{q}, Na)}
	\\
	&\gelabel{a}  \sum_{r = 0}^{T} {T \choose r} \pB^{T - r} (1- \pB)^r \left( 1 + \frac{s  {T \choose r}^{-1} \sum_{j = 1}^{T \choose r}\xibt^{\sum_{i \in T_j^r}w_{t_i}}}{\beta_0(\bold{q}, \Na)} \right)^{-\alpha_0(\bold{q}, \Na)}.
	\\
	&= \LaplaceL_{\GM(\bold{F}, \boldsymbol{\theta})}(s).
\end{align*}
Where (a) follows from Jensen's Inequality. $\qedwhite$

\subsection{Proof of Lemma 12}
We now prove the lower bound for the  Laplace Transform of $\AM(\bold{F}, \boldsymbol{\theta})$. We first establish the following intermediate lemma

{\bf Lemma F: (Minimization of Product of Weighted Combination of Log Convex Functions)} {\em Let $f: \G \rightarrow \R_+$ be a log convex function and let $\{\theta_n\}_{n = 1}^{N}$ be a series of convex weights (where $N \in \N$). Then, for any $x \in \G$
\begin{align}
	\prod_{n = 1}^{N} f(\theta_n x) \ge f\left(\frac{x}{N}\right)^{N}
\end{align}}
{\em Proof:} Follows from a contradiction argument. Assume the claim doesn't hold and pick two weights $\theta_i < \frac{1}{N}$ and $\theta_j > \frac{1}{N}$. Construct new weights $\theta^{'}_i = \theta^{'}_j = \frac{\theta_i + \theta_j}{2}$. By log convexity, replacing $\theta_i$ and $\theta_j$ with $\theta^{'}_i$ and $\theta^{'}_j$ yields a lower bound. Repeating this process iteratively yields $\theta_i = \frac{1}{N}$ $\forall i \in [N]$. $\qedwhite$

Using Lemma F, we establish the lower bound in the following manner.
\begin{align}
	\Laplace_{\AM(\bold{F}, \boldsymbol{\theta})}(s) &= \E\left[\exp\left(-s \sum_{(t,n) \in \Srt} \theta_{t,n} \lvert H_n \rvert ^2 B_t\right) \right]
	\\
	&= \E \left[\E\left[\exp\left(-s \sum_{n = 1}^{N} \left( \sum_{(t,n) \in \Srt} \theta_{t,n}  B_t \right) \lvert H_n \rvert ^2 \right) \bigg\vert \{B_t\}_{t \in [T]} \right] \right]
	\\
	&= \E \left[\prod_{n = 1}^{N} \Laplace_{\lvert H \rvert^2}\left(s \left( \sum_{t = 1}^{T} w_t  B_t \right)\frac{\left( \sum_{(t,n) \in \Srt} \theta_{t,n}  B_t \right)}{\sum_{t = 1}^{T} w_t  B_t} \right) \right]
	\\
	&\gelabel{a} \E \left[\Laplace_{\lvert H \rvert^2}\left(s \frac{\sum_{t = 1}^{T} w_t  B_t }{N} \right)^{N} \right]
	\\
	&= \sum_{r = 0}^{T} \pB^{T - r} (1- \pB)^r \sum_{j = 1}^{T \choose r} \Laplace_{\lvert H \rvert^2}\left(s \frac{\left(\sum_{i \in T_j^r} w_t\right) ( \xibt - 1) + 1 }{N} \right)^{N}
	\\
	&\gelabel{b} \sum_{r = 0}^{T} {T \choose r} \pB^{T - r} (1- \pB)^r \left( 1 + \frac{s  {T \choose r}^{-1}\sum_{j = 1}^{T \choose r}\left((\xibt - 1){\sum_{i \in T_j^r}w_{t_i}} + 1\right)}{N \Na} \right)^{-N \Na}
	\\
	&= \LaplaceL_{\AM(\bold{F}, \boldsymbol{\theta})}(s).
\end{align}
Where (a) follows from Lemma F and (b) follows from Jensen's Inequality. $\qedwhite$

\subsection{Proof of Lemma 13}
We begin by establishing part i) where $N_a = 1 \implies \lvert H_n \rvert^2 \sim Exp(1)$ . Consider $\HM(\bold{F}, \boldsymbol{\theta})$
\begin{align*}
	\HM(\bold{F}, \boldsymbol{\theta}) &= \left( \sum_{(t,n) \in \Srt} \theta_{t,n}\left( \lvert H_n \rvert^2 B_t \right)^{-1} \right)^{-1}
	\\
	&\ge \left( \sum_{(t,n) \in \Srt} \theta_{t,n}\left( \lvert H_n \rvert^2 \min_{t} \{ B_t \} \right)^{-1} \right)^{-1}
	\\
	&\ge  \min_{t} \{ B_t \} \HM(\bold{H}, \bold{q}).
\end{align*}
Now, under the assumption that $\lvert H_n \rvert^2$ is an exponential random variable, it may equivalently be expressed as
\begin{align*}
	&\lvert H_n \rvert^2 = \frac{1}{2} X_n +  \frac{1}{2} Y_n &X_n, Y_n \sim {\rm Gamma}\left(\frac{1}{2}, \frac{1}{2}\right) \quad X_n \indep Y_n.
\end{align*}
Moreover, by concavity of $\HM(\bold{H}, \bold{q})$ we with respect to $\bold{H}$ we have
\begin{align*}
	&\HM(\bold{H}, \bold{q}) \ge \frac{1}{2} \HM(\bold{X}, \bold{q}) +  \frac{1}{2} \HM(\bold{Y}, \bold{q}).
\end{align*}
Now, by the definition of $\HM(\bold{X}, \bold{q})$, we have
\begin{align}
	HM(\bold{X}, \bold{q}) = \left( \sum_{n = 1}^{N} q_n X_n^{-1} \right)^{-1}
\end{align}
Since $X_n \sim {\rm  Gamma}(\frac{1}{2}, \frac{1}{2})$, we have that $X_n^{-1} \sim Levy(0, 1)$. Whence,
\begin{align}
	\sum_{n = 1}^{N} q_n X_n^{-1} \sim Levy\left(0, \left( \sum_{n = 1}^{N} q_n^{\frac{1}{2}} \right)^{2} \right).
\end{align}
Therefore, $\HM(\bold{X}, \bold{q}) \sim {\rm Gamma}\left(\frac{1}{2}, \frac{1}{2} \left( \sum_{n = 1}^{N} q_n^{\frac{1}{2}} \right)^{2}\right)$.

Putting these observations together, we have
\begin{align*}
	\Laplace_{\HM(\bold{F}, \boldsymbol{\theta})}(s) &= \E\left[ \exp\left( -s \HM(\bold{F}, \boldsymbol{\theta})  \right) \right]
	\\
	&\le  \E\left[ \exp\left( -s \min_{t} \{ B_t \} \left( \frac{1}{2} \HM(\bold{X}, \bold{q}) +  \frac{1}{2} \HM(\bold{Y}, \bold{q}) \right) \right) \right]
	\\
	&\eqlabel{a} \E\left[ \left( 1 + \frac{s \min_{t} \{ B_t \}}{\left( \sum_{n = 1}^{N} q_n^{\frac{1}{2}} \right)^{2}}  \right)^{-1} \right]
	\\
	&= \pB^{T}\left( 1 + \frac{s }{\left( \sum_{n = 1}^{N} q_n^{\frac{1}{2}} \right)^{2}}  \right)^{-1} + (1 - \pB^{T})\left( 1 + \frac{s \xibt}{\left( \sum_{n = 1}^{N} q_n^{\frac{1}{2}} \right)^{2}}  \right)^{-1}
	\\
	&= \LaplaceU_{\HM(\bold{F}, \boldsymbol{\theta})}(s).
\end{align*}
Where (a) follows from the Laplace Transform of the sum of two gamma random variables.

Part ii) leverages bounds on the first and second moments of $\HM(\bold{F}, \boldsymbol{\theta})$ in the case where $N_a > 1$. We establish these first before proceeding with the proof.

First, we establish a lower bound on the first moment.
\begin{align*}
\E\left[ \HM(\bold{F}, \boldsymbol{\theta}) \right] &= \E \left[ \left( \sum_{(t,n) \in \Srt} \theta_{t,n}\left( \lvert H_n \rvert^2 B_t \right)^{-1} \right)^{-1} \right]
\\
&\gelabel{a} \E \left[ \sum_{(t,n) \in \Srt} \theta_{t,n}\left( \lvert H_n \rvert^2 B_t \right)^{-1} \right]^{-1}
\\
&= \left( \E \left[\lvert H \rvert^{-2}\right] \E \left[B^{-1}\right]\right)^{-1}
\\
&= \left(\left(\pB + (1 - \pB)\xibt^{-1}\right)\frac{\Na}{\Na - 1}\right)^{-1}
\\
&= m_1.
\end{align*}
Where (a) follows from Jensen's Inequality. 

Next, we establish an upper bound on the second moment. First note that
\begin{align*}
	\HM(\bold{F}, \boldsymbol{\theta}) &= \left( \sum_{(t,n) \in \Srt} \theta_{t,n}\left( \lvert H_n \rvert^2 B_t \right)^{-1} \right)^{-1}
	\\
	&\lelabel{a} \left( \left( \sum_{n = 1}^{N} q_n \lvert H_n \rvert^{-2} \right) \left( \sum_{t = 1}^{T} w_n B_t^{-1} \right)  \right)^{-1}
	\\
	&= \HM(\bold{H}, \bold{q}) \HM(\bold{B}, \bold{w})
	\\
	&\lelabel{b} \GM(\bold{H}, \bold{q}) \HM(\bold{B}, \bold{w}).
\end{align*}
Where (a) follows from the FKG inequality \cite[Theorem 6.2.1]{Alon2016} and (b) follows from the AM-GM-HM Inequality. Thus, we may upper bound the second moment as
\begin{align*}
\E\left[ \HM(\bold{F}, \boldsymbol{\theta})^2 \right] &\le \E\left[ \GM(\bold{H}, \bold{q})^2 \right] \E\left[ \HM(\bold{B}, \bold{w})^2 \right]
\\
&= \left(\sum_{r = 0}^{T} \pB^{T - r} (1- \pB)^r \sum_{j = 1}^{T \choose r} \left( 1 - \left(\sum_{i \in T_j^r}w_{t_i}\right) +  \left(\sum_{i \in T_j^r}w_{t_i}\right) \xibt^{-1} \right)^{-2} \right) \Na^{-2}\prod_{n = 1}^{N} \frac{\Gamma(2q_n + \Na)}{\Gamma(\Na)}
\\
&= m_2.
\end{align*}

To prove part ii), we employ the following upper bound on the Laplace Transform for a non-negative random variable for $s \in \R_+$. The bound is taken from \cite{Eckberg77} and follows along the lines of Lemma C. The bound is given as
\begin{align}
	\Laplace_{\HM(\bold{F}, \boldsymbol{\theta})}(s;N_a) \le \frac{\E\left[ \HM(\bold{F}, \boldsymbol{\theta})^2 \right] - \E\left[ \HM(\bold{F}, \boldsymbol{\theta}) \right]^2}{\E\left[ \HM(\bold{F}, \boldsymbol{\theta})^2 \right]} + \frac{\E\left[ \HM(\bold{F}, \boldsymbol{\theta}) \right]^2}{\E\left[ \HM(\bold{F}, \boldsymbol{\theta})^2 \right]} \exp\left(-s\frac{\E\left[ \HM(\bold{F}, \boldsymbol{\theta})^2 \right]}{\E\left[ \HM(\bold{F}, \boldsymbol{\theta}) \right]} \right).
\end{align}
One may show that the bound is decreasing in $\E\left[ \HM(\bold{F}, \boldsymbol{\theta}) \right]$ and increasing in $\E\left[ \HM(\bold{F}, \boldsymbol{\theta})^2 \right]$, whence
\begin{align*}
	\Laplace_{\HM(\bold{F}, \boldsymbol{\theta})}(s;N_a) &\le \frac{m_2 - m_1^2}{m_2} + \frac{m_1^2}{m_2} \exp\left(-s\frac{m_2}{m_1} \right)
	\\
	&= \LaplaceU_{\HM(\bold{F}, \boldsymbol{\theta})}(s;N_a) 
\end{align*}

$\qedwhite$

\section{Proofs of Some Technical Results}
\label{app:tech_res}
\subsection{Proof of Lemma B}
\textbf{Lemma B: (A H{\"o}lder-Like Inequality for Harmonic Means of Sets of Functions)} {\em Let $p$ be a PMF on $[n]$ for some $n \in \N$, and let $(\G, \mathcal{G}, \mu)$ be a measure space. Let $\{f_k\}_{k \in [n]}$ be a sequence measurable functions from $\G$ to $\R_+$. Then
\begin{align}
	\norm{\left(\sum_{k = 1}^n p_k f_k^{-1}\right)^{-1}}_1 \le \left(\sum_{k = 1}^{n} p_k\norm{f_k}_{1}^{-1}\right)^{-1}.
\end{align}
Or, alternatively
\begin{align}
	\norm{HM(\{f_k\}_{k \in [n]}, p)}_1 \le HM(\{\norm{f_k}_{1}\}_{k \in [n]}, p).
\end{align}
Where the above norms are taken with respect to $\mu$.

Moreover, equality is obtained iff $\{f_k\}_{k \in [n]}$ are linearly dependent. }

{\em Proof:} We shall first consider the case where $n = 2$ and then prove the general case by induction. Hence, with some abuse of notation, we shall take our PMF to be $(p, 1-p)$.

Before proceeding, we first deal with the cases where $\norm{f_k}_1 \in \{0, \infty\}$ for some $k \in \{1, 2\}$. If $\norm{f_1}_1 = 0$ then $f_1(x) = 0$ $\mu$-almost surely, and
\begin{align*}
	&HM(\{f_1(x), f_2(x)\}, (p, 1-p)) = 0 &\mu-a.s.,
	\\
	&HM(\{\norm{f_1}_{1}, \norm{f_2}_{1}\}, (p, 1-p)) = 0
\end{align*}
A similar argument may be made for the case where $\norm{f_2}_1 = 0$.

If both $\norm{f_1}_1$ and $\norm{f_2}_1$ are infinite, then $HM(\{\norm{f_1}_1, \norm{f_2}_1\}, (p, 1-p)) = \infty$, in which case the bound is trivial. 

Having established these results, we may assume that at least one  of $\norm{f_k}_1 \in (0, \infty)$ for $k \in \{1, 2\}$. Without loss of generality, we shall assume $\norm{f_1}_1 \in (0, \infty)$.  

First, consider the function $g:\R_+ \rightarrow \R_+$ such that
\begin{align}
	g(x) = \left( p  + (1-p) x^{-1}\right)^{-1}.
\end{align}
One may confirm that for all $p \in (0, 1)$ the second derivative of $g$ is strictly negative, and equal to zero for $p \in \{0, 1\}$. Thus,  by second order concavity conditions, $g$ is strictly concave for $p \in (0,1)$. We then obtain
\begin{align*}
	\norm{HM(\{f_k\}_{k \in [2]}, (p, 1-p))}_1 &= \int_{\G}\left( p f_1(x)^{-1} + (1-p) f_2(x)^{-1}\right)^{-1} \mu(dx)
	\\
	&= \norm{f_1}_1 \int_{\G}\left( p  + (1-p) \frac{f_1(x)}{f_2(x)}\right)^{-1} \frac{f_1(x)\mu(dx)}{\norm{f_1}_1}
	\\
	&= \norm{f_1}_1 \int_{\G}g\left(\frac{f_1(x)}{f_2(x)}\right) \frac{f_1(x)\mu(dx)}{\norm{f_1}_1}
	\\
	&\lelabel{a} \norm{f_1}_1 g\left(\int_{\G} \frac{f_2(x)}{f_1(x)}\frac{f_1(x)\mu(dx)}{\norm{f_1}_1}\right)
	\\
	&= HM(\{\norm{f_1}_1, \norm{f_2}_1\}, (p, 1-p)).
\end{align*}
Where (a) follows from Jensen's Inequality and noting that $ \nu = f_1 \mu /\norm{f_1}_1$ is a probability measure. Therefore the inequality holds for the case where $n = 2$. Moreover, equality is obtained if and only if $f_1 /f_2$ is $\nu$-almost-surely a constant. Equivalently, we may say that equality is obtained if and only if $f_1$ and $f_2$ are linearly dependent $\mu$-almost surely.

We now show the general case for $n \in \N$ by induction. Assume that
\begin{align*}
	\norm{HM(\{f_k\}_{k \in [n]}, p)}_1 \le HM(\{\norm{f_k}_{1}\}_{k \in [n]}, p).
\end{align*}
We will then show it holds for $n+1$ functions. Let $\{f_{k}\}_{k \in [n+1]}$ be an ensemble of measurable functions from $\G$ to $\R_+$ and let $p$ be a PMF on $[n+1]$. Define the PMF $q$ and $[n]$ as follows
\begin{align*}
	q_k = \begin{cases}
	p_k &k \in [n-1]
	\\
	p_n + p_{n+1}	&k = n.
 \end{cases}
\end{align*}
Further, define the function $g_n = HM(\{f_n, f_{n+1}\}, (p_n/q_n, p_{n+1}/q_n))$. Then by the induction hypothesis
\begin{align*}
\norm{HM(\{f_k\}_{k \in [n+1]}, p)}_1 &= \norm{HM(\{f_1, \dots, f_{n-1}, g_n\}, q)}_1
	\\
	&\lelabel{a} HM(\{\norm{f_1}_1, \dots, \norm{f_{n-1}}_1, \norm{g_n}_1\}, q)
	\\
	&= \left(\sum_{k = 1}^{n-1} p_k \norm{f_k}_1^{-1} + q_n\norm{g_n}_1^{-1}\right)^{-1}
	\\
	&\lelabel{b} \left(\sum_{k = 1}^{n-1} p_k \norm{f_k}_1^{-1} + q_n\left(\frac{p_n}{q_n}\norm{f_n}_1^{-1} + \frac{p_{n+1}}{q_n}\norm{f_{n+1}}_1^{-1} \right)\right)^{-1}
	\\
	&= HM(\{\norm{f_k}_1\}_{k \in [n+1]}, p).
\end{align*}
Where (a) follows from the induction hypothesis, and (b) follows from applying the induction hypothesis to $g_n$. $\qedwhite$

\subsection{Proof of Proposition 3}

We prove Proposition 3 in the case where $X$ admits a PDF, which is stated separately in the following lemma.

{\bf Lemma G: (Relation between Mellin and Laplace Transforms)} {\em Let $f \in L^2_{\C}(\R_+)$. Then
\begin{align}
	\Laplace_f(s) = \Mellin^{-1}\{\Mellin_f(1-p)\Gamma(p)\}(s).
\end{align}}
{\em Proof:} Using the fact that the family $\{E_\beta(r) = r^{-2\pi j \beta - \frac{1}{2}}\}_{\beta \in \R}$ form an orthonormal basis on $L^2_{\C}(\R_+)$ \cite[Theorem-11.3.1.1]{bertrand:hal-03152634} we may express $f$ as
\begin{align}
	f(r) = \int_{-\infty}^{\infty} \Mellin_f\left(2 \pi j \beta + \frac{1}{2}\right) r^{-2 \pi j \beta - \frac{1}{2}} d\beta. 
\end{align}
Hence, we may express the Laplace Transform of $f$ as 
\begin{align*}
	\Laplace_f(s) &= \int_{0}^{\infty} f(r) e^{-sr}dr
	\\
	&= \int_{0}^{\infty} \int_{-\infty}^{\infty} \Mellin_f\left(2 \pi j \beta + \frac{1}{2}\right) r^{-2 \pi j \beta - \frac{1}{2}}e^{-sr} d\beta  dr
	\\
	&\eqlabel{a}  \int_{-\infty}^{\infty} \Mellin_f\left(2 \pi j \beta + \frac{1}{2}\right) \int_{0}^{\infty}  r^{-2 \pi j \beta + \frac{1}{2} - 1} e^{-sr}  dr d\beta.
\end{align*}
Where (a) follows from Fubini's theorem. Now, note that 
\begin{align}
	&\int_{0}^{\infty}  r^{-2 \pi j \beta + \frac{1}{2} - 1} e^{-sr}  dr = s^{-2 \pi j \beta - \frac{1}{2}} \Gamma\left(\frac{1}{2} -2 \pi j \beta\right) &\Re\{s\} \ge 0.
\end{align}

Therefore, for $\Re\{s\} \ge 0$ we have that $\Laplace_f(s) = \hat{\Laplace}_f(s)$, where
\begin{align*}
	\hat{\Laplace}_f(s) &= \int_{-\infty}^{\infty} \Mellin_f\left(2 \pi j \beta + \frac{1}{2}\right) \Gamma\left(\frac{1}{2} -2 \pi j \beta\right) s^{-2 \pi j \beta - \frac{1}{2}}  d\beta
	\\
	&= \int_{-\infty}^{\infty} \Mellin_f\left(1 - 2 \pi j \beta - \frac{1}{2}\right) \Gamma\left(2 \pi j \beta + \frac{1}{2}\right) s^{-2 \pi j \beta - \frac{1}{2}}  d\beta.
\end{align*}

Finally, note that $\hat{\Laplace}_f(s)$ is analytic, and we may define an analytic continuation for $\Re\{s\} < 0$ using the inverse Mellin Transform with respect to the Mellin-Barnes Integral:
\begin{align*}
	\tilde{\Laplace}_f(s) &= \oint \Mellin_f\left(1 - p\right) \Gamma\left(p\right) s^{-p}  dp
	\\
	&\Mellin^{-1}\{\Mellin_f(1-p)\Gamma(p)\}(s)
\end{align*}
On the other hand $\Laplace_f(s)$ is also an analytic continuation of $\hat{\Laplace}_f(s)$, whence, by uniqueness of analytic continuation $\Laplace_f(s) = \tilde{\Laplace}_f(s)$ for all $s \in \C$. $\qedwhite$

We may extend Lemma G to encompass general probability measures over $(\R_+, \mathcal{B}(\R_+))$  using methods referenced in \cite{bertrand:hal-03152634}.

{\bf Proposition 3: (Relation between Mellin and Laplace Transforms of Nonnegative Random Variables)} {\em Let $X$ be a non-negative random variable. Then
\begin{align}
	\Laplace_X(s) = \Mellin^{-1}\{\Mellin_X(1-p)\Gamma(p)\}(s).
\end{align}}

\nocite{}
\bibliographystyle{IEEEtran}
\bibliography{ccs_refs.bib}

\begin{thebibliography}{10}
\providecommand{\url}[1]{#1}
\csname url@samestyle\endcsname
\providecommand{\newblock}{\relax}
\providecommand{\bibinfo}[2]{#2}
\providecommand{\BIBentrySTDinterwordspacing}{\spaceskip=0pt\relax}
\providecommand{\BIBentryALTinterwordstretchfactor}{4}
\providecommand{\BIBentryALTinterwordspacing}{\spaceskip=\fontdimen2\font plus
\BIBentryALTinterwordstretchfactor\fontdimen3\font minus
  \fontdimen4\font\relax}
\providecommand{\BIBforeignlanguage}[2]{{%
\expandafter\ifx\csname l@#1\endcsname\relax
\typeout{** WARNING: IEEEtran.bst: No hyphenation pattern has been}%
\typeout{** loaded for the language `#1'. Using the pattern for}%
\typeout{** the default language instead.}%
\else
\language=\csname l@#1\endcsname
\fi
#2}}
\providecommand{\BIBdecl}{\relax}
\BIBdecl

\bibitem{DeLima2021}
C.~De~Lima, D.~Belot \emph{et~al.}, ``{Convergent Communication, Sensing and
  Localization in 6G Systems: An Overview of Technologies, Opportunities and
  Challenges},'' \emph{IEEE Access}, vol.~9, pp. 26\,902--26\,925, Jan. 2021.

\bibitem{Liu2021}
F.~Liu, Y.~Cui \emph{et~al.}, ``{Integrated Sensing and Communications: Towards
  Dual-functional Wireless Networks for 6G and Beyond},'' Mar. 2021.

\bibitem{Kumari2021}
P.~Kumari, N.~J. Myers, and R.~W. Heath, ``Adaptive and fast combined
  waveform-beamforming design for mmwave automotive joint
  communication-radar,'' \emph{IEEE Journal of Selected Topics in Signal
  Processing}, vol.~15, no.~4, pp. 996--1012, 2021.

\bibitem{Husheng22}
H.~Li, ``Mac scheduling in joint communications and sensing networks based on
  virtual queues,'' in \emph{GLOBECOM 2022 - 2022 IEEE Global Communications
  Conference}, 2022, pp. 4069--4074.

\bibitem{Wild2021}
T.~Wild, V.~Braun, and H.~Viswanathan, ``{Joint Design of Communication and
  Sensing for Beyond 5G and 6G Systems},'' \emph{IEEE Access}, vol.~9, pp.
  30\,845--30\,857, Feb. 2021.

\bibitem{Nguyen2021}
N.~C. Luong, X.~Lu \emph{et~al.}, ``{Radio Resource Management in Joint Radar
  and Communication: A Comprehensive Survey},'' \emph{IEEE Communications
  Surveys Tutorials}, vol.~23, no.~2, pp. 780--814, April 2021.

\bibitem{Zhang2021}
A.~Zhang, M.~L. Rahman \emph{et~al.}, ``{Perceptive Mobile Networks: Cellular
  Networks With Radio Vision via Joint Communication and Radar Sensing},''
  \emph{IEEE Vehicular Technology Magazine}, vol.~16, no.~2, pp. 20--30, Jun.
  2021.

\bibitem{Kumari2018}
P.~Kumari, J.~Choi, N.~Gonz{\'a}lez-Prelcic, and R.~W. Heath, ``{IEEE
  802.11ad-Based Radar: An Approach to Joint Vehicular Communication-Radar
  System},'' \emph{IEEE Transactions on Vehicular Technology}, vol.~67, no.~4,
  pp. 3012--3027, Nov. 2018.

\bibitem{Andrews2011}
J.~G. {Andrews}, F.~{Baccelli}, and R.~K. {Ganti}, ``{A Tractable Approach to
  Coverage and Rate in Cellular Networks},'' \emph{IEEE Transactions on
  Communications}, vol.~59, no.~11, pp. 3122--3134, October 2011.

\bibitem{Bai2015}
T.~{Bai} and R.~W. {Heath}, ``{Coverage and Rate Analysis for Millimeter-Wave
  Cellular Networks},'' \emph{IEEE Transactions on Wireless Communications},
  vol.~14, no.~2, pp. 1100--1114, February 2015.

\bibitem{Ping2018}
P.~Ren, A.~Munari, and M.~Petrova, ``{Performance Tradeoffs of Joint
  Radar-Communication Networks},'' \emph{IEEE Wireless Communications Letters},
  vol.~8, no.~1, pp. 165--168, Aug. 2019.

\bibitem{Ren2020}
------, ``{Performance Analysis of a Time-sharing Joint Radar-Communications
  Network},'' in \emph{2020 International Conference on Computing, Networking
  and Communications (ICNC)}, Mar. 2020, pp. 908--913.

\bibitem{Ram2022}
S.~{Sundar Ram} and G.~{Ghatak}, ``{Optimization of Network Throughput of Joint
  Radar Communication System Using Stochastic Geometry},'' \emph{arXiv
  e-prints}, p. arXiv:2201.03221, Jan. 2022.

\bibitem{Ghozlani2021}
D.~Ghozlani, A.~Omri \emph{et~al.}, ``{Stochastic Geometry-based Analysis of
  Joint Radar and Communication-Enabled Cooperative Detection Systems},'' in
  \emph{2021 17th International Conference on Wireless and Mobile Computing,
  Networking and Communications (WiMob)}, Nov. 2021, pp. 325--330.

\bibitem{Fang2019}
Z.~Fang, Z.~Wei \emph{et~al.}, ``{Performance of Joint Radar and Communication
  Enabled Cooperative Detection},'' in \emph{2019 IEEE/CIC International
  Conference on Communications in China (ICCC)}, Oct. 2019, pp. 753--758.

\bibitem{Munari2018}
A.~Munari, S.~Ljiljana, and M.~Petrova, ``{Stochastic Geometry Interference
  Analysis of Radar Network Performance},'' \emph{IEEE Communications Letters},
  vol.~22, no.~11, pp. 2362--2365, Sept. 2018.

\bibitem{Park2018}
J.~Park and R.~W. Heath, ``{Analysis of Blockage Sensing by Radars in Random
  Cellular Networks},'' \emph{IEEE Signal Processing Letters}, vol.~25, no.~11,
  pp. 1620--1624, Sept. 2018.

\bibitem{Skouroumounis2021}
C.~Skouroumounis, C.~Psomas, and I.~Krikidis, ``{FD-JCAS Techniques for mmWave
  HetNets: Ginibre Point Process Modeling and Analysis},'' \emph{IEEE
  Transactions on Mobile Computing}, pp. 1--1, Apr. 2021.

\bibitem{Woddward1951}
P.~M. Woodward, ``Information theory and the design of radar receivers,''
  \emph{Proceedings of the IRE}, vol.~39, no.~12, pp. 1521--1524, Dec. 1951.

\bibitem{Bell1993}
M.~Bell, ``{Information theory and radar waveform design},'' \emph{IEEE
  Transactions on Information Theory}, vol.~39, no.~5, pp. 1578--1597, Sept.
  1993.

\bibitem{Bliss2014}
D.~W. Bliss, ``{Cooperative Radar and Communications Signaling: The Estimation
  and Information theory Odd Couple},'' in \emph{2014 IEEE Radar Conference},
  Aug. 2014, pp. 0050--0055.

\bibitem{Paul2015}
B.~Paul and D.~W. Bliss, ``{Extending joint radar-communications bounds for
  FMCW radar with Doppler estimation},'' in \emph{2015 IEEE Radar Conference
  (RadarCon)}, Jun. 2015, pp. 0089--0094.

\bibitem{Chryath2017}
A.~R. Chiriyath, B.~Paul, and D.~W. Bliss, ``{Radar-Communications Convergence:
  Coexistence, Cooperation, and Co-Design},'' \emph{IEEE Transactions on
  Cognitive Communications and Networking}, vol.~3, no.~1, pp. 1--12, Feb.
  2017.

\bibitem{Ouyang23}
C.~Ouyang, Y.~Liu, H.~Yang, and N.~Al-Dhahir, ``Integrated sensing and
  communications: A mutual information-based framework,'' \emph{IEEE
  Communications Magazine}, vol.~61, no.~5, pp. 26--32, 2023.

\bibitem{Bolcskei2002}
H.~Bolcskei, D.~Gesbert, and A.~Paulraj, ``{On the capacity of OFDM-based
  spatial multiplexing systems},'' \emph{IEEE Transactions on Communications},
  vol.~50, no.~2, pp. 225--234, 2002.

\bibitem{Ahsan2019}
A.~Ali, R.~Vesilo, and M.~Di~Renzo, ``{Stochastic Geometry Analysis of
  Multi-User Asynchronous OFDM Wireless Networks},'' \emph{IEEE Wireless
  Communications Letters}, vol.~8, no.~3, pp. 845--848, 2019.

\bibitem{Lin2015}
X.~Lin, L.~Jiang, and J.~G. Andrews, ``Performance analysis of asynchronous
  multicarrier wireless networks,'' \emph{IEEE Transactions on Communications},
  vol.~63, no.~9, pp. 3377--3390, 2015.

\bibitem{baccelliSG}
\BIBentryALTinterwordspacing
F.~Baccelli, B.~B{\l}aszczyszyn, and M.~Karray, \emph{{Random Measures, Point
  Processes, and Stochastic Geometry}}.\hskip 1em plus 0.5em minus 0.4em\relax
  {Inria}, Jan. 2020. [Online]. Available:
  \url{https://hal.inria.fr/hal-02460214}
\BIBentrySTDinterwordspacing

\bibitem{Bai2014}
T.~{Bai}, R.~{Vaze}, and R.~W. {Heath}, ``{Analysis of Blockage Effects on
  Urban Cellular Networks},'' \emph{IEEE Transactions on Wireless
  Communications}, vol.~13, no.~9, pp. 5070--5083, Sept. 2014.

\bibitem{Journet2011}
J.~M. {Jornet} and I.~F. {Akyildiz}, ``{Channel Modeling and Capacity Analysis
  for Electromagnetic Wireless Nanonetworks in the Terahertz Band},''
  \emph{IEEE Transactions on Wireless Communications}, vol.~10, no.~10, pp.
  3211--3221, October 2011.

\bibitem{Canan2020}
C.~Aydogdu, M.~F. Keskin \emph{et~al.}, ``{Radar Interference Mitigation for
  Automated Driving: Exploring Proactive Strategies},'' \emph{IEEE Signal
  Processing Magazine}, vol.~37, no.~4, pp. 72--84, Jun. 2020.

\bibitem{Haenggi2012}
M.~Haenggi, \emph{{Stochastic Geometry for Wireless Networks}}, 1st~ed.\hskip
  1em plus 0.5em minus 0.4em\relax USA: Cambridge University Press, 2012.

\bibitem{Braun2014ofdm}
\BIBentryALTinterwordspacing
K.~Braun and F.~Jondral, \emph{{OFDM Radar Algorithms in Mobile Communication
  Networks}}, ser. Forschungsberichte aus dem Institut f{\"u}r
  Nachrichtentechnik des Karlsruher Instituts f{\"u}r Technologie.\hskip 1em
  plus 0.5em minus 0.4em\relax KIT-Bibliothek, Jan. 2014. [Online]. Available:
  \url{https://books.google.com/books?id=8FXVzQEACAAJ}
\BIBentrySTDinterwordspacing

\bibitem{Hmamouche2021}
Y.~Hmamouche, M.~Benjillali \emph{et~al.}, ``{New Trends in Stochastic Geometry
  for Wireless Networks: A Tutorial and Survey},'' \emph{Proceedings of the
  IEEE}, vol. 109, no.~7, pp. 1200--1252, Jul. 2021.

\bibitem{Roberts22}
\BIBentryALTinterwordspacing
I.~P. Roberts, S.~Vishwanath, and J.~G. Andrews, ``{LoneSTAR: Analog
  Beamforming Codebooks for Full-Duplex Millimeter Wave Systems},'' Jun. 2022.
  [Online]. Available: \url{https://arxiv.org/abs/2206.11418}
\BIBentrySTDinterwordspacing

\bibitem{Olson2022}
N.~R. Olson, J.~G. Andrews, and R.~W. Heath, ``{Coverage and Capacity of
  Terahertz Cellular Networks With Joint Transmission},'' \emph{IEEE
  Transactions on Wireless Communications}, vol.~21, no.~11, pp. 9865--9878,
  2022.

\bibitem{Rappaport2014}
M.~R. Akdeniz, Y.~Liu \emph{et~al.}, ``{Millimeter Wave Channel Modeling and
  Cellular Capacity Evaluation},'' \emph{IEEE Journal on Selected Areas in
  Communications}, vol.~32, no.~6, pp. 1164--1179, June 2014.

\bibitem{ITUR2013}
{International Telecommunications Union Radiocommunication Sector (ITU-R)},
  ``{Recommendation ITU-R P.676-10 Attenuation by Atmospheric Gases},'' ITU-R,
  Tech. Rep., Sept. 2013.

\bibitem{Petrov2017}
V.~Petrov, M.~Komarov \emph{et~al.}, ``{Interference and SINR in Millimeter
  Wave and Terahertz Communication Systems With Blocking and Directional
  Antennas},'' \emph{IEEE Transactions on Wireless Communications}, vol.~16,
  no.~3, pp. 1791--1808, Jan. 2017.

\bibitem{dahlman2021}
E.~Dahlman, S.~Parkvall, and J.~Sk{\"o}ld, \emph{{5G NR The Next Generation
  Wireless Access Technology}}.\hskip 1em plus 0.5em minus 0.4em\relax London:
  Elsevier, 2021.

\bibitem{AlAmmouri2019}
A.~AlAmmouri, J.~G. Andrews, and F.~Baccelli, ``{A Unified Asymptotic Analysis
  of Area Spectral Efficiency in Ultradense Cellular Networks},'' \emph{IEEE
  Transactions on Information Theory}, vol.~65, no.~2, pp. 1236--1248, Jun.
  2019.

\bibitem{Mahler2013}
R.~Mahler, ``{``Statistics 102'' for Multisource-Multitarget Detection and
  Tracking}, year={2013},'' \emph{IEEE Journal of Selected Topics in Signal
  Processing}, vol.~7, no.~3, pp. 376--389.

\bibitem{Clark2021}
D.~E. Clark, ``{Multi-Sensor Network Information for Linear-Gaussian
  Multi-Target Tracking Systems},'' \emph{IEEE Transactions on Signal
  Processing}, vol.~69, pp. 4312--4325, 2021.

\bibitem{VanTrees2006}
H.~L. Van~Trees, K.~L. Bell, and Y.~Wang, ``{Bayesian Cramer-Rao bounds for
  multistatic radar},'' in \emph{2006 International Waveform Diversity {\&}
  Design Conference}, 2006, pp. 1--4.

\bibitem{Ohto2017}
T.~Ohto, K.~Yamamoto \emph{et~al.}, ``{Stochastic Geometry Analysis of
  Normalized SNR-Based Scheduling in Downlink Cellular Networks},'' \emph{IEEE
  Wireless Communications Letters}, vol.~6, no.~4, pp. 438--441, 2017.

\bibitem{Strum2011}
C.~Sturm and W.~Wiesbeck, ``{Waveform Design and Signal Processing Aspects for
  Fusion of Wireless Communications and Radar Sensing},'' \emph{Proceedings of
  the IEEE}, vol.~99, no.~7, pp. 1236--1259, May 2011.

\bibitem{Graff2021}
A.~M. Graff, W.~N. Blount \emph{et~al.}, ``{Analysis of OFDM Signals for
  Ranging and Communications},'' in \emph{{Proceedings of the 34th
  International Technical Meeting of the Satellite Division of The Institute of
  Navigation (ION GNSS+ 2021)}}, Sept. 2021, pp. 2910--2924.

\bibitem{Griffiths2015}
H.~Griffiths, L.~Cohen \emph{et~al.}, ``{Radar Spectrum Engineering and
  Management: Technical and Regulatory Issues},'' \emph{Proceedings of the
  IEEE}, vol. 103, no.~1, pp. 85--102, 2015.

\bibitem{Gaudio2019}
L.~Gaudio, M.~Kobayashi, B.~Bissinger, and G.~Caire, ``{Performance Analysis of
  Joint Radar and Communication using OFDM and OTFS},'' in \emph{{2019 IEEE
  International Conference on Communications Workshops (ICC Workshops)}}, Jul.
  2019, pp. 1--6.

\bibitem{Payan2019}
J.~Payan, C.~Jauffret, and A.-C. P{\'e}rez, ``{Maximization of the Fisher
  Information in PDA},'' in \emph{2019 IEEE 8th International Workshop on
  Computational Advances in Multi-Sensor Adaptive Processing (CAMSAP)}, 2019,
  pp. 11--15.

\bibitem{van2013detection}
\BIBentryALTinterwordspacing
H.~Van~Trees, Z.~Tian, and K.~Bell, \emph{{Detection Estimation and Modulation
  Theory, Part I: Detection, Estimation, and Filtering Theory}}, ser. Detection
  Estimation and Modulation Theory.\hskip 1em plus 0.5em minus 0.4em\relax
  Wiley, 2013. [Online]. Available:
  \url{https://books.google.com/books?id=dnvaxqHDkbQC}
\BIBentrySTDinterwordspacing

\bibitem{Eckberg77}
\BIBentryALTinterwordspacing
A.~E. Eckberg, ``{Sharp Bounds on Laplace-Stieltjes Transforms, with
  Applications to Various Queueing Problems},'' \emph{Mathematics of Operations
  Research}, vol.~2, no.~2, pp. 135--142, May 1977. [Online]. Available:
  \url{http://www.jstor.org/stable/3689650}
\BIBentrySTDinterwordspacing

\bibitem{Olson2021}
N.~R. Olson, J.~G. Andrews, and R.~W. Heath, ``{Single Channel Equivalent Point
  Processes of Poisson Networks With Multiple Channel Laws},'' \emph{IEEE
  Communications Letters}, vol.~26, no.~3, pp. 711--715, Dec. 2022.

\bibitem{Alon2016}
N.~Alon and J.~H. Spencer, \emph{{The Probabilistic Method}}.\hskip 1em plus
  0.5em minus 0.4em\relax John Wiley \& Sons, 2016.

\bibitem{bertrand:hal-03152634}
\BIBentryALTinterwordspacing
J.~Bertrand, P.~Bertrand, and J.-P. Ovarlez, ``{The Mellin Transform},'' in
  \emph{{The Transforms and Applications Handbook}}, Aug. 1995. [Online].
  Available: \url{https://hal.archives-ouvertes.fr/hal-03152634}
\BIBentrySTDinterwordspacing

\end{thebibliography}

\end{document}